\documentclass[a4paper,11pt]{article}
\usepackage{jheppub}
\pdfoutput=1
\usepackage[most]{tcolorbox}
\newtcbox{\mymath}[1][]{%
    nobeforeafter, math upper, tcbox raise base,
    enhanced, colframe=blue!30!black,
    colback=blue!30, boxrule=1pt,
    #1}

\usepackage{blindtext}
\usepackage{braket}
\usepackage[normalem]{ulem}
\usepackage[most]{tcolorbox}

\def\lsim{\raise0.3ex\hbox{$\;<$\kern-0.75em\raise-1.1ex\hbox{$\sim\;$}}}
\def\gsim{\raise0.3ex\hbox{$\;>$\kern-0.75em\raise-1.1ex\hbox{$\sim\;$}}}

\def\bea{\begin{eqnarray}}
\def\eea{\end{eqnarray}}

\newcommand{\ogw}{\Omega_\text{GW}}
\newcommand{\ahcRD}{a^\text{hc}_{\text{RD}}}
\newcommand{\ahcRH}{a^\text{hc}_{\text{RH}}}
\newcommand{\hhcRD}{H^\text{hc}_{\text{RD}}}
\newcommand{\hhcRH}{H^\text{hc}_{\text{RH}}}

\newcommand{\hRH}{H_\text{rh}}
\newcommand{\rSM}{\rho_\text{SM}}

\newcommand{\arh}{a_\text{rh}}
\newcommand{\gss}{g_{\star s}}
\newcommand{\gsr}{g_{\star\rho}}
\newcommand{\Trh}{T_\text{rh}}
\newcommand{\Tmax}{T_\text{max}}

\newcommand{\gvp}{g_{V\phi}}
\newcommand{\gap}{h_{a\phi}}
\newcommand{\fmax}{f_\text{max}}

\newcommand{\mdm}{m_\text{DM}}
\newcommand{\ndm}{n_\text{DM}}
\newcommand{\adm}{a_\text{DM}}
\newcommand{\lNP}{\Lambda_\text{UV}}

\newcommand{\DNeff}{\Delta N_\text{eff}}
\newcommand{\rGW}{\rho_\text{GW}}

\newcommand{\gsp}{g_{S\phi}}
\newcommand{\gpp}{g_{\psi\phi}}
\newcommand{\hap}{h_{a\phi}}

\newcommand{\ahc}{a^\text{hc}}

\def\gev{\;\hbox{GeV}}

\newcommand{\rhosm}{\rho_{\text{\tiny SM}}}

\newcommand{\nc}{\newcommand}
\nc{\non}{\nonumber}
\nc{\noi}{\noindent}
\nc{\hsp}{\hspace{0.5cm}}
\nc{\lsp}{\hspace{1cm}}
\nc{\Lsp}{\hspace{2cm}}
\nc{\LLsp}{\lsp\lsp}
\nc{\with}{\hsp\text{with}\hsp}
\begin{document}
%%%%%%%%%%%%%%%%%%
\title{Measuring Inflaton Couplings via Primordial Gravitational Waves}
%%%%%%%%%%%%%%%%%%%%%%%%%%%%%%%
\author[a]{Basabendu Barman,}
\author[a]{Anish Ghoshal,}
\author[a]{Bohdan Grzadkowski,}
\author[a]{and Anna Socha}

\affiliation[a]{Institute of Theoretical Physics, Faculty of Physics, University of Warsaw\\
ul. Pasteura 5, 02-093 Warsaw, Poland}

\emailAdd{basabendu88barman@gmail.com}
\emailAdd{anish.ghoshal@fuw.edu.pl}
\emailAdd{bohdan.grzadkowski@fuw.edu.pl}
\emailAdd{anna.socha@fuw.edu.pl}
%%%%%%%%%%%%%%%%%
\abstract{
We investigate the reach of future gravitational wave (GW) detectors in probing inflaton couplings with visible sector particles that can either be bosonic or fermionic in nature. Assuming reheating takes place through perturbative quantum production from vacuum in presence of classical inflaton background field, we find that the spectral energy density of the primordial GW generated during inflation becomes sensitive to inflaton-matter coupling. We conclude, obeying bounds from Big Bang Nucleosysthesis and Cosmic Microwave Background, that, e.g., inflaton-scalar couplings of the order of $\sim\mathcal{O}(10^{-20})$~GeV fall within the sensitivity range of several proposed GW detector facilities. However, this prediction is sensitive to the size of the inflationary scale, nature of the inflaton-matter interaction and shape of the potential during reheating. Having found the time-dependent effective inflaton decay width, we also discuss its implications for dark matter (DM) production from the thermal plasma via UV freeze-in during reheating. It is shown, that one can reproduce the observed DM abundance for its mass up to several PeVs, depending on the dimension of the operator connecting DM with the thermal bath and the associated scale of the UV physics. Thus we promote primordial GW to observables sensitive to feebly coupled inflaton, which is very challenging if not impossible to test in conventional particle physics laboratories or astrophysical measurements.
}
%%%%%%%%%%%%%%%%%
\maketitle
%%%%%%%%%%%%%%%%%

%%%%%%%%%%%
\section{Introduction}
\label{sec:intro}
%%%%%%%%%%%
Inflation stands as one of the most fundamental pillars of contemporary cosmology, explaining several puzzles of the early Universe, for example, the horizon or the flatness problem~\cite{Guth:1980zm,Linde:1981mu}. In its simplest form, inflation can be described by a single, slowly-rolling scalar field with an approximately flat potential that dominates the energy density of the primordial Universe. The flatness of the potential is generally expressed in terms of the so-called {\it slow roll parameters}, which are connected to cosmological observables measured by the spectrum of the cosmic microwave background (CMB). Cosmic inflation dilutes any pre-existing matter and radiation and thus requires a {\it reheating} mechanism to eventually result in the Universe dominated by radiation. Perturbative reheating can be realized through interactions between the classical, coherently-oscillating inflaton and thermal bath. In this framework, the Standard Model (SM) particles are produced in a quantum process from the vacuum in the homogeneous inflaton background. Hereafter, this process is dubbed as the {\it inflaton decay}. The common assumption of perturbative reheating, a constant decay width of the inflaton, cannot be justified in a generic scenario, e.g., if inflaton $\phi$ oscillates in a potential of the form $V(\phi)\propto\phi^{2n}$, with $n \neq1$~\cite{Garcia:2020wiy, Barman:2022tzk} or a daughter field acquires a vast mass due to its coupling with $\phi$, which generates kinematic suppression \cite{Ahmed:2021fvt, Ahmed:2022tfm}. 

The primordial gravitational wave (GW) is one of the most crucial predictions of inflationary paradigm. During inflation, quantum fluctuations inevitably give rise to a scale-invariant spectrum of tensor metric perturbations at super-Hubble scales. In a standard post-inflationary scenario, tensor modes become sub-horizon during the radiation-dominated (RD) stage. It is, however, well established that the presence of \textit{non-standard cosmologies}, e.g., an early {\it stiff} era before radiation domination, breaks such a scale invariance. In this case, the GW spectrum becomes significantly blue-tilted in the frequency range corresponding to the modes crossing the horizon during the stiff period~\cite{Giovannini:1998bp,Giovannini:1999bh,Riazuelo:2000fc,Seto:2003kc,Boyle:2007zx,Stewart:2007fu,Li:2021htg,Artymowski:2017pua,Caprini:2018mtu,Bettoni:2018pbl,Figueroa:2019paj,Opferkuch:2019zbd,Bernal:2020ywq,Ghoshal:2022ruy,Caldwell:2022qsj,Gouttenoire:2021jhk}\footnote{Effects of axion kination eras on primordial
GW spectral shape can be found in~\cite{PhysRevD.107.064071}.}. In addition, the amplitude of the tensor power spectrum becomes considerably enhanced, as compared to the standard scenario, where the Universe immediately transitioned from the inflationary into radiation-dominated phase. For instance, the onset of the RD epoch can be delayed if the reheating process is not instantaneous. This happens in a class of reheating models with a time-dependent inflaton decay rate.

The fact that Dark Matter (DM) constitutes about 24\% of the matter-energy budget of the Universe, has been unequivocally established from several cosmological and astrophysical observations~\cite{Jungman:1995df, Bertone:2004pz, Feng:2010gw}. As an alternative to the weakly interacting massive particle (WIMP) paradigm~\cite{Roszkowski:2017nbc, Arcadi:2017kky}, feebly interacting massive particles (FIMPs) have gained quite an attention~\cite{McDonald:2001vt,Hall:2009bx, Bernal:2017kxu}. Freeze-in, as opposed to freeze-out, requires very suppressed interaction rates between the dark and visible sectors, which can be achieved either via small couplings, known as IR freeze-in~\cite{Hall:2009bx}, or via non-renormalizable operators, suppressed by a high mass scale, called UV freeze-in~\cite{Hall:2009bx,Elahi:2014fsa}. The latter scenario is particularly interesting, as the DM yield is sensitive to the highest temperature reached by the SM plasma, controlled by the dynamics of the inflaton decay. The highest temperature of the bath, on the other hand, depends on the nature of the inflaton coupling or equivalently, the decay rate. 

Motivated by the above arguments, we explore a generic reheating scenario with the time-dependent inflaton decay width whose dependence on the scale factor is parameterized by a power-law dependence~\cite{Co:2020xaf,Ahmed:2021fvt,Barman:2022tzk,Banerjee:2022fiw}.
The non-standard evolution of the inflaton decay rate results in the non-trivial behavior of the inflaton and SM energy densities in the post-inflationary phase. We consider several types of inflaton interactions with bosonic and fermionic fields, and for each case, we obtain the size of the inflaton-matter coupling that could be probed by the proposed GW detectors, satisfying bounds from the Big Bang Nucleosynthesis (BBN) and CMB on GW energy density. We finally discuss DM production via UV freeze-in under the influence of time-dependent inflaton decay width and show that it is possible to have DM over a wide mass range $\sim\mathcal{O}(\text{MeV})-\mathcal{O}(\text{PeV})$, satisfying the PLANCK observed relic abundance. Interestingly, since both DM dynamics and GW energy density are controlled by the inflaton decay width, the proposed GW detectors can probe inflaton couplings that can give rise to the correct relic abundance for the DM, making a direct correspondence between DM and GW detection prospects.

The paper is organized as follows. In Sec.~\ref{sec:framework}, we discuss the thermal history of the Universe in the post-inflationary regime, elaborating on the underlying inflaton-matter interaction. The derivation of the spectral energy density of the primordial gravitational wave from inflation is concerned in Sec.~\ref{sec:gw}. Subsequent results of the GW analysis are furnished in Sec.~\ref{sec:result}. The impact of time-dependent inflaton decay on dark matter abundance is discussed in Sec.~\ref{sec:uv-fi}. Finally, in Sec.~\ref{sec:concl}, we summarize our findings. Detailed calculations of inflaton decay are presented in Appendix.~\ref{sec:app-inf-decay}, while Appendix.~\ref{sec:app-inf} discusses the bound on the scale of inflation from inflationary observables.

%%%%%%%%%%%
\section{The Framework}
\label{sec:framework}
%%%%%%%%%%%
Let us consider the following action of a system composed of the inflaton $\phi$ and the SM sector:
\begin{align}\label{eq:lgrng}
& S = \int d^4x\,\sqrt{-g}\,\left[-\frac{M_P^2}{2}\,\mathcal{R}+\mathcal{L}_\phi+\mathcal{L}_\text{SM}+\mathcal{L}_\text{int}\right]\,,
\end{align}
where $\mathcal{L}_{\phi\,,\text{SM}\,,\text{int}}$ are respectively the Lagrangian densities corresponding to the inflaton, the SM field, and relevant interactions, which we will elaborate in a moment. We assume that all matter fields are minimally coupled to gravity so that their kinetic terms are canonically normalized.  We denote the Ricci scalar via $\mathcal{R}$, and $M_P\!\equiv\!1/\sqrt{8 \pi G}\!=\!2.435 \!\times\! 10^{18}\gev$ is the reduced Planck mass. The background metric is assumed to be in the FLRW form. In the above expression, $g$ denotes the determinant of the metric tensor, whose line element is 
\begin{align}
ds^2 = dt^2-a^{2}(t)\,d\vec{x}^2\,,
\end{align}
with $a(t)$ being the scale factor. The inflaton Lagrangian reads
\begin{align}
\mathcal{L}_\phi \supset  \frac{1}{2}\,\partial_\mu\phi\,\partial^\mu\phi-V(\phi)\,,
\end{align}
with $V(\phi)$ being the inflaton potential. In this work, we focus on the $\alpha$-attractor T-model of inflation~\cite{Kallosh:2013hoa},
where $V(\phi)$ is given by
\begin{align}\label{eq:inf-pot}
V(\phi) = \Lambda^4\,\tanh^{2n}\,\left(\frac{|\phi|}{M}\right) \simeq
\Lambda^4\,
     \begin{cases}
     1 & \; \text{for}\; |\phi| \gg M,\\
     \left(\frac{|\phi|}{M}\right)^{2n} & \; \text{for}\; |\phi|\ll M\,,
    \end{cases}
\end{align}
where $\Lambda$ is the scale of inflation, and $M=\sqrt{6\,\alpha}\,M_P$. Hereafter the parameter $\alpha$ will be fixed at $1/6$. In the following discussion of the inflationary period $V(\phi) = \Lambda^4$ will be adopted, while during reheating we will use $V(\phi) = \Lambda^4 (|\phi|/M)^{2n}$. The Lagrangian $\mathcal{L}_\text{int}$, describing the interaction between the inflaton and matter shall be specified shortly.

%%%%%%%%%%%%%
\subsection{Post-inflationary evolution of the Universe}
\label{sec:beq}
\begin{figure}[htb!]
    \centering
     \includegraphics[scale=0.6]{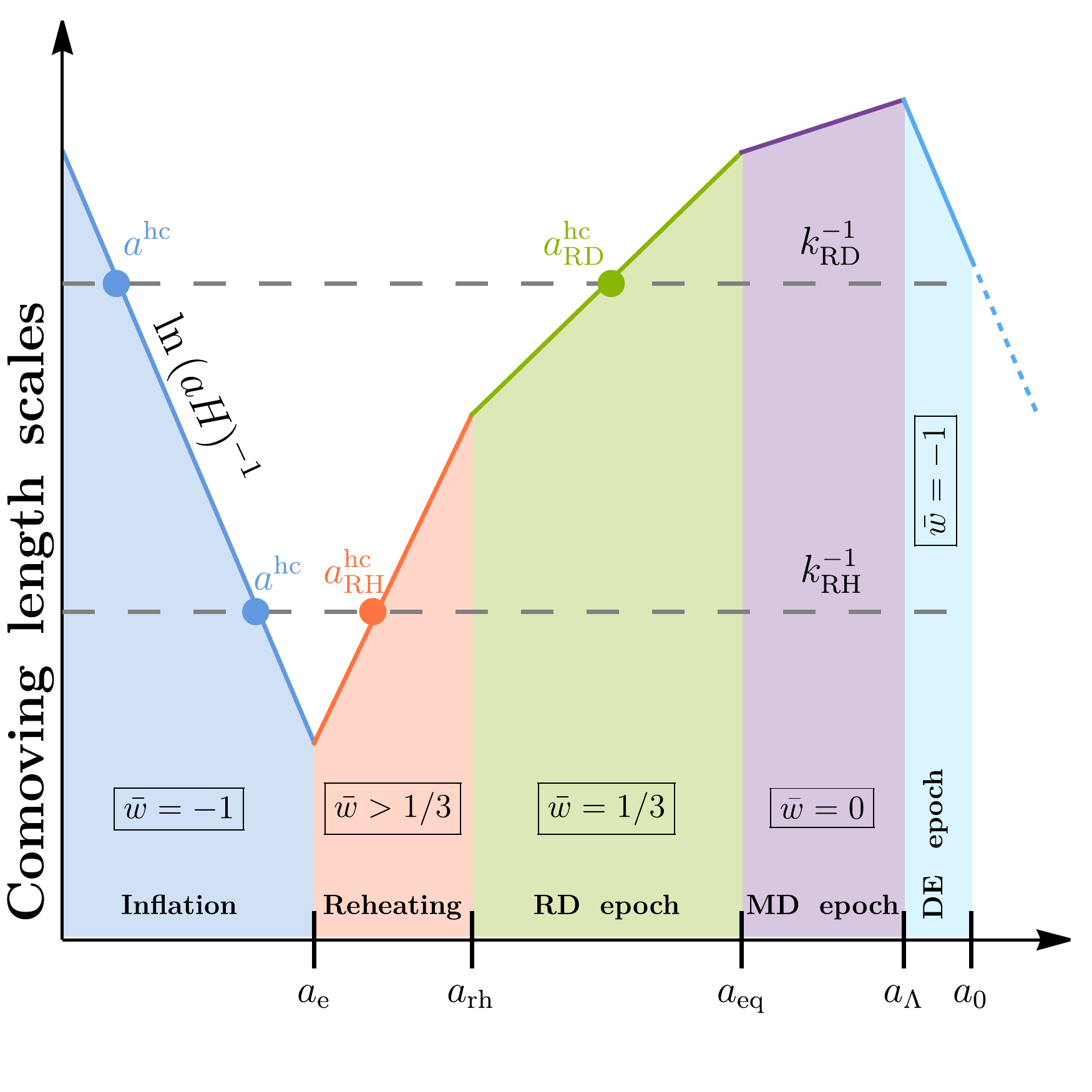}
    \caption{Evolution of cosmological comoving length scales with the scale factor at different epochs of the evolution of the universe. Here ``RD" stands for radiation domination, ``MD" implies matter domination, and ``DE" indicates dark energy domination. In each case corresponding equation of state is also mentioned.}
    \label{fig:stiff}
\end{figure}

%%%%%%%%%%%%%%%%
We first discuss the evolution of energy density of inflaton and radiation during the era of reheating, when the massive scalar field $\phi$ (inflaton) governs the evolution of the SM radiation energy density. The dynamics of such a system is controlled by the following set of coupled Boltzmann equations (BEQ) for time-averaged (over one period of inflaton oscillations~\cite{Ahmed:2022tfm}) energy densities of inflaton $(\rho_\phi)$ and radiation $(\rho_{\text{\tiny SM}})$ as
\begin{align}\label{eq:rhophi_rhord}
& \dot \rho_\phi + 3(1+\bar{w}) H\rho_\phi=-\left(1+\bar{w}\right)\, \Gamma_\phi\,\rho_\phi,
\nonumber\\&
\dot \rho_{\text{\tiny SM}} + 4 H\rho_{\text{\tiny SM}}=+\left(1+\bar{w}\right)\,\Gamma_\phi\,\rho_\phi\,,	\end{align}
where $\pm(1+\bar{w}) \Gamma_\phi \rho_\phi$ accounts for the energy gain (loss) of the SM bath (the inflaton field). Note that above, we have assumed that $\phi$ mainly transfers its energy into the SM sector. Here the {\it{dots}} denote derivatives with respect to the cosmic time $t$, and $\bar{w}$ is the time-averaged equation-of-state (EoS) parameter for the inflaton
\begin{align}
    \bar w \equiv \frac{\langle p_\phi\rangle}{\langle \rho_\phi\rangle}\,,
\end{align}
where $\langle ... \rangle$ denotes the time average, with $p$ being the inflaton pressure. For a given inflaton potential, the EoS parameter is calculable, e.g., for monomial potentials, it can be related to the index $n$ through the following formula: $\bar w=(n-1)/(n+1)$. At the onset of the reheating phase, the total energy density of the Universe is mainly in the form of the inflaton, while its end is defined as a moment when the inflaton and radiation energy densities become equal. In this work, we limit ourselves to the class of reheating models characterized by a stiff EoS parameter, i.e., $\bar w >1/3$. The evolution with $\bar w >1/3$ is referred to as \textit{the stiff epoch}. In Fig.~\ref{fig:stiff}, we have schematically shown different periods of the evolution of the Universe and their corresponding equation of states.
%%%%%%%%%%%%%%%%%%%%%%%%%%
%%%%%%%%%%%%%%%%%%%%%%%%%%%%%%%%

We adopt the following~\cite{Ahmed:2021fvt} parametrization of the inflaton width $\Gamma_\phi$:
\begin{align}\label{eq:decay-gen}
\Gamma_\phi = \Gamma_\phi^e \left( \frac{a_e}{a} \right)^\beta\,,
\end{align}
where $a_e$ is the initial value of the scale factor, indicating the end of inflation, $\Gamma_\phi^e$ denotes the inflaton width at $a=a_e$, while $\beta$ is assumed to be a constant parameter. The above parametrization of the inflaton width can be systematically derived for a homogeneous inflaton  condensate
with a power-law evolution of the inflaton energy density, assuming massless final states, as elaborated in Appendix.~\ref{sec:app-inf-decay} adopting the $\alpha$-attractor T-model of inflation. More generically, such a parametrization is valid for osciallation of the inflaton in any monomial potential during reheating. In addition, as shown in \cite{Garcia:2020wiy,Ahmed:2022tfm},  the generic power-law form of the inflaton width is also well justified if the time dependence of $\Gamma_\phi$ arises due to kinematical suppression effects for non-zero final state masses. Further, a general time (temperature) dependent dissipation rate can have several physical origins as discussed in Sec.3 of Ref.~\cite{Co:2020xaf}. It is worth to realize and emphasize that corrections that originate from kinematical mass effects do not change the form of $a$-dependence of the inflaton width, it turns out they merely modify the power $\beta$ in \eqref{eq:decay-gen}. In the following part, we will show how $\beta$ can be connected to the underlying inflationary potential.

The Hubble rate, $H$, evolves according to the Friedmann equation
\begin{align}
&H^2 = \frac{\rho_{\rm{tot}}}{3 M_P^2}\, &&\rho_{\rm{tot}} =\rho_\phi + \rhosm\,.
\end{align}
The initial conditions for the inflaton and SM radiation are given by
\begin{align}
&\rho_\phi(a_e) \equiv \rho_e\,, &&\rhosm(a_e) = 0\,. \label{eq:Friedmann}
\end{align}
At early times, the term associated with the expansion, i.e., \mbox{$3\,(1+\bar{w})\,H \rho_\phi,$} typically dominates over the interaction term $(1+\bar{w})\,\Gamma_\phi\,\rho_\phi$. The right-hand side of the first Boltzmann equation \eqref{eq:rhophi_rhord} can then be neglected, and straightforward integration gives a solution for the inflaton energy density during the oscillatory phase
\begin{align}\label{eq:rhoInf}
&\rho_\phi^{\text{\tiny{RH}}}(a) \simeq \rho_e \left( \frac{a_e}{a}\right)^\frac{6\,n}{n+1},~\text{for}~a \in [a_e, \arh)\,,
\end{align}
where the value of the inflaton energy density at the end of inflation can be expressed in terms of the initial value of the Hubble rate, $H_e$, though the Friedmann equation, i.e., $\rho_e= 3\,M_P^2\,H_e^2$. During the early stages of reheating, one can neglect the radiation contribution to the total energy density, i.e., $\rho_{\rm{tot}}(a_e)\simeq\rho_e$, which allows us to find the evolution of the Hubble rate
\begin{align}
& H^{\text{\tiny{RH}}}(a) \simeq H_e \left(\frac{a_e}{a} \right)^{\frac{3\,n}{n+1}}\,,~\text{for}~a \in [a_e, \arh)\,. \label{eq:Hubblerh}
\end{align}
Substituting Eq.~\eqref{eq:rhoInf} and~\eqref{eq:Hubblerh} into the second Boltzmann equation in Eq.~\eqref{eq:rhophi_rhord}, leads to
\begin{align}
&\rhosm^{\text{\tiny{RH}}}(a) &\simeq\left(\frac{6\,n}{n+1}\right)\cdot\Gamma_\phi^e\,H_e\,M_P^2 
\begin{cases}
  \frac{n+1}{n+4-\beta\,(n+1)}
  \bigg[\left(\frac{a_e}{a}\right)^{\beta +\frac{3\,n}{n+1}}-\left(\frac{a_e}{a}\right)^{4}\bigg], & \beta \neq \frac{n+4}{n+1},\\
  \left( \frac{a_e}{a} \right)^4 \ln{\left( \frac{a}{a_e}\right)}, & \beta = \frac{n+4}{n+1},
  \end{cases}
  \,,~a_e < a \leq \arh. \label{eq:rhoRrh}
\end{align}
which can further be written as
\begin{align}\label{eq:rhoRrh2}
\rhosm^{\text{\tiny{RH}}}(a) &\approx \rho_\text{rh}\,\left(\frac{\arh}{a}\right)^4\, \begin{cases}
    \left[\frac{1-\left(a_e/a\right)^\frac{\beta\,(n+1)-(n+4)}{n+1}}{1-\left(a_e/\arh\right)^\frac{\beta\,(n+1)-(n+4)}{n+1}}\right], &\beta \neq \frac{n+4}{n+1},\\
    \frac{\ln{(a/a_e)}}{\ln{(\arh/a_e)}}, & \beta = \frac{n+4}{n+1},
    \end{cases}
    \,,~a_e < a \leq \arh,
    % \label{eq:rhoRrh}
\end{align}
where we have used the value of the radiation energy density at the end of reheating, defined by the inflaton-radiation equality
\begin{align}
\rho_\phi^{\text{\tiny{RH}}}(\arh) = \rhosm^{\text{\tiny{RH}}}(\arh)\equiv \rho_\text{rh}.
\end{align}

There is a comment here regarding continuity of $\rhosm^{\text{\tiny{RH}}}(a)$ as a function of $n$ at $\beta=(n+4)/(n+1)$. Within any given inflaton model, $\beta$ is a fixed function of $n$, e.g., in the case of the inflaton Yukawa interactions $\beta=\beta_\psi=3(n-1)/(n+1)$. Then, $\rhosm^{\text{\tiny{RH}}}(a)$ should be a continuous function of $n$, including the values of $n$ that follows from the condition $\beta=(n+4)/(n+1)$. Indeed $\rhosm^{\text{\tiny{RH}}}(a)$ given by Eq.~\eqref{eq:rhoRrh}, does satisfy this condition. To prove the continuity, it is essential to keep both terms in the bracket of Eq.~\eqref{eq:rhoRrh}. However, to perform further analytical calculations, we will be dropping one of those terms at a time. Therefore cases $\beta \ll (n+4)/(n+1)$ and $\beta \gg (n+4)/(n+1)$ will appear, and it is important to note that such approximations are not applicable in the close vicinity of $\beta=(n+4)/(n+1)$.

As the inflaton energy density dominates the early stages of reheating, the background dynamics is determined by the value of $\bar{w}$, whereas the behavior of the SM sector depends on both $\bar{w}$ and $\beta$. Note that, for $\beta < (n+4)/(n+1)$, the first term in the square bracket in \eqref{eq:rhoRrh} dominates, whereas if $\beta > (n+4)/(n+1)$ the energy density of the radiation bath decreases as $a^{-4}$. Thus, depending on the value of $\beta + 3\,n/(n+1)$, one obtains a very different evolution of the radiation sector, which implies a non-trivial scaling of a thermal bath temperature $T$ with the scale factor during reheating. In particular, the temperature of the SM sector is measured by the radiation energy density
\begin{align}
T(a) = \left( \frac{30}{\pi^2 g_{\star \rho}(T)} \right)^{1/4} \rhosm^{1/4}(a)\,,
\end{align}
where $g_\star(T)$ counts the effective number of relativistic degrees of freedom at temperature $T$. Utilizing \eqref{eq:rhoRrh}, one finds that during the non-standard phase of reheating, $T$ behaves for $a_e < a \leq \arh$ as
\begin{align}
&T^{\text{\tiny{RH}}}(a) \approx  \left[\frac{180\,n\; \Gamma_\phi^e H_e M_P^2}{(n+1)\,\pi^2 g_{\star \rho}(T)} \right]^{1/4} \begin{cases}
 \frac{n+1}{n+4-\beta\,(n+1)}\,\bigg[\left(\frac{a_e}{a}\right)^{\beta +\frac{3\,n}{n+1}}-\left(\frac{a_e}{a}\right)^{4}\bigg]^{1/4}, & \beta\neq\frac{n+4}{n+1},\\
\frac{a_e}{a} [\ln{(a/a_e)}]^{1/4},  &\beta = \frac{n+4}{n+1}
  \end{cases}\,. 
\label{eq:Trh}
\end{align}

The scale factor at the end of reheating can be determined from Eqs.~\eqref{eq:rhoInf} and \eqref{eq:rhoRrh} as follows
\begin{align}\label{eq:arh}
&\arh = a_e 
\begin{cases} \left(\frac{n+4-\beta\,(n+1)}{2\,n}\,\frac{H_e}{\Gamma_\phi^e} \right)^\frac{n+1}{n\,(3-\beta)-\beta},  &\beta \ll \frac{n+4}{n+1},\\
\left( \frac{H_e}{\Gamma_\phi^e}\,\frac{n-2}{n} \right)^{\frac{n+1}{2\,(n-2)}} \mathcal{W}^{\frac{n+1}{2\,(2-n)}}\,\left( \frac{H_e}{\Gamma_\phi^e}\frac{n-2}{n}\right), &\beta = \frac{n+4}{n+1},\\
\left(\frac{\beta-4+n\,(\beta-1)}{2\,n} \frac{H_e}{\Gamma_\phi^e}\right)^{\frac{n+1}{2\,(n-2)}}, &\beta \gg \frac{n+4}{n+1}\,,
\end{cases}
\end{align}
where $\mathcal{W}[z]$ denotes the Lambert $\mathcal{W}$-function. Note that in order to obtain the above result, we had to drop either the first or the second term in the bracket in Eq.~\eqref{eq:rhoRrh}.

During non-instantaneous decay of the inflaton, the bath temperature can rise to several orders of magnitude above the reheating temperature~\cite{Giudice:2000ex}. This happens at $a=a_{\rm{max}}$, defined via
\begin{align}
a_{\rm {max}} &\equiv a_e \begin{cases}
 \left[\frac{1}{4} \left(\beta + \frac{3\,n}{n+1}\right)\right]^\frac{n+1}{\beta\,(n+1)-(n+4)} , & \beta \neq\frac{n+4}{n+1} \,,\\
 \exp(1/4), & \beta = \frac{n+4}{n+1}\,,
 \end{cases}
  \label{eq:aMAX}
\end{align}
so that the maximum temperature of the  radiation bath is
\begin{align}
T_{\tiny{\rm{max}}}\equiv  \, T^{\text{\tiny{RH}}}(a_{\rm{max}}) &\approx   \left(\frac{180\,n\,M_P^2}{(n+1)\,\pi^2\,\gsr(T)}\, \Gamma_\phi^e\,H_e\right)^{1/4}  \nonumber \\
&\times 
\begin{cases}
\left[ \frac{n+1}{n+4-\beta\,(n+1)}\right]^{1/4}\,\bigg[\frac{1}{4} \left(\beta + \frac{3\,n}{n+1}\right)\bigg]^\frac{\beta+n\,(\beta+3)}{4\,\left((n+4)-\beta\,(n+1)\right)}, &\beta \ll \frac{n+4}{n+1},\\
\frac{1}{\sqrt{2}}e^{-1/4}, & \beta = \frac{n+4}{n+1},\\
\left[\frac{n+1}{\beta-4+n\,(\beta-1)}\right]^{1/4}\,\bigg[\frac{1}{4}\,\left(\beta + \frac{3\,n}{n+1}\right)\bigg]^\frac{n+1}{n+4-\beta\,(n+1)}, &  \beta \gg \frac{n+4}{n+1}\,.
\end{cases} 
\label{eq:Tmax}
\end{align}
From the discussion so far, it is understandable that once one specifies the inflaton potential $V(\phi)$ and the form of the inflaton-SM interactions (the ``model"), $\bar{w}, H_e, \beta, \Gamma_\phi^e$ can be determined. 

Before proceeding further, let us briefly discuss the details of inflaton dynamics during the oscillatory phase. In the FLRW background, the classical equation of motion (EoM) for the inflaton is given by
\begin{align}\label{eq:inf-eom}
\ddot \phi + 3\,H\,\dot\phi + \frac{\partial V(\phi)}{\partial\phi} = 0\,,
\end{align}
where we have assumed that $\phi$ is spatially homogeneous\footnote{Note that here we are neglecting contributions from inflaton-matter interactions, which, in the non-instantaneous reheating scenario, are typically irrelevant at the onset of reheating.}. During the period of reheating, the oscillating inflaton field with a time-dependent amplitude can be parametrized as~\cite{Ichikawa:2008ne,Kainulainen:2016vzv,Clery:2021bwz,Co:2022bgh,Garcia:2020wiy,Ahmed:2022qeh}.
\begin{align}\label{eq:varphi}
& \phi(t) = \varphi(t)\,.\,\mathcal{P}(t)\,.
\end{align}
Here, $\mathcal{P}(t)$ is a quasi-periodic, fast-oscillating function, and $\varphi(t)$ is a slowly-varying envelope. It is instructive to introduced the effective mass of the inflaton
\begin{align}\label{eq:inf-mass}
m_\phi^2 \equiv \frac{\partial^2 V(\phi)}{\partial\phi^2}\Bigg|_{\phi=\varphi}  = \frac{n\, \left(2n-1\right)}{3\,\alpha}\,\frac{\Lambda^4}{M_P^2}\,\left(\frac{\rho_\phi}{\Lambda^4}\right)^\frac{n-1}{n}\,.
\end{align}

The slow-roll evolution of the inflaton field during cosmic inflation can be parameterized by the so-called potential slow-roll parameters
\begin{align}\label{eq:sr-param}
& \epsilon_V = \frac{M_P^2}{2}\,\left(\frac{V'(\phi)}{V(\phi)}\right)^2, &&\eta_V = M_P^2\,\frac{V''(\phi)}{V(\phi)}\,,
\end{align}
that are related to the cosmological observables, namely the spectral index $n_s$ and the tensor-to-scalar ratio $r$ via
\begin{align}
& r \simeq 16\,\epsilon_V\,, &&n_s \simeq 1-6\,\epsilon_V + 2\eta_V\,.
\end{align}
At the very end of inflation, $\ddot a=0$, and the first potential slow-roll parameter is $\epsilon_V \simeq 1$. This condition determines the field at $a=a_e$, i.e., 
\begin{align}\label{eq:ph-end}
& \varphi_e =  \sqrt{\frac{3}{2}}\,M_P\,\sinh^{-1}\left[\frac{2\,n}{\sqrt{3\,\alpha}}\right]\,.
\end{align}
Moreover, at $a=a_e$, the potential energy of $\phi$ matches with its kinetic energy, so that 
\begin{align}\label{eq:rho-end}
& \rho_e = \frac{3}{2}\,V(\varphi_e)\,,
\end{align}
where we have assumed that $\mathcal{P}(a_e) = 1$.
From the above expression, we see that  $\rho_e$ has a very mild dependence on the index $n$ for a fixed $\alpha$ and $\Lambda$. For example, with $\alpha=1/6$ and $\Lambda=3\times 10^{-3}\,M_P$, we find $\rho_e\simeq 2\times 10^{63}\,\text{GeV}^4$ for $n\in [2\,,10]$.

%%%%%%%%%%%%%%%%%%%
\subsection{Models of inflaton-matter interactions}
\label{sec:CMB-coupling}
%%%%%%%%%%%%%%%%%%%%
Let us now specify interactions between the inflaton and other matter particles. Here, we assume that perturbative expansion is justified while calculating the quantum production of radiation (reheating) from the vacuum in the classical inflaton background. Once the inflaton couples to matter fields, its oscillations are severely damped by the decay. Without going into the details of the UV completion, we consider four possible interaction vertices\footnote{We assume vertices like $\phi^2\,S^2$ or $\phi^2\,V_\mu\,V^\mu$ are absent by assumption, and we focus only on the tri-linear interactions.}
\begin{align}
& \mathcal{L}_\text{int}\supset\mathcal{L}_{SS\phi}+\mathcal{L}_{\psi\psi\phi}+\mathcal{L}_{VV\phi}+\mathcal{L}_{aa\phi}\,,
\end{align}
and parametrize them as
\begin{align}
& \mathcal{L}_{SS\phi} \supset \gsp\,SS\,\phi\,,
\nonumber\\&
\mathcal{L}_{\psi\psi\phi} \supset
\gpp\,\overline{\psi}\,\psi\,\phi\,,
\nonumber\\&
\mathcal{L}_{VV\phi} \supset \gvp\,V_\mu\,V^\mu\,\phi\,,
\nonumber\\&
\mathcal{L}_{aa\phi} \supset \hap\,\partial_\mu a\,\partial^\mu a\,\phi\,,
\end{align}
where $S\,,V\,,\psi, a$ are generic scalar field, vector boson, Dirac fermion, and particles with derivative interactions (e.g., axion-like particles), respectively. We refer to scenarios with an inflaton-matter coupling specified as to {\it models} and consider them individually. It is important to stress that in the adopted parametrization, the coupling $\gsp$ and $\gvp$ have dimensions of mass, $\gpp$ is dimensionless, while $\hap$ has inverse mass dimension. Let us also notice that once the model of reheating is specified, $\Gamma_\phi^e$ can be calculated, see Eq.~\eqref{eq:Gammae}, in terms of the model coupling $g_{i\phi}$ ($i\in SS\,,\psi\psi\,,\mathcal{VV}\,,aa$) and parameters of the inflaton potential, which also determines the value of the $\beta$ parameter. Moreover, the time-averaged EoS parameter $\bar w$ is predicted solely by the shape of the inflaton potential Eq.~\eqref{eq:inf-pot}.
Within a given model of inflaton-matter interaction, we adopt the following set of independent parameters~\footnote{Instead of the parameter set~\eqref{par_mod} one could use $\{g_{i\phi}\,,n\,,\Lambda\}$ with $i\in S\,,\psi\,,\mathcal{V}\,,a$. Since the couplings $g_{i\phi}$ are of different mass dimensions, it is more convenient to adopt the width $\Gamma_\phi^e$ instead of the couplings when comparing the models.}
\begin{align}
& \{ \Gamma_{\phi \to f}^e\,,\Lambda, \, n\}\,,   
\label{par_mod}
\end{align}
where $f=SS,\,\psi\psi,\,\mathcal{VV}\,,aa$.

%%%%%%%%%%%%%%%%%%%%%%%%%%%%%%%%%%%%
\section{Spectrum of Primordial Gravitational Waves}
\label{sec:gw}
%%%%%%%%%%%%%%%%%%%%%%%%%%%%%%%%%%%%
Here we briefly summarize formalism to calculate the spectrum of a stochastic GW background of primordial origin. The complete derivations can be found in, for example, Refs.~\cite{Boyle:2005se, Watanabe:2006qe, Saikawa:2018rcs, Caprini:2018mtu}, hence without going into the details, here we highlight the salient points that help in obtaining the final expression for primordial GW spectrum. We shall mainly stick to the convention followed in Ref.~\cite{Saikawa:2018rcs} while elaborating on the definition of different relevant quantities.

GWs are described as the tensor metric perturbations in a spatially-flat FLRW Universe
\begin{align}
ds^2 = dt^2 - a^2(t)\,\left( \delta_{ij} + h_{ij} (t,{\bf x})\right)\,dx^i\,dx^j\,.
\end{align}
Tensor fluctuations are assumed to be small perturbations, i.e., $|h_{ij}|\ll 1$, satisfying the transverse-traceless conditions, $h_i^i=\partial^i\,h_{ij}=0$. The GWs equation of motion follows the Einstein equations linearized to first order in $h_{ij}$ over the FRLW background\footnote{In presence of a viscous background the mode equation receives correction due to (bulk and shear) viscosity, that causes damping of the GW amplitude~\cite{Ghiglieri:2015nfa,Saikawa:2018rcs}. However, such a damping is inefficient in an expanding background as long as the interaction rate of particles in the viscous medium is much faster compared to the cosmic expansion rate~\cite{Saikawa:2018rcs}, which is satisfied in the present analysis via instantaneous thermalization.}
\begin{align}
& \ddot h_{ij} + 3\,H\,\dot h_{ij} - \frac{\nabla^2}{a^2}\,h_{ij} = 16\,\pi\,G\,\Pi_{ij}^\text{TT}\,,
\end{align}
with $\Pi_{ij}^\text{TT}$ being the transverse and traceless part of the anisotropic stress tensor. It is instructive to decompose the perturbation $h_{ij}$ over two polarization states $\lambda$ as
\begin{align}\label{eq:hij-eom}
& h_{ij}(t\,,{\bf x}) = \sum_{\lambda=+\,,\times}\int\,\frac{d^3 {\bf k}}{(2\,\pi)^3}\, h^\lambda(t\,,{\bf k})\,\epsilon_{ij}^\lambda({\bf k})\,\exp\left(i\,{\bf k}\cdot{\bf x}\right)\,,
\end{align}
where $\epsilon_{ij}^\lambda({\bf k})$ are spin-2 polarization tensors satisfying orthonormality and completeness relations
\begin{align}
& \sum_{ij}\epsilon_{ij}^\lambda\,\left(\epsilon_{ij}^{\lambda'}\right)^\star = 2\,\delta^{\lambda\lambda'}\,,
\nonumber\\&
\sum_{\lambda=+\,,\times}\,\epsilon_{ij}^\lambda({\bf \hat k})\,\epsilon_{lm}^{\lambda}\,({\bf \hat k}) = P_{il}\,P_{jm}+P_{im}\,P_{jl}-P_{ij}\,P_{lm}\,,
\end{align}
with $P_{ij}=\delta_{ij}-\hat k_i\,\hat k_j$ being the projection operator and  $\bf\hat k$ being the unit vector along the direction of $\bf k$. 

The mode function $h^\lambda(t, \bf{k})\equiv h_{\bf k}^\lambda$ satisfies the following equation:
\begin{align}\label{eq:hk-eom}
& \ddot h_{\bf k}^\lambda + 3\,H\,\dot h_{\bf k}^\lambda+\frac{k^2}{a^2}\,h_{\bf k}^\lambda=16\,\pi\,G\,\Pi_{\bf k}^\lambda\,,
\end{align}
where $\Pi_{\bf k}^\lambda$ is the Fourier components of $\Pi_{ij}^\text{TT}$. It is convenient rewrite the above equation in conformal coordinates
\begin{align}\label{eq:hkc-eom}
& \left(h_{\bf k}^\lambda\right)'' + 2\,\frac{a'}{a} \left(h_{\bf k}^{\lambda}\right)' + k^2\,h_{\bf k}^{\lambda} = 16\,\pi\,G\,a^2\,\Pi_{\bf k}^\lambda\,,
\end{align}
where \textit{prime} denotes a derivative with respect to the conformal time coordinate, defined by $dt= a d\tau$. Eq.~\eqref{eq:hkc-eom} admits approximate analytical solutions in two extreme regimes as follows
\begin{itemize}
\item {\it \underline{Super-Hubble scale:}} \\ For modes far outside the Hubble horizon, i.e., $k\ll aH$, one can write Eq.~\eqref{eq:hkc-eom} as
\begin{align}
& \left(h_{\bf k}^\lambda\right)'' + 2\,\frac{a'}{a} \left(h_{\bf k}^{\lambda}\right)' \approx 0\,\implies a^{-2}\,\left(a^2\,h_{\bf k}^{\lambda \prime}\right)^\prime\approx 0\,,
\end{align}
where we have ignored the source term. This equation has a solution of the form
\begin{align}\label{eq:kllaH}
&  h_{\bf k}^\lambda(\tau) = \mathcal{C}_1 + \mathcal{C}_2\int\,\frac{d\tau'}{a(\tau')^2}\,,
\end{align}
where $\mathcal{C}_{1,2}$ are constants of integration. The second term in the above expression decays with time\footnote{In particular, for super-Hubble modes generated from quantum fluctuations of the tensor perturbation during inflation, the decaying mode becomes quickly negligible due to the exponential expansion of the Universe.}. Therefore, ignoring the decaying term, one concludes that $h_{\bf k}^\lambda$ stays constant for super-Hubble modes~\cite{Watanabe:2006qe, Saikawa:2018rcs}. At some point after inflation, these modes become sub-Hubble, i.e., $k>a\,H$ and re-enter the horizon.

\item {\it \underline{Sub-Hubble scale:}} \\ 
After the end of inflation, modes eventually re-enter the horizon $(k>a\,H)$ and start to oscillate. In this case, Eq.~\eqref{eq:hkc-eom} can be solved by assuming a solution of the form
\begin{align}\label{eq:kggaH}
& h_{\bf k}^\lambda = A(\tau)\,\exp\left(i\,B(\tau)\right)\,,
\end{align}
where $A$ and $B$ are real functions.
Substituting this in Eq.~\eqref{eq:hk-eom}, and comparing the real and imaginary components, we obtain
\begin{align}
& A''-A\,\left(B'\right)^2+\frac{2\,a'}{a}\,A'+k^2\,A=0\,,
\nonumber\\&
2\,A'\,B'+A\,B''+\frac{2\,a'}{a}\,A\,B'=0\,,
\end{align}
where again, we have dropped the source term. Now, considering the oscillation to be very rapid compared to the time variation of the amplitude, and the modes to be well inside the horizon $k \gg a'/a$, we obtain from the first equation
\begin{align}
& B(\tau) = \pm k\,\tau+\text{constant}\,,
\end{align}
which on substitution in the second equation brings us to
\begin{align}
& A(\tau)\propto a(\tau)^{-1}\,.
\end{align}
The behaviour of the mode function $h_{\bf k}^\lambda$ is thus described by a WKB solution of the from 
\begin{align}
    h_{\bf k}^\lambda\simeq \left(C/a\right)\,\exp\left(\pm i k\,\tau\right),
    \end{align}
    where $C$ is some arbitrary constant.
\end{itemize}

The energy density carried by the GWs is given~\cite{Maggiore:1999vm, Watanabe:2006qe, Saikawa:2018rcs, Caprini:2018mtu}  by
\begin{align}\label{eq:rho-GW}
& \rGW = \frac{1}{64 \pi G a^2} \left\langle  \left( \partial_\tau h_{ij}  \right)^2+  \left( \nabla h_{ij}  \right)^2 \right\rangle\,,
\end{align}
where $\langle...\rangle$ denotes the spatial average. As it could be seen from Eq.~\eqref{eq:hk-eom}, once a particular mode re-enters (in other words crosses) the horizon, i.e., $k> aH$, the corresponding mode function obeys $|\partial h_{\bf k}^\lambda(\tau)/\partial\tau|^2=k^2\,|h_{\bf k}^\lambda(\tau)|^2$, therefore the energy density could be written as
\begin{equation}
\rGW = \frac{1}{32\,\pi\,G}\,\int d\,\ln k\,\left(\frac{k}{a}\,\right)^2\,\left[\frac{k^3}{\pi^2} \sum_\lambda \left| h_{\bf k}^\lambda \right|^2\right].
\end{equation}
It is useful to define the spectral GW density per logarithmic wavenumber interval, normalized to the critical density $\rho_c=3\,H^2/(8\,\pi\,G)$ as follows
\begin{align}\label{eq:omega-gw}
& \ogw(\tau,k) = \frac{1}{\rho_c}\,\frac{d\rGW}{d\log k}\,.
\end{align}
Then, starting with Eq.~\eqref{eq:rho-GW}, one can obtain~\cite{Watanabe:2006qe, Saikawa:2018rcs}\footnote{In deriving Eq.~\eqref{eq:ogw-k} we have ignored the effect of neutrino free-streaming, that leads to damping of GW spectral energy density by $\sim 35\%$ typically in the frequency range of $10^{-17}\lesssim f\lesssim 10^{-11}$ Hz~\cite{Saikawa:2018rcs,Caprini:2018mtu}, i.e., for the modes that enter the horizon between the epoch of neutrino decoupling and the matter-radiation equality. Those modes are not the focus of the present analysis.}
\begin{align}\label{eq:ogw-k}
& \ogw(\tau\,,k) = \frac{1}{12}\,\left(\frac{k}{a(\tau)\,H(\tau)}\right)^2 \mathcal{P}_{T,\text{prim}}\,\mathcal{T}(\tau,k)\,,
\end{align}
where the primordial power spectrum has been defined as
\begin{align}
\mathcal{P}_{T,\text{prim}} \equiv \frac{k^3}{\pi^2}\,\sum_\lambda \left|h_{{\bf k}\,,\text{prim}}^\lambda\right|^2\,,
\end{align}
with the primordial mode function $h_{{\bf k}\,,\text{prim}}^\lambda$  defined as the mode function $h_{{\bf k}}^\lambda(\tau)$ at some moment shortly after the end of inflation, when all the modes have already exited the horizon.

The primordial tensor power spectrum $\mathcal{P}_{T,\text{prim}}$ is determined by the Hubble parameter at the time when the corresponding mode crosses the horizon during inflation $(k = aH)$~\cite{Watanabe:2006qe, Saikawa:2018rcs, Caprini:2018mtu}
\begin{align}
\mathcal{P}_{T,\text{prim}} = \frac{2\,H^2}{\pi^2\,M_P^2}\Bigg|_{k=a\,H}\,,
\end{align}
where we have used the mode solutions by matching the sub-Hubble modes (during inflation) with the super-Hubble modes (at the end of inflation) at $k=a\,H$~\cite{Boyle:2005se,Caprini:2018mtu}. The transfer function, $\mathcal{T}(\tau\,,k)$, adopted in Eq.(\ref{eq:ogw-k}), connects primordial mode functions with mode functions at some later time $h_{\bf k}^\lambda(\tau)$ as~\cite{Boyle:2005se,Saikawa:2018rcs}
\begin{align}
& \mathcal{T}(\tau\,,k)=\left|\frac{h_{\bf k}^\lambda
(\tau)}{h_{{\bf k},\text{prim}}^\lambda}\right|^2\,.
\end{align}

As shown earlier, the modes $h_{{\bf k}}^\lambda$ remain constant on super-horizon scales, while they decrease as $a^{-1}$ once they re-enter the Hubble horizon. In other words, in the sub-horizon regime, i.e., $k>aH$, $h_{{\bf k}}^\lambda=\left(a^\text{hc}/a\right)\,h_{{\bf k}}^{\text{hc},\lambda}$; therefore, the transfer function could be written~\cite{Kuroyanagi:2008ye,Opferkuch:2019zbd,Allahverdi:2020bys} as
\begin{align}\label{eq:T-fun}
& \mathcal{T}(\tau\,,k) = \frac{1}{2}\,\left(\frac{a^\text{hc}}{a}\right)^2\,,
\end{align}
where the prefactor 1/2 appears due to oscillation-averaging the tensor mode functions~\cite{Saikawa:2018rcs,Choi:2021lcn,Figueroa:2018twl} and $\ahc$ is scale factor at the horizon crossing. As evident from Eq.~\eqref{eq:T-fun}, the transfer function characterizes the expansion history between the moment of horizon crossing $\tau=\tau^\text{hc}$ of a given mode $k$ and some later moment $\tau>\tau^\text{hc}$. From Eq.~\eqref{eq:ogw-k}, one can see that the spectral GW energy density at the present time is given by
\begin{align}
\ogw^{(0)}(k)=\frac{1}{24}\left(\frac{k}{a_0 H_0}\right)^2 \mathcal{P}_{T,\text{prim}} \left(\frac{\ahc}{a_0}\right)^2\,,
\label{eq:ogw-0}
\end{align}
where $a_0=1$ and $H_0$ are respectively the scale factor and the Hubble parameter measured today.

One anticipates that $\ogw^{(0)}(k)$ has a power-law dependence on $k$. Using the fact that $k=a^\text{hc} H(a^\text{hc})$, and the equation-of-state parameter after the horizon crossing is $\bar w$, one obtains 
\begin{align}
a^\text{hc} \propto k^{\frac{n+1}{1-2\,n}}\,
\end{align}
Hence, from Eq.~\eqref{eq:ogw-0} we find the following scaling applies
\begin{align} \label{ogw-scal}
    \ogw^{(0)}(k)\propto k^\frac{2\,n-4}{2\,n-1}\,.
\end{align}
Let us now find the coefficient of proportionality in Eq.~\eqref{ogw-scal}. This coefficient might, in particular, depend on parameters such as, e.g., the inflaton width $\Gamma_\phi^e$ at the end of inflation, which, in turn, is determined by the inflaton-matter couplings. The sensitivity of the spectral GW energy density upon the inflaton couplings is one of the central tasks of this study. We will determine the present value of the spectral GW energy density for the modes re-entering the horizon at different epochs; after the end of reheating, during RD, and prior to the onset of RD, that is, during inflaton domination (reheating). At first, let us assume that the horizon re-entry happens during the RD epoch after the end of reheating, i.e., $\ahcRD \in (a_{\rm rh}, a_{\rm eq})$. In this case, we reproduce the standard {\it scale-invariant} result for the spectral GW energy density
\begin{align}\label{eq:ogw-rad}
& \ogw^{(0), \rm{RD}} =  \frac{\mathcal{P}_{T,\text{prim}}}{24}\,\left(\frac{\ahcRD\,\hhcRD}{a_0\,H_0}\right)^2\,\left(\frac{\ahcRD}{a_0}\right)^2
=\Omega_{\gamma}^{(0)}\,\frac{\mathcal{P}_{T,\text{prim}}}{24}\,\mathcal{F}(g_\star)\,,
\end{align}
where the function $\mathcal{F}(g_\star)\equiv \frac{\gsr^\text{RH}}{\gsr^0}\,\left(\frac{\gss^0}{\gss^\text{RH}}\right)^\frac{4}{3}$ tracks the number of degrees of freedom from the end of reheating till present, and $\Omega_{\gamma}^{(0)} =\rho_{\gamma,0}/\rho_c=2.47\times 10^{-5}\,h^{-2}$ is the fraction of the energy density of photons at the present epoch. Here we have assumed entropy conservation from the moment of horizon crossing (RD) till today, implying $T\propto a^{-1}\,\gss^{-1/3}$.
On the other hand, if the horizon crossing happens during reheating, before the radiation era, i.e., $\ahcRH \in (a_e, a_{\rm rh})$, one obtains
\begin{align}\label{eq:ogw-md}
& \ogw^{(0), \rm{RH}} = \frac{\mathcal{P}_{T,\text{prim}}}{24}\,\left(\frac{\hhcRH}{H_0}\right)^2\,\left(\frac{\ahcRH}{a_0}\right)^4
=\ogw^{(0),\text{RD}}\, \left[\left(\frac{g_{\star\rho}^\text{rh}}{g_{\star\rho,\text{RD}}^{\text{hc}}}\right)^\frac{3}{4}\,\frac{g_{\star s, \text{RD}}^{\text{hc}}}{g_{\star s}^{\text{rh}}} \right]^\frac{4}{3}\,\left( \frac{\ahcRH}{\arh}\right)^\frac{4-2\,n}{n+1}\,,
\end{align}
where we have used the following relations:
\begin{align}
& \left(\frac{\arh}{\ahcRD} \right)^4 \left(\frac{\hRH}{\hhcRD} \right)^2  = \frac{g_{\star\rho}^\text{rh}}{g_{\star \rho, \text{RD}}^{\text{hc}}}\,\left(\frac{g_{\star s, \text{RD}}^{\text{hc}}}{g_{\star s}^{\text{rh}}} \right)^{4/3}\,,
\nonumber\\&
\left(\frac{\ahcRH}{\arh} \right)^4 \left(\frac{\hhcRH}{\hRH} \right)^2 = \left( \frac{\ahcRH}{\arh}\right)^\frac{4-2\,n}{n+1}\,.
\end{align}
Now, for the re-entry of the modes happening during the period of reheating, redshifting the GW frequency $f$ to present, we obtain~\cite{Saikawa:2018rcs}
\begin{align}\label{eq:freq}
&  f = \frac{k}{2\pi}\frac{1}{a_0} = \frac{\hhcRH}{2\,\pi}\,\frac{\ahcRH}{a_0}  = \frac{\rho_\text{rh}^{1/2}}{2\,\pi\,\sqrt{3\,M_P^2}}\,\frac{\ahcRH}{a_0}\,\left(\frac{\arh}{\ahcRH}\right)^\frac{3\,n}{n+1}\,,
\end{align}
where $\rho_\text{rh}$ in the total energy density of the Universe at the end of reheating.

There exists an upper bound on frequencies, dictated by the modes that re-entered the horizon after the end of inflation
\begin{align}\label{eq:fmax}
& f_\text{max} = \frac{H_e}{2\pi}\,\frac{a_e}{a_0} \simeq \frac{\Lambda^2\,(n\,\sqrt{2})^n}{2\,\sqrt{2}\pi\,M_P}\,\left(\frac{\gss(T_0)^\frac{1}{3}\,T_0}{\gss(\Trh)^\frac{1}{3}\,\Trh}\right)
\nonumber\\&
\times
\begin{cases}
\left[\left(n\sqrt{2}\right)^n\,\left(\frac{(n+4)-\beta\,(n+1)}{2\,n}\right)\,\frac{\Lambda^2}{\sqrt{2}\,M_P\,\Gamma_\phi^e} \right]^{\frac{n+1}{n\,(\beta-3)+\beta}},  &\beta \ll \frac{n+4}{n+1}\,,\\
\left[\left(n\sqrt{2}\right)^n\,\frac{\Lambda^2}{\sqrt{2}\,M_P\,\Gamma_\phi^e} \,\frac{n-2}{n}\right]^{\frac{n+1}{2\,(2-n)} } \mathcal{W}^{\frac{n+1}{2\,(2-n)}}\left[ \left(n\sqrt{2}\right)^n\,\frac{\Lambda^2}{\sqrt{2}\,M_P\,\Gamma_\phi^e}\,\frac{n-2}{n}\right], & \beta=\frac{n+4}{n+1}\,,\\
\left[\left(n\sqrt{2}\right)^n \left(\frac{\beta-4+n\,(\beta-1)}{2\,n}\right) \,\frac{\Lambda^2}{\sqrt{2}\,M_P\,\Gamma_\phi^e}\right]^{\frac{1+n}{4-2\,n}}, & \beta \gg \frac{n+4}{n+1}\,,
\end{cases}
\end{align}
while modes with frequencies $f>f_\text{max}$ are never produced.  It is worth mentioning here that we are working in the small-field limit, where $\phi_e\simeq\sqrt{2}\,n\,M_P$ and $H_e=\sqrt{\rho_e/\left(3\,M_P^2\right)}\simeq \Lambda^2\,\left(n\,\sqrt{2}\right)^n/\left(\sqrt{2}\,M_P\right)$. Using Eq.~\eqref{eq:freq} we can relate the ratio of the scale factors and the frequency $f$, arriving at
\begin{align}
  & \ogw^{(0), \rm{RH}} =
 \ogw^{(0), \rm{RD}} \,\widetilde{\mathcal{F}}(g_\star) \left[\frac{2\pi}{\sqrt{3}}\frac{T_0}{M_P}\,\left(\frac{30}{\pi^2\,g_{\star \rho}^\text{rh}}\right)^{-\frac{1}{4}}\right]^{\frac{4-2\,n}{2\,n-1}} \rho_\text{rh}^\frac{2-n}{4\,n-2}\,f^\frac{2\,n-4}{2\,n-1}\,,
\label{eq:ogw-rho}
\end{align}
with
\begin{align}
  \widetilde{\mathcal{F}}(g_\star) \equiv \left(\frac{\gss^0}{\gss^\text{rh}}\right)^{\frac{4-2\,n}{6\,n-3}} \frac{g_{\star
\rho}^\text{rh}}{g_{\star \rho, \text{RD}}^{\text{hc}}} \left(\frac{g_{\star s, \text{RD}}^{\text{hc}}}{g_{\star s}^{\text{rh}}} \right)^{4/3}\,,
\end{align}
which shows the blue-tilted behaviour of the spectrum for $\bar{w}>1/3$, which is equivalent to $n>2$. One can then, exploiting Eq.~\eqref{eq:rhoRrh} and Eq.~\eqref{eq:arh}, rewrite $\rho_\text{rh}$ in Eq.~\eqref{eq:ogw-rho} in terms of $\Gamma_\phi^e$ and remaining parameters  ($\Lambda, \beta, n$), so that the spectral GW energy density as a function of the frequency measured today reads
\begin{align}\label{eq:gw-decay}
& \ogw^{(0),\text{RH}}(f) \simeq \ogw^{(0), \rm{RD}}\,
\widetilde{\mathcal{F}}(g_\star) \left[\frac{2\pi}{\sqrt{3}}\frac{T_0}{M_P}\,\left(\frac{30}{\pi^2\,g_{\star \rho}^\text{rh}}\right)^{-\frac{1}{4}}\right]^{\frac{4-2\,n}{2\,n-1}}\,f^{\frac{2\,n-4}{2\,n-1}}\,
\left[\frac{3\,n}{n+1}\,(2\,n^2)^n\,\Lambda^4\right]^\frac{2-n}{4\,n-2}
\nonumber \\
& \times 
\begin{cases}
\left(\frac{2\,n}{n+1}\right)^\frac{n-2}{4\,n-2}\,\left[\frac{n+4-\beta\,(n+1)}{2\,n}\right]^\frac{\beta\,(n+1)\,(2-n)}{(2\,n-1)\,(\beta\,(n+1)-3\,n)}\,\left[\frac{(n\sqrt{2})^n\,\Lambda^2}{\sqrt{2}\,M_P\,\Gamma_\phi^e}\right]^\frac{3\,n\,(n-2)}{(1-2\,n)\,(\beta\,(1+n)-3n)}\,, &\beta \ll \frac{n+4}{n+1} \,,\\[8pt]
\vartheta^{\frac{n+1}{2n-1}} \mathcal{W}^{\frac{n+1}{1-2\,n}}(\vartheta)  \left[\ln{\left(\vartheta^{\frac{n+1}{2n-4}} \mathcal{W}^{\frac{n+1}{4-2\,n}}(\vartheta) \right)}\right]^{\frac{2-n}{4\,n-2}}\,\left[\frac{(n\sqrt{2})^n\,\Lambda^2}{\sqrt{2}\,M_P\,\Gamma_\phi^e}\right]^\frac{n-2}{4\,n-2}\,,  &\beta=\frac{n+4}{n+1}\,\\
\left(\frac{n+1}{2\,n}\right)^\frac{2-n}{4\,n-2}\,\left[\frac{\beta\,(n+1)-(n+4)}{2\,n}\right]^\frac{3\,n}{4\,n-2}\,\left[\frac{(n\sqrt{2})^n\,\Lambda^2}{\sqrt{2}\,M_P\,\Gamma_\phi^e}\right]^\frac{3\,n}{4\,n-2}\,, &\beta \gg \frac{n+4}{n+1}\,,
\end{cases}
\end{align}
with
\begin{align}
    \vartheta \equiv \left[\frac{(n\sqrt{2})^n\,\Lambda^2}{\sqrt{2}\,M_P\,\Gamma_\phi^e}\right]\,\frac{n-2}{n}\,.
\end{align}
The above expression explicitly shows how the proportionality coefficient of Eq.~\eqref{ogw-scal} depends on the inflaton decay width $\Gamma_\phi^e$ at the end of inflation. Since the only frequency dependence in Eq.~\eqref{eq:gw-decay} enters through $f^\frac{2\,n-4}{2\,n-1}$, one can write the full spectrum today in an approximate piecewise form as
\begin{align} \label{omega_res}
& \ogw^{(0)}(f)\sim \ogw^{(0),\text{RD}}\,
\begin{cases}
1\,, & f < f_c \\[8pt]
f^\frac{2\,n-4}{2\,n-1}\,, &  f_c < f < \fmax \\[8pt]
0\,, & f > \fmax
\end{cases}
\end{align}
where $f_c$ can be analytically computed as
\begin{align}
& f_c\simeq 
\left[\frac{n\,(2\,n^2)^n}{n+1}\,\frac{\sqrt{3}\,\pi}{20}\,g_{\star\rho}^\text{rh}\,\frac{\Lambda^4\,M_P}{T_0}\,\widetilde{\mathcal{F}}(g_\star)^\frac{8\,n-4}{4-2\,n}\right]^{1/4}
\nonumber \\
& \times 
\begin{cases}
\left(\frac{2\,n}{n+1}\right)^\frac{2\,n-1}{4-8\,n}\,\left[\frac{n+4-\beta\,(n+1)}{2\,n}\right]^\frac{3\,n}{2\,\left(\beta\,(n+1)-3\,n\right)}\,\left[\frac{(n\sqrt{2})^n\,\Lambda^2}{\sqrt{2}\,M_P\,\Gamma_\phi^e}\right]^\frac{3}{2\,(\beta\,(1+n)-3n)}\,, &\beta \ll \frac{n+4}{n+1} \,,\\[8pt]
\vartheta^{\frac{n+1}{4-2\,n}} \mathcal{W}^{\frac{n+1}{2\,n-4}}(\vartheta)  \left[\ln{\left(\vartheta^{\frac{n+1}{2n-4}} \mathcal{W}^{\frac{n+1}{4-2\,n}}(\vartheta) \right)}\right]^{1/2}\,\left[\frac{(n\sqrt{2})^n\,\Lambda^2}{\sqrt{2}\,M_P\,\Gamma_\phi^e}\right]^{-1/4}\,,  &\beta=\frac{n+4}{n+1}\,\\
\left(\frac{n+1}{2\,n}\right)^\frac{1}{4}\,\left[\frac{\beta\,(n+1)-(n+4)}{2\,n}\right]^\frac{3\,n}{8-4\,n}\,\left[\frac{(n\sqrt{2})^n\,\Lambda^2}{\sqrt{2}\,M_P\,\Gamma_\phi^e}\right]^\frac{3\,n}{4\,(1-2\,n)}\,, &\beta \gg \frac{n+4}{n+1}\,.
\end{cases}
\end{align}
Here we would like to emphasize that Eq.~\eqref{eq:gw-decay} is a general expression for GW energy density today, agnostic to the underlying inflaton coupling, and only relies on the fact that the inflaton decay width depends on the scale factor the way it has been assumed in Eq.~\eqref{eq:decay-gen}. In Sec.~\ref{sec:result} we present plots of $\ogw^{(0)}(f)$ calculated according to  Eq.~\eqref{omega_res} as a function of $f$ for different inflaton models and various choices of parameters  $\{g_{i\phi}\,,n\,,\Lambda\}$. It is worth noting the presence of the constant contribution for $f\lsim f_c$.

%%%%%%%%%%%%%%%%%%%%%
\section{Results and Discussion}
\label{sec:result}
%%%%%%%%%%%%%%%%%%%
So far, we have considered primordial gravitational waves (PGW) and the connection between the spectral energy density of PGW and the inflaton decay width at the end of inflation, $\Gamma_\phi^e$, i.e., we have found a correspondence between the particle physics model and GW spectrum. Hereafter we assume that all the inflaton decay products belong exclusively to the thermal bath, and they thermalize instantaneously after being produced\footnote{We have checked that this is indeed a viable assumption since it turned out that the process $SS \to VV$ provides fast thermalization of all SM decay products.}. In particular, we emphasize that the inflaton does not decay into dark matter (DM) pairs. Below, first, we review existing experimental constraints on the parameter space. Then, we present and discuss results for GWs spectral energy density within each inflaton model defined in Sec.~\ref{sec:CMB-coupling}.

%%%%%%%%%%%%%%%%%%%%
\subsection{Constraints on the model parameters}
%%%%%%%%%%%%%%%%%%%%
There exist experimental constraints on the parameter space $\{g_{i\phi}\,,\Lambda\,,n \}$  or, equivalently, upon 
$\{ \Gamma_{\phi \to f}^e\,,\Lambda, \, n\}$. They will be reviewed shortly below.
%%%%%%%%%%%%%%%%%%%%%
\subsubsection{$\DNeff$}
\label{sec:neff}
%%%%%%%%%%%%%%%%%%%%%%
As it has been shown in Sec.~\ref{sec:gw}, the energy density of GW background scales as $\rho_\text{GW}\propto a^{-4}$, i.e., the same way as that of radiation energy density. This implies that the GWs act as an additional source of radiation. Any observable capable of probing the background evolution of the Universe (and hence its energy content) has, therefore, the potential ability to constrain the GW energy density. In fact, it is possible to put an upper limit on the GWs energy density at the time of BBN and CMB decoupling. As it is well known, an upper bound on any extra radiation component, in addition to those of the SM, can be expressed in terms of the $\DNeff$. The number of effective relativistic degrees of freedom $N_\text{eff}$ is defined through the expression for the radiation energy density in the late Universe as
\begin{equation}
    \rho_\text{rad}(T\ll m_e) = \rho_\gamma + \rho_\nu + \rho_\text{GW} = \left[1 + \frac78 \left(\frac{T_\nu}{T_\gamma}\right)^4 N_\text{eff}\right] \rho_\gamma\,,
\end{equation}
where $\rho_\gamma$, $\rho_\nu$, and $\rho_\text{GW}$ correspond to the photon, SM neutrino, and GW energy densities, respectively, with $T_\nu/T_\gamma = (4/11)^{1/3}$. Note that, the relevant temperature here is the photon temperature after $e^+\,e^-$ annihilation. Within the SM, the prediction taking into account the non-instantaneous neutrino decoupling is $N_\text{eff}^\text{SM} = 3.044$~\cite{Dodelson:1992km, Hannestad:1995rs, Dolgov:1997mb, Mangano:2005cc, deSalas:2016ztq, EscuderoAbenza:2020cmq, Akita:2020szl, Froustey:2020mcq, Bennett:2020zkv}, whereas the presence of GW implies
\begin{align}
& \DNeff = N_\text{eff}-N_\text{eff}^\text{SM} = \frac{8}{7}\,\left(\frac{11}{4}\right)^\frac{4}{3}\,\left(\frac{\rho_\text{GW}(T)}{\rho_\gamma(T)}\right)\,.
\end{align}
Since $\rho_\text{GW}\propto a^{-4}$ and $\rho_\gamma\propto T^4$, the above relation can be utilized to put a constraint on the GW energy density redshifted to today via
\begin{align}\label{eq:GW-neff}
& \left(\frac{\rho_\text{GW}}{\rho_c}\right)\Bigg|_0\leq\Omega_\gamma^{(0)}\,\left(\frac{4}{11}\right)^\frac{4}{3}\,\frac{7}{8}\,\DNeff \simeq 5.62\times 10^{-6}\,\DNeff\,,
\end{align}
where, as before, $\Omega_\gamma^{(0)}\,h^2\simeq 2.47\times 10^{-5}$. The bound in Eq.~\eqref{eq:GW-neff} is applicable for an integrated energy density as follows
\begin{align}\label{eq:GW-neff2}
&  \left(\frac{h^2\,\rho_\text{GW}}{\rho_c}\right)\Bigg|_0 = \int_{f_\text{BBN}}^{\fmax}\,\frac{df}{f}\,h^2\,\ogw^{(0)}(f)\lesssim 5.62\times 10^{-6}\,\DNeff\,, 
\end{align}
where $f_\text{max}$ is given by Eq.~\eqref{eq:fmax}. Note that the upper limit of the integration is $\fmax$ for modes that match the size of the horizon at the end of inflation, while the lower limit can be $f_\text{BBN}$, which corresponds to the modes that match the horizon size at the time of BBN. One can then perform the integration analytically using Eq.~\eqref{eq:gw-decay}, and for all $f_\text{BBN}< f<\fmax$ show that $\ogw\,h^2(f)\lsim 2\times5.62\times10^{-6}\,\left(\frac{n-2}{2\,n-1}\right)\,\DNeff \lesssim 5.62 \times 10^{-6}\,\DNeff$~\cite{Boyle:2007zx,Kuroyanagi:2014nba,Caprini:2018mtu,Figueroa:2019paj}, for $n>2$. From Eq.~\eqref{eq:ogw-rho}, we note that, for $n>2$, $\ogw^{(0), \rm{RH}}$ is inversely proportional to a positive power of the radiation energy density at the end of reheating. This implies, an increase in $\rho_\text{rh}$ should relax the $\DNeff$ bound on $\ogw^{(0)}$. In the following subsections we will see, for instance, a larger inflaton-matter coupling remains safe from $\DNeff$ constraint as it leads to more radiation production, diluting the GW energy density.

Hereafter we use orange color to denote regions of the parameter space excluded by $\DNeff$ measurements. Within the framework of $\Lambda$CDM, the Planck legacy data produces $N_\text{eff} = 2.99 \pm 0.34$ at 95\% CL~\cite{Planck:2018vyg}. In our plots, this is shown by the solid orange horizontal line. Once the baryon acoustic oscillation (BAO) data are included, the measurement becomes more stringent: $N_\text{eff} = 2.99 \pm 0.17$. As computed in Ref.~\cite{Yeh:2022heq}, a combined BBN+CMB analysis shows $N_\text{eff} = 2.880 \pm 0.144$. We denote this by the dashed horizontal line. Upcoming CMB experiments like CMB-S4~\cite{Abazajian:2019eic} and CMB-HD~\cite{CMB-HD:2022bsz} will be sensitive to a precision of $\DNeff \simeq 0.06$ and $\DNeff \simeq 0.027$, respectively. These are denoted by dot-dashed and dotted lines in subsequent figures. The next generation of satellite missions, such as COrE~\cite{COrE:2011bfs} and Euclid~\cite{EUCLID:2011zbd}, shall improve the limit even further up to $\DNeff \lesssim 0.013$, as shown by the large dashed line. The orange-shaded region is thus disallowed, depending on the sensitivity of a particular experiment. We collectively named them ``$\DNeff$ bounds" in all the subsequent figures.

%%%%%%%%%%%%%
\subsubsection{\text{BBN}: $\Trh$}
\label{sec:trh-bbn}
%%%%%%%%%%%%%%%%
The reheating temperature can be obtained by evaluating Eq.\eqref{eq:Trh} at $a=\arh$ and utilizing Eq.\eqref{eq:arh}
\begin{align}\label{eq:Treh}
& \Trh \simeq  \left[\frac{45\,M_P}{\pi^2 \gsr}\,\frac{ 2^{\frac{n+3}{2}}\,n^{n+1}}{(n+1)}\,\Gamma_\phi^e\,\Lambda ^2\right]^\frac{1}{4}\,
\nonumber\\&\times
\begin{cases}
\left[\frac{n+4-\beta\,(n+1)}{n+1}\right]^\frac{3\,n}{2\,\left(\beta\,(n+1)-3\,n\right)}\,\left(2^\frac{n-3}{2}\,n^{n-1}\,\frac{\Lambda^2\,(n+1)}{\Gamma_\phi^e\,M_P}\right)^\frac{\beta\,(n+1)+3\,n}{4\,\left(\beta\,(n+1)-3\,n\right)}\,, & \beta \ll \frac{n+4}{n+1},\\
\left[2^{\frac{n-1}{2}}\,(n-2)\,n^{n-1}\,\frac{\Lambda ^2 }{\Gamma_\phi^e\,M_P}\right]^{\frac{n+1}{2 (n-2)}}\,\mathcal{W}^\frac{n+1}{4-2\,n}\ln\left[...\right]\,, & \beta = \frac{n+4}{n+1}\,,\\
\left[\frac{\beta\,(n+1)-(n+4)}{n+1}\right]^\frac{n+4}{4\,n-8}\,\left(2^\frac{n-3}{2}\,n^{n-1}\,\frac{\Lambda^2\,(n+1)}{\Gamma_\phi^e\,M_P}\right)^\frac{n+1}{2\,n-4}\,,    & \beta \gg \frac{n+4}{n+1}\,,
\end{cases}
\end{align}
where $[...]=\left[2^{\frac{n-1}{2}} (n-2) n^{n-1}\,\frac{\Lambda ^2 }{\Gamma_\phi^e\,M_P}\right]^{\frac{n+1}{2 (n-2)}}$. Note that when the reheating scenario is specified, $\beta$ becomes a given function of $n$. For instance, for inflaton-fermion Yukawa coupling, $\beta=\beta_\psi =3(n-1)/(n+1)$. However, regardless the explicit form of $\beta(n)$, in the middle case $\Trh$ does not depend on $\beta$. When a particular model is chosen, corresponding $n$ for the middle case is fixed, e.g., for Yukawa interaction, this corresponds to $n=7/2$, while the most upper and most lower cases correspond to $n \ll 7/2$ and $n \gg 7/2$, respectively.

To avoid conflict with BBN predictions, we require $\Trh\gtrsim 4$ MeV~\cite{Sarkar:1995dd,Kawasaki:1994af,Kawasaki:2000en,Hannestad:2004px,DeBernardis:2008zz,deSalas:2015glj,Hasegawa:2019jsa}. As one can understand from Eq.~\eqref{eq:Treh}, this requirement puts a constraint on the inflaton couplings (through $\Gamma_\phi^e$) and potential parameters $\Lambda$ and $n$. Here we would like to mention that for $\bar w\gtrsim 0.65$ (equivalently $n\gtrsim 5$), purely gravitational reheating (PGR) becomes important~\cite{Haque:2022kez,Clery:2022wib,Co:2022bgh,Haque:2023yra} and may dominate over perturbative reheating as shown in Ref.~\cite{Haque:2022kez,Haque:2023yra}. Namely, for $\bar w\lesssim 0.65$, PGR alone cannot successfully reheat the Universe because i) the energy density of the SM radiation redshifts faster than the inflaton energy ($\bar{w} \leq 0$), and ii) the reheating temperature is below the BBN bound ($\bar w\lesssim 0.65$). Since our goal is to predict constraints on perturbative inflaton-matter coupling, hereafter, we consider $n\in\left[2\,,5\right]$, where the gravitational reheating could be neglected. For completeness, we would like to mention that there exist several other kinds of reheating mechanisms. For example, in {\it minimal preheating} scenario, discussed in~\cite{Brandenberger:2023zpx}, the
oscillating inflaton condensate directly transfers its energy to the SM Higgs field, even in the absence of a direct inflaton-Higgs coupling. Such a process, which relies typically on the tachyonic instablities~\cite{Kofman:1994rk}, turns out to be very efficient compared to the perturbative reheating scenario. Also, preheating arising from tri-linear vertex can feature {\it tachyonic resonance}~\cite{Kofman:1997yn}, that can lead to non-perturbative copious particle production immediately after inflation. As it has recently been shown in~\cite{Garcia:2023eol}, the
presence of inflaton self-interactions can lead to inflaton {\it fragmentation} that can severely influence the reheating dynamics. A complete analysis including all such collective effects requires a dedicated lattice simulation~\cite{Figueroa:2017vfa,Cosme:2022htl,Garcia:2023eol}, which we plan to address in a future work.

%%%%%%%%%%%%%
\subsubsection{\text{Inflation}: $\Lambda$}
\label{sec:inf-lambda}
%%%%%%%%%%%%%%%%
CMB measurement of the inflationary observables $r$ (the tensor-to-scalar ratio) and $n_s$ (the spectral index) puts an upper limit on the Hubble scale at the end of inflation, that can be recast into a limit on the scale of inflation $\Lambda$ so that $\Lambda\lesssim 10^{16}$ GeV, see Appendix.~\ref{sec:app-inf} for details.
In addition, adopting the end-of-inflation condition $\epsilon_V(\phi_e) \simeq 1$, one can determine the inflaton field strength at the end of inflation $\phi_e$.  

%%%%%%%%%%%%%%%%%%%%%%%%%%%%%
\begin{figure}[htb!]
    \centering
    \includegraphics[scale=0.32]{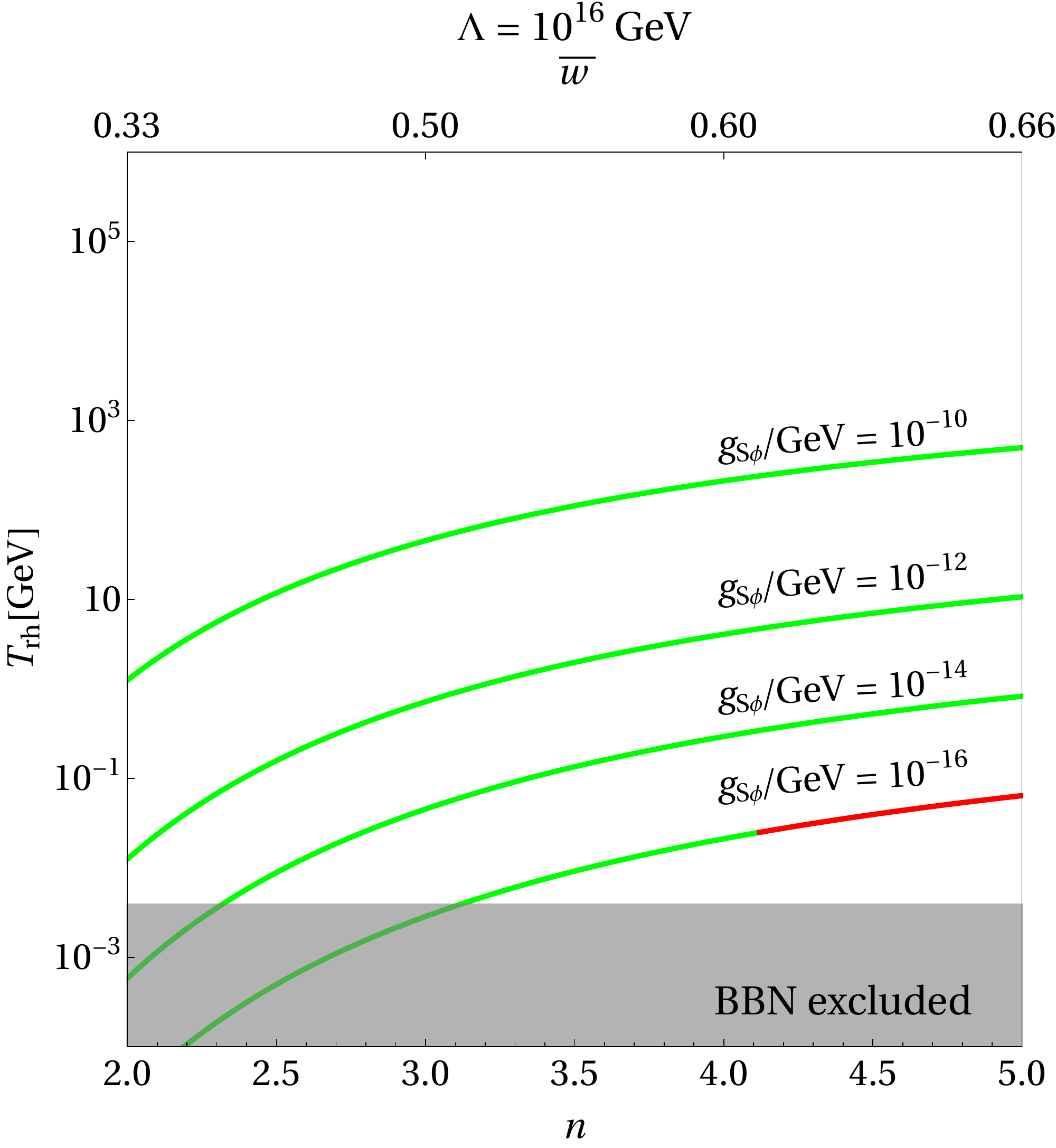}~~~~\includegraphics[scale=0.32]{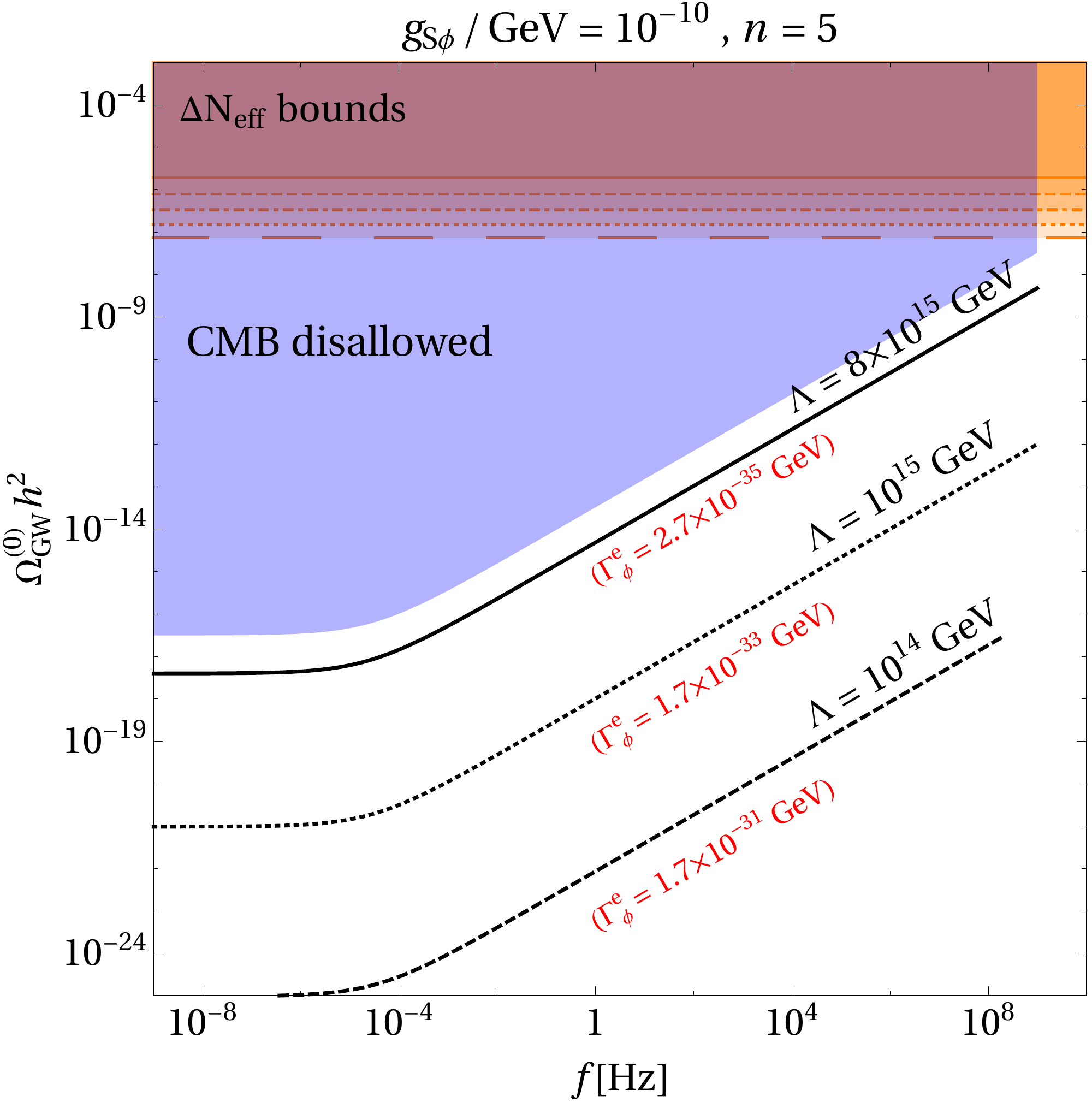}\\[10pt]
    \includegraphics[scale=0.32]{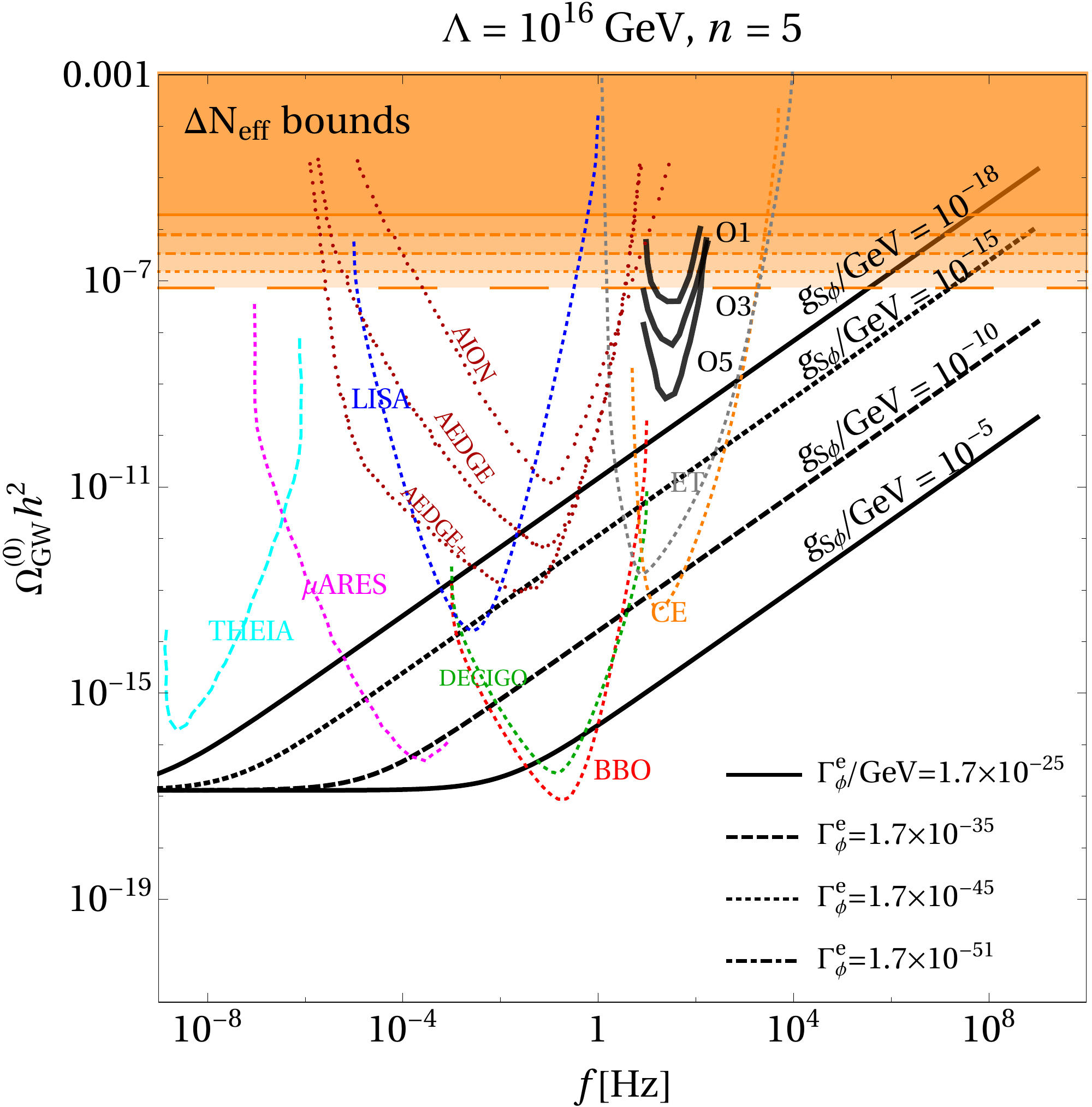}~~~~\includegraphics[scale=0.32]{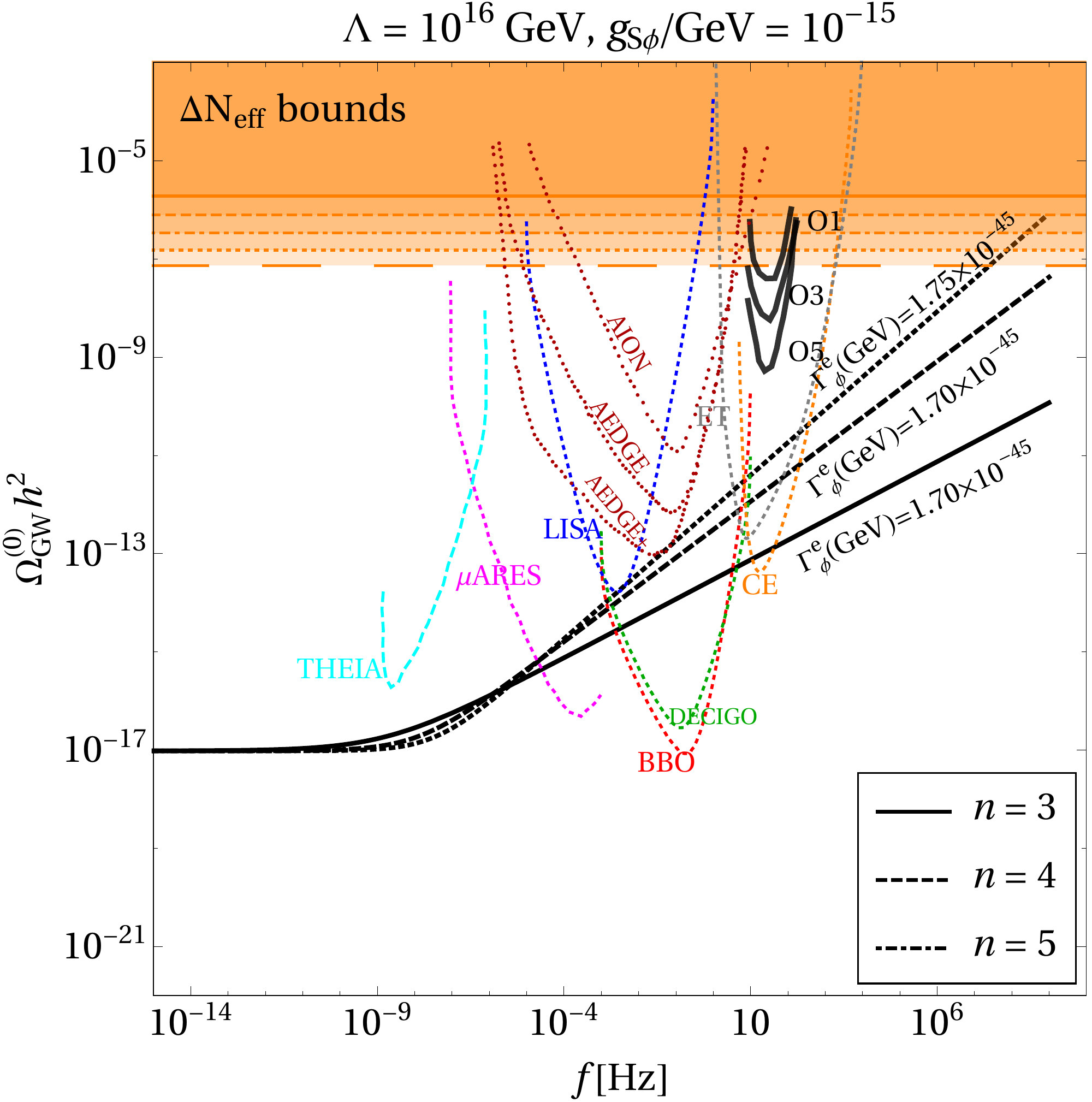}
    \caption{ $\phi\to SS$ scenario. {\it Top Left:} Reheating temperature $\Trh$ as a function of $n$. The red part of the curve is discarded from present bound on $\DNeff$ due to PLANCK~\cite{Planck:2018vyg}. The grey-shaded region is in conflict with the BBN limit on $\Trh$.{\it Top Right:} $\ogw^{(0)}$ as a function of frequency, where different curves correspond to different choices of $\Lambda$ for a fixed $\gsp$ and $n$.The orange-shaded regions are discarded from $\DNeff$ due to overproduction of GWs.  {\it Bottom Left:} Same as top right, but for fixed $\Lambda$ and $n$, with different choices of $\gsp$ specified below the curves. {\it Bottom Right:} Same as bottom left but for a fixed $\Lambda$ and $\gsp$, different curves correspond to different $n$ values. In the bottom and upper-right panels the parameters $\Lambda$, $n$ and $g_{S\phi}$ are chosen so that the BBN limit $\Trh > 4$~MeV is satisfied [cf. Fig.~\ref{fig:summary}]. Moreover, in the bottom panels, we show sensitivities of future GWs experiments. }
    \label{fig:gw-ss}
\end{figure}
%%%%%%%%%%%%%%%%%%%%%%%%%%%

%%%%%%%%%%%%%%%%%%
\subsection{Scenario I: $\phi\to SS$}
%%%%%%%%%%%%%%%%%%
Now, we are ready to discuss various reheating models.
We begin with the tri-linear scalar interaction involving the inflaton and a pair of real scalars $S$. Adopting Eq.~\eqref{eq:decay-msv}, the following result for $\phi\to SS$ decay width has been obtained
\begin{align}
\Gamma_{\phi \rightarrow SS}^e \simeq
\frac{1}{\sqrt{6\,\pi}}\,\left(\frac{3}{2}\right)^\frac{1}{2\,n}\,\frac{(n+1)\,n^{1-n}}{2^{n/2}}\,\left(\frac{\Gamma\left[\frac{n+1}{2n}\right]}{\Gamma\left[\frac{1}{2n}\right]}\right)\,\frac{g_{S\phi}^2\,M_P}{\Lambda^2}\,\sum \ell\,\left|\mathcal{P}_\ell\right|^2\,,
\end{align}
with $\beta_S= \frac{3(1-n)}{n+1}$. 
Also note that $\beta_S=(n+4)/(n+1)$ for $n = 1/2$, so it happens to be outside of our region of variation for $2<n<5$. Unlike the model-independent analysis before, here, for each $n$, the exponent $\beta_S$ of the decay width is fixed by construction. Now, assuming reheating takes place through the decay of inflaton into scalars, we can compute the temperature $\Trh$ at the end of reheating from Eq.~\eqref{eq:Treh}. This is, of course, a function of the decay width itself, which in turn makes it dependent on the inflaton-scalar coupling $\gsp$. Note that the same coupling strength also controls GW spectral energy density [cf. Eq.~\eqref{eq:gw-decay}]. The question that one can therefore ask is, for what size of $\gsp$ is it possible to have GW energy density that may have some detection prospects satisfying $\DNeff$ constraint, together with successful reheating, keeping the BBN predictions unharmed.

In the top left panel of Fig.~\ref{fig:gw-ss}, we choose different coupling strengths and compute corresponding $\Trh$ as a function of $n$ for fixed inflationary scale $\Lambda=10^{15}$ GeV. From this figure, one can see that for inflaton-scalar coupling $\gsp \gtrsim 10^{-12} \; \rm{GeV}$, the reheating temperature exceeds the BBN bound $\Trh\gtrsim 4$ MeV for all considered values of n. Moreover, the strongest the $\gsp$ coupling, the highest $\Trh$. This can be understood from Eq.~\eqref{eq:Treh} noting that $\Trh\propto \left[\left(\gsp\right)^n/\Lambda\right]^{1/(2\,n-1)}$, hence a large coupling leads to higher reheating temperature for $n>2$. Following Eq.~\eqref{eq:gw-decay}, the GW spectral energy density goes as $\ogw^{(0)}\propto \left(\Lambda^{8\,n-7}/\gsp^{n-2}\right)^\frac{2n}{(1-2\,n)^2}$. Therefore, for $n>7/8$ and given $\gsp$, a large scale of inflation $\Lambda$ implies the overproduction of GWs. This is evident from the top right panel of Fig.~\ref{fig:gw-ss}, where we see that a large $\Lambda$ can easily be in conflict with $\DNeff$ constraint (albeit there exists an upper bound from CMB). Also, in the top left panel, the red curves correspond to the spectral energy density of GW that is ruled out from the present bound on $\DNeff$ due to PLANCK~\cite{Planck:2018vyg}. As expected, this bound is stronger for smaller $\gsp$, following the argument adopted above. The same feature is also visible from the bottom left panel. Finally, in the bottom right panel, we show the shape of  $\ogw^{(0)}(f)$ for different choices of $n$, where we see the spectral energy density corresponding to $n=3\,,4$ lies within reach of the proposed GWs detectors while remaining beyond the reach of the future $\DNeff$ constraints. We show existing and expected sensitivity reaches\footnote{Here we have used the sensitivity curves derived in Ref.~\cite{Schmitz:2020syl}.} from LIGO~\cite{LIGOScientific:2016aoc,LIGOScientific:2016sjg,LIGOScientific:2017bnn,LIGOScientific:2017vox,LIGOScientific:2017ycc,LIGOScientific:2017vwq}, LISA~\cite{2017arXiv170200786A,Baker:2019nia}, CE~\cite{LIGOScientific:2016wof,Reitze:2019iox}, ET~\cite{Punturo:2010zz,Hild:2010id}, BBO~\cite{Crowder:2005nr,Corbin:2005ny}, DECIGO~\cite{Seto:2001qf,Kudoh:2005as,Nakayama:2009ce,Yagi:2011wg,Kawamura:2020pcg}, $\mu$-ARES~\cite{Sesana:2019vho}, THEIA~\cite{10.3389/fspas.2018.00011}, AEDGE~\cite{AEDGE:2019nxb,Badurina:2021rgt} and AION~\cite{Badurina:2021rgt,Graham:2016plp,Graham:2017pmn,Badurina:2019hst} in the bottom panel, and in  the subsequent figures in this section.

%%%%%%%%%%%%%%%%%%%%%%%%
\subsection{Scenario II: $\phi\to \psi\psi$}
%%%%%%%%%%%%%%%%%%%%
For the fermionic final state, the inflaton decay width at the end of inflation is given by
\begin{align}
\Gamma_{\phi \rightarrow \psi \bar{\psi}}^e \simeq
g_{\psi\phi}^2\,\frac{n\,(1+n)}{4}\,\sqrt{\frac{3\,\pi}{2}}\,\left(\frac{2}{3}\right)^\frac{1}{2\,n}\,\left(\sqrt{2}\,n\right)^n\,\left(\frac{\Gamma\left[\frac{n+1}{2n}\right]}{\Gamma\left[\frac{1}{2n}\right]}\right)^3\,\frac{\Lambda^2}{M_P}\,\sum \ell^3\,\left|\mathcal{P}_\ell\right|^2\,, \end{align}
%%%%%%%%%%%%%%%%%%%%%%%%%%%%%
\begin{figure}[htb!]
    \centering
    \includegraphics[scale=0.32]{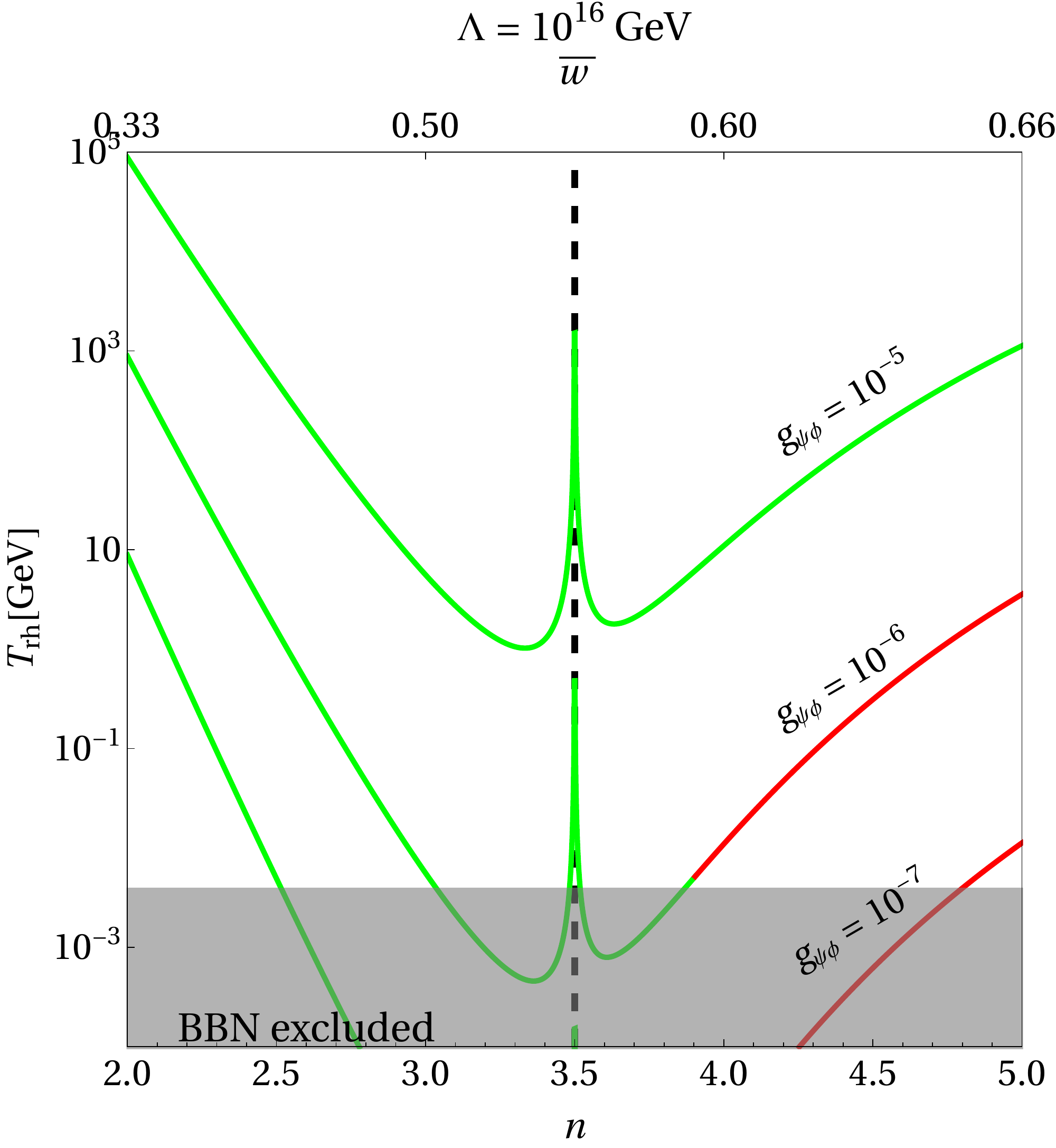}~~~~\includegraphics[scale=0.32]{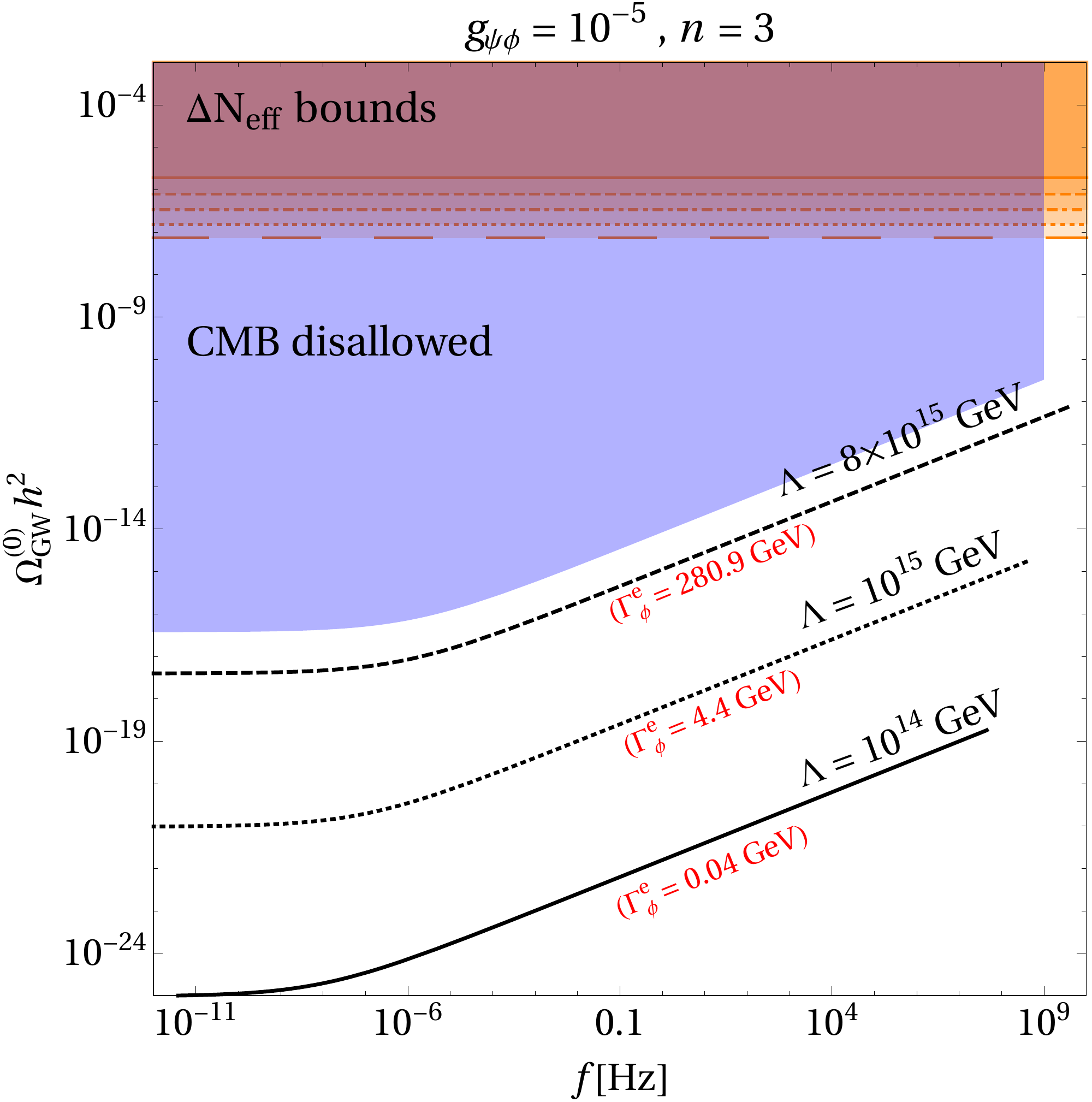}\\[10pt]
    \includegraphics[scale=0.32]{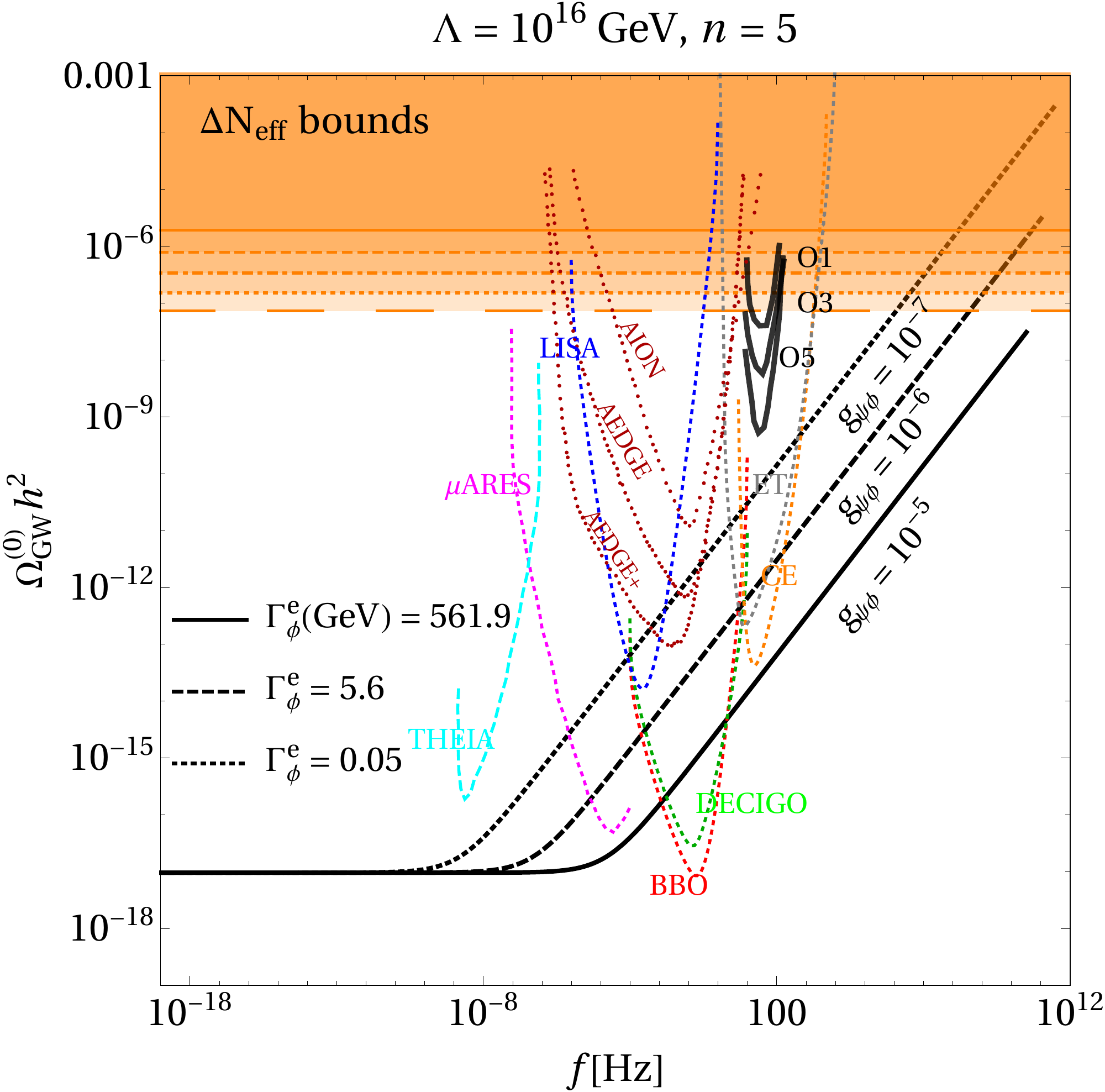}~~~~\includegraphics[scale=0.32]{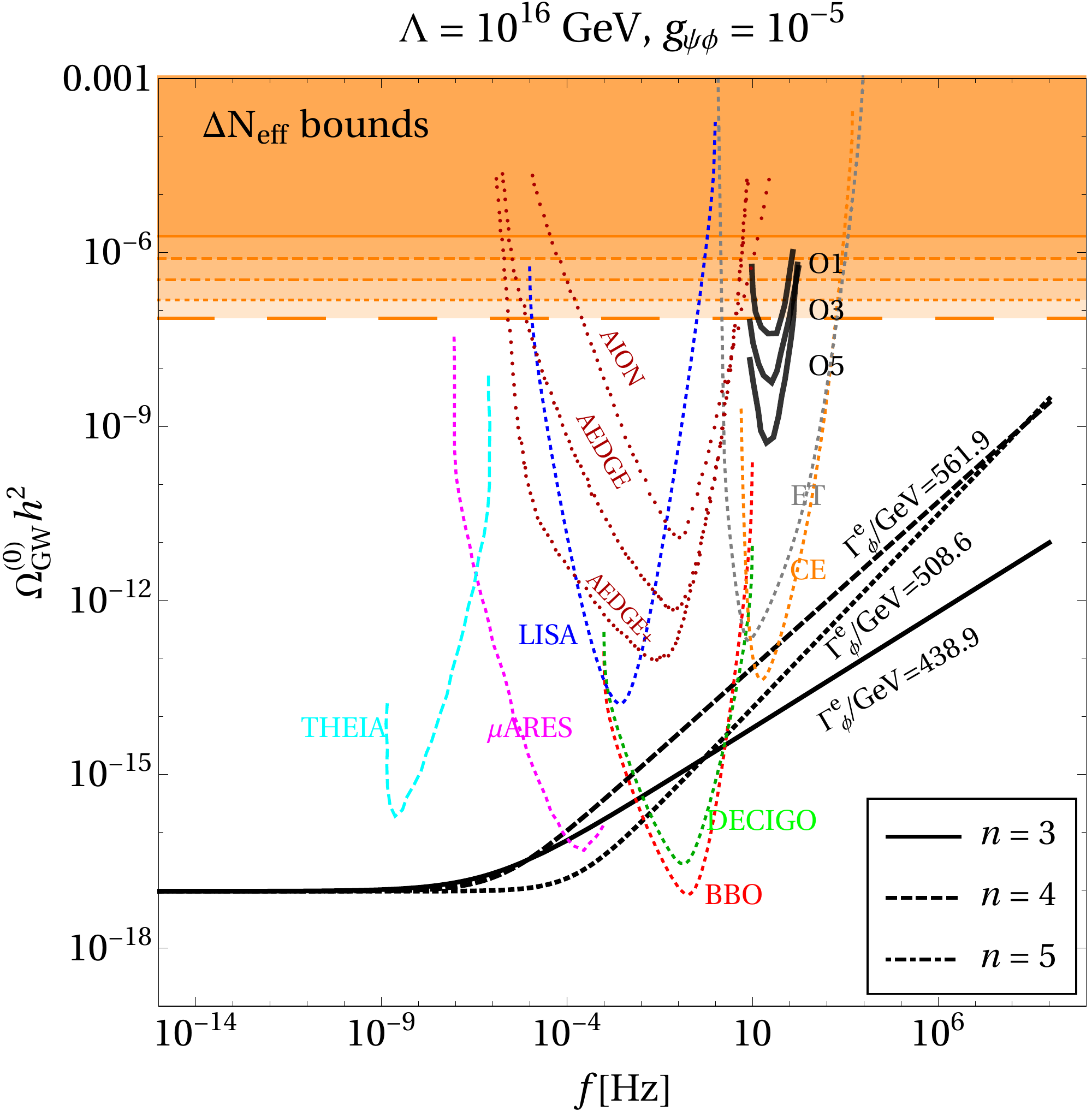}
    \caption{Same as Fig.~\ref{fig:gw-ss}, but for $\phi\to \psi\psi$ scenario. In the top left panel the vertical dashed line corresponds to $n=7/2$, for which $\beta_\psi=(n+4)/(n+1)$ (see text for details).}
    \label{fig:gw-ff}
\end{figure}
%%%%%%%%%%%%%%%%%%%%%%%%%%%
following Eq.~\eqref{eq:decay-massls}, with $\beta_\psi=3\,(n-1)/(n+1)$. Note that, in this case, $\beta_\psi\gtrless (n+4)/(n+1)$ for $n \gtrless 7/2$. We see the effect of this transition in the top left panel of Fig.~\ref{fig:gw-ff}, where the evolution of the reheating temperature has different behaviour around $n=7/2$. This can be understood from Eq.~\eqref{eq:Treh}, where, for a fixed $\Lambda$ and $\gpp$, one finds that
\begin{align}
& \Trh\propto
\begin{cases}
\frac{2^{\frac{n}{4}\,(4-n)}}{\left[\left(7-2\,n\right)\,n^{n-2}\right]^{n/2}} \,,&\,n \ll 7/2\,,\\
\left[\frac{2^{n\,(5-n)}}{n^{2\,n\,(n+1)}\,(2\,n-7)^{6\,n}}\right]^\frac{1}{8\,n-16}\,,&\,n \gg 7/2\,.
\end{cases}
\end{align}
This shows, $\Trh$ decreases with $n$ for $2<n\ll 7/2$, while for $7/2 \gg n<5$, an increase is observed. As before, we see from the bottom left panel, for fixed $n=3$, small couplings are disallowed from the present $\DNeff$ constraint on $\ogw^{(0)}$. This is understandable from Eq.~\eqref{eq:gw-decay}, as for $n=3$, the GW spectral energy density behaves as $\ogw^{(0)}\propto\left(\Lambda^6/\gpp\right)^{3/5}$, implying, a too small coupling may result in $\ogw^{(0)}$ overproduction for a given $\Lambda$. On the other hand, it also shows that large $\Lambda$ may overproduce GW energy density for a fixed $n$ and coupling strength, as seen in the top right panel. In the bottom right panel, we show different values of $n$ that are within reach of several future experiments for $\gpp=10^{-5}$ and $\Lambda=10^{16}$ GeV.     

%%%%%%%%%%%%%%%%%%%%%%
\subsection{Scenario III: $\phi\to aa$}
%%%%%%%%%%%%%%%%%%%%%%%%
The derivative interaction between inflaton and a pair of scalar bosons give rise to a decay width of
\begin{align}\label{eq:game-aa}
& \Gamma_{\phi \rightarrow aa}^e \simeq
\frac{h_{a\phi}^2}{3}\,\left(\frac{2}{3}\right)^\frac{3}{2\,n}\,\left(\frac{3\,\pi}{8}\right)^\frac{3}{2}\,2^{3n/2}\,\left(\frac{\Gamma\left[\frac{n+1}{2n}\right]}{\Gamma\left[\frac{1}{2n}\right]}\right)^5\,\frac{\Lambda^{6}}{M_P^3}\,\sum \ell^5\,\left|\mathcal{P}_\ell\right|^2\,,
\end{align}
with $\beta_a=\frac{9(n-1)}{1+n}$. In this case $\beta_a=(n+4)/(n+1)$ for $n=13/8$, so it is beyond our range of variation for $n$.
%%%%%%%%%%%%%%%%%%%%%%%%%%%%%
\begin{figure}[htb!]
    \centering
    \includegraphics[scale=0.32]{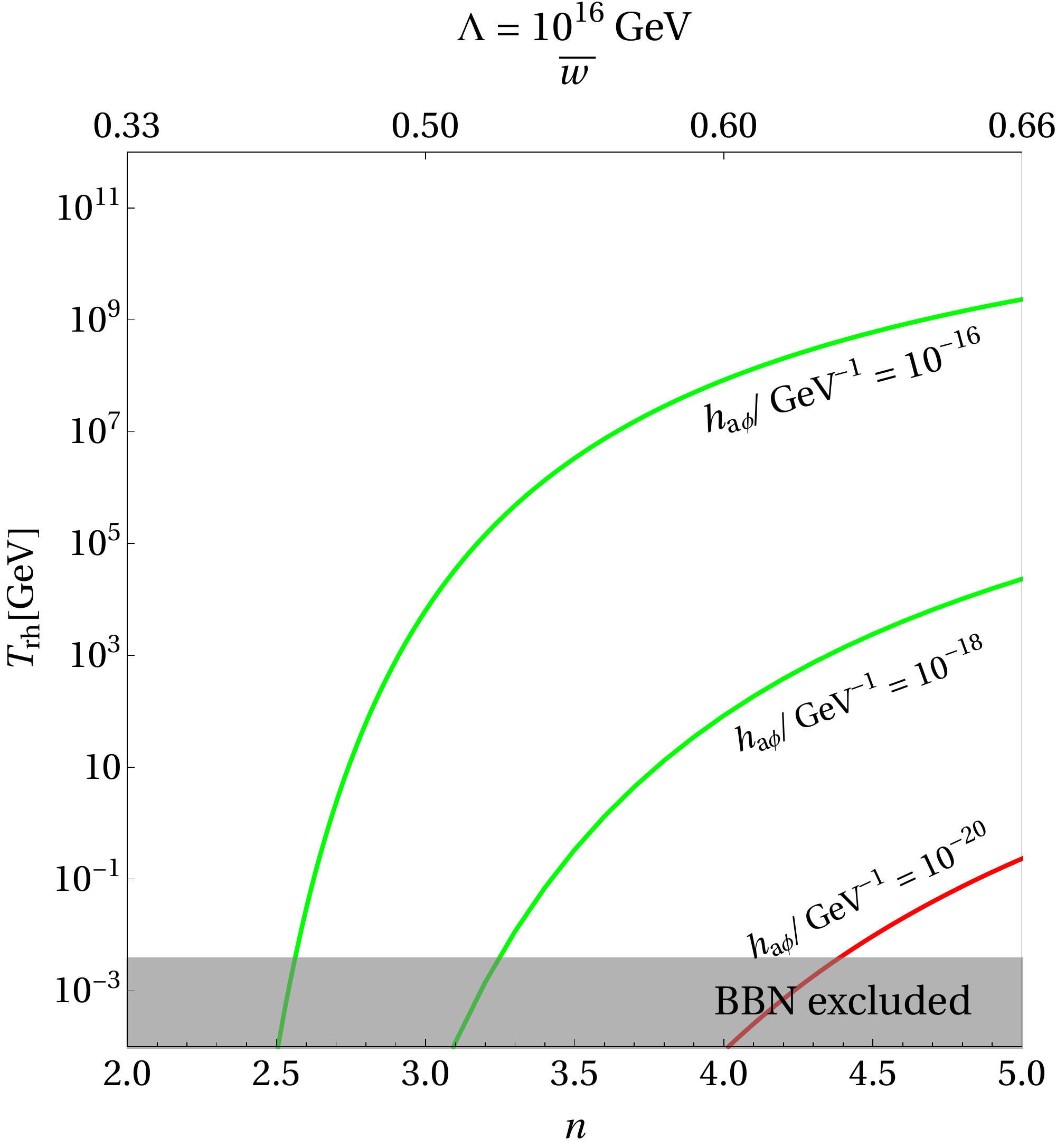}~~~~\includegraphics[scale=0.32]{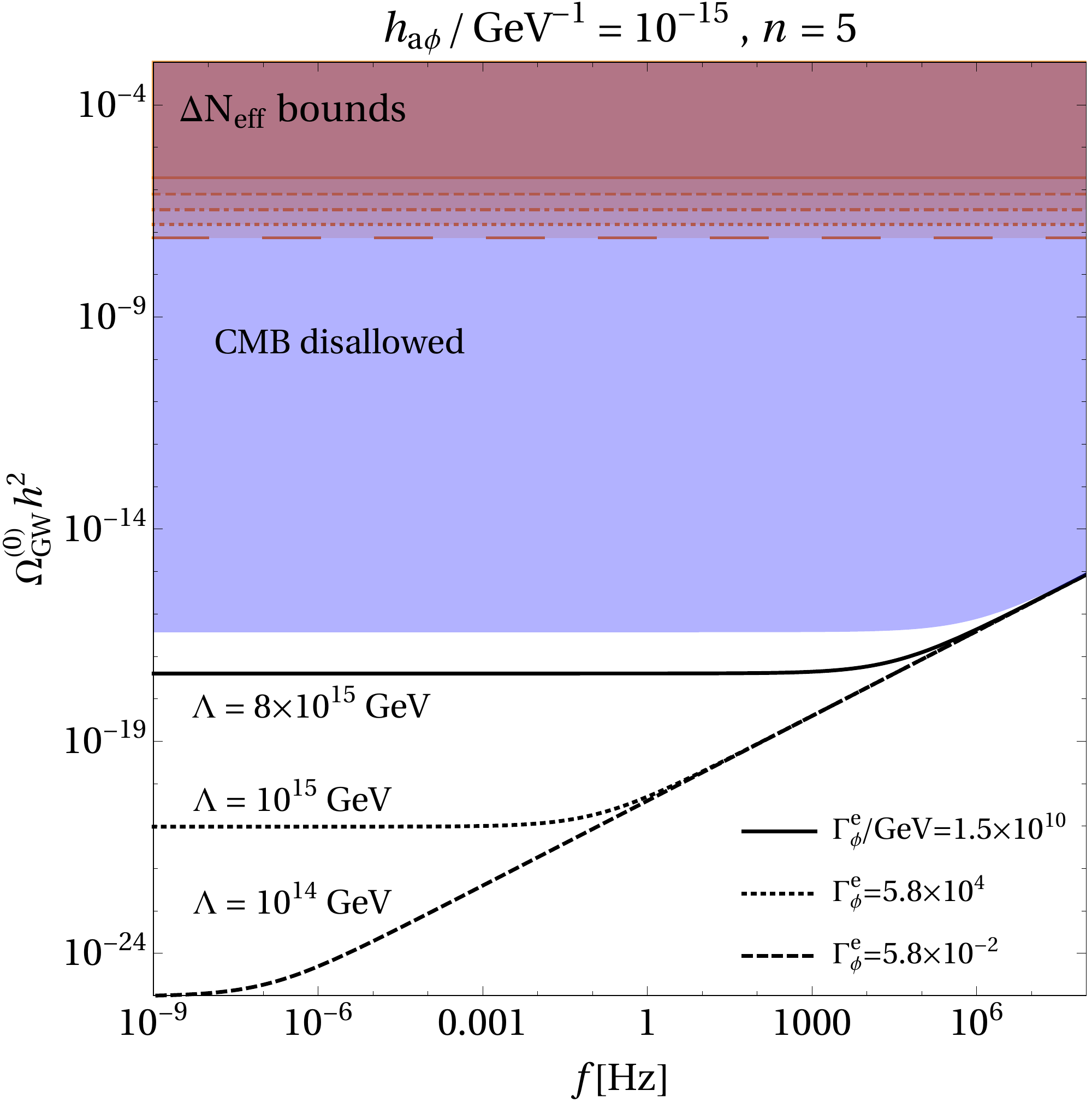}\\[10pt]
    \includegraphics[scale=0.32]{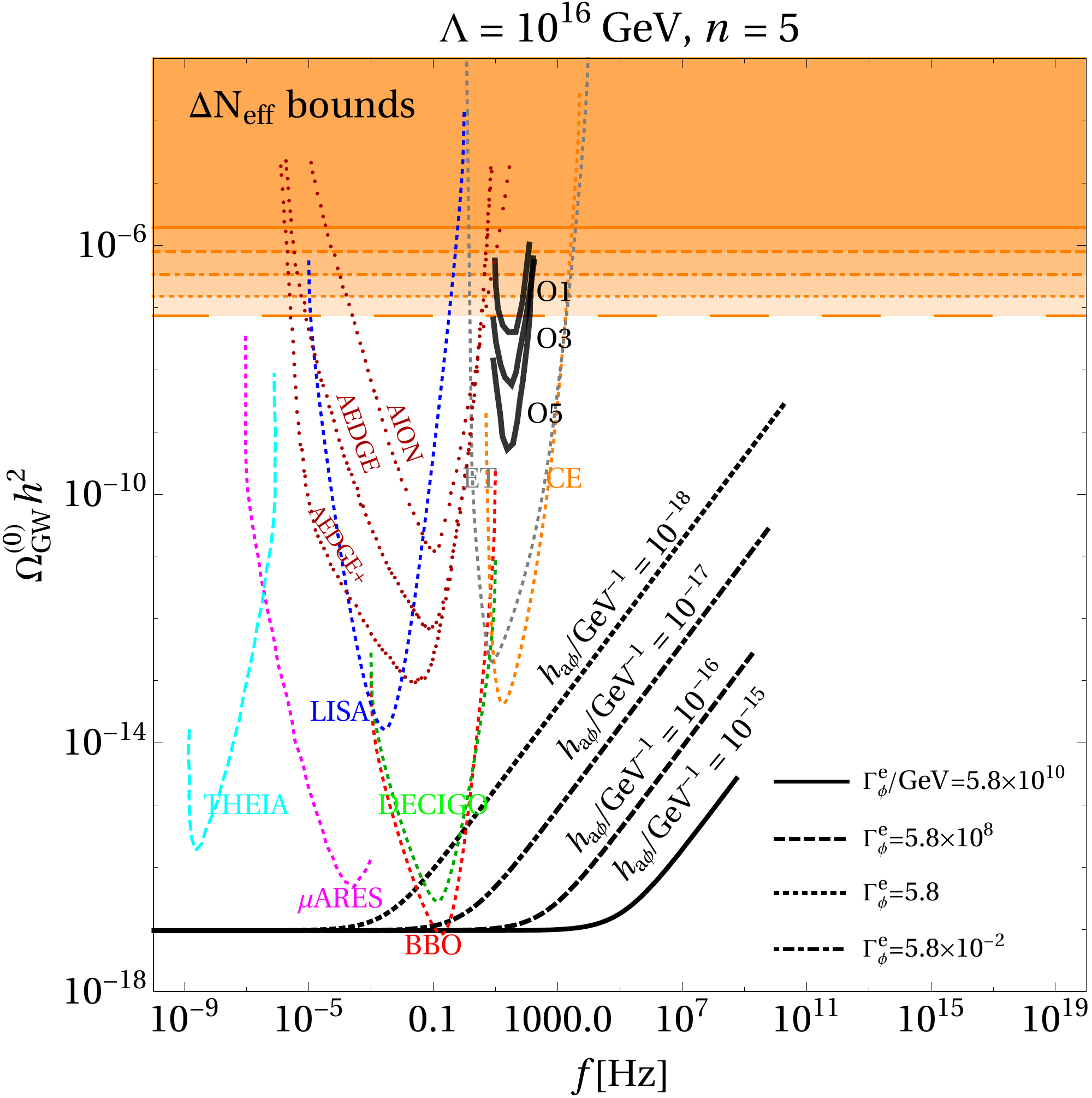}~~~~\includegraphics[scale=0.32]{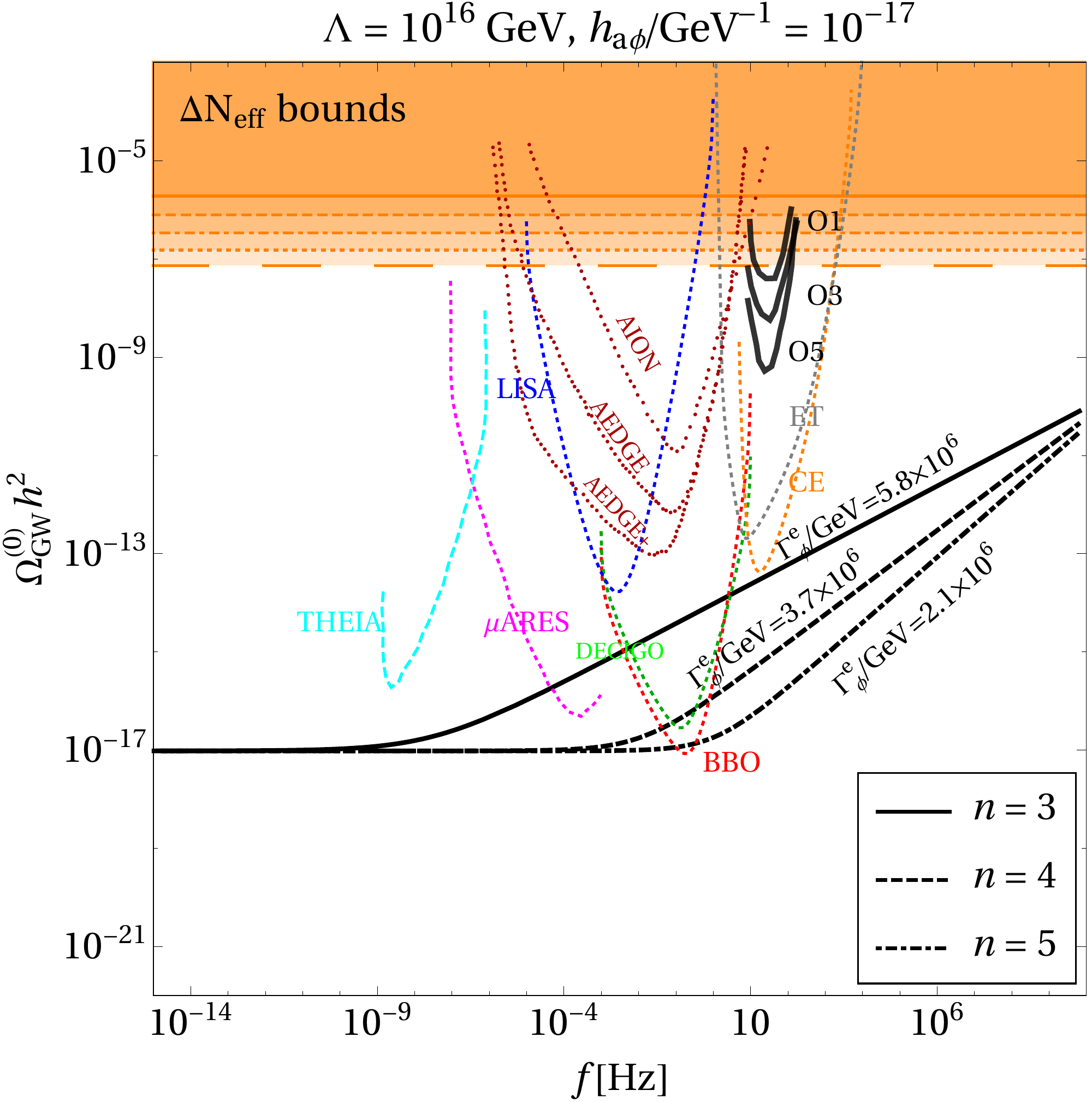}
    \caption{Same as Fig.~\ref{fig:gw-ss}, but for $\phi\to aa$ scenario. }
    \label{fig:gw-aa}
\end{figure}
%%%%%%%%%%%%%%%%%%%%%%%%%%%%%%%%%
Note that the $h_{a\phi}$ coupling has an inverse mass dimension. We find that for a fixed $\Lambda=10^{16}$ GeV, couplings weaker than $\gap\lesssim 10^{-20}\,\text{GeV}^{-1}$ are entirely forbidden from present $\DNeff$ constraint on $\ogw^{(0)}$. This is shown via the red curve in the top left panel of Fig.~\ref{fig:gw-aa}. One can understand this behaviour from the fact that for $n> 2$, smaller couplings lead to GW overproduction, as $\ogw^{(0)}\propto\gap^\frac{3\,n}{1-2\,n}$. We fix $n=5$ as a benchmark value and show the dependence of $\ogw^{(0)}(f)$ on the scale of inflation $\Lambda$ for a given $\gap$ in the top right panel. It is interesting to note here that the cut-off frequency $\fmax$ is independent of the choice of the scale of inflation $\Lambda$. As mentioned, for $n>2$, one cannot go to a very small coupling since it results in GW overproduction and contradicts $\DNeff$ bounds. This can be realized from the bottom left panel. However, for $\gap= 10^{-17}\,\text{GeV}^{-1}$ the GW amplitude falls within reach of the future detector sensitivities, still satisfying $\DNeff$ constraint, as shown in the bottom right panel. 
%%%%%%%%%%%%%
\subsection{Combined limits}
\begin{figure}[htb!]
    \centering
    \includegraphics[scale=0.34]{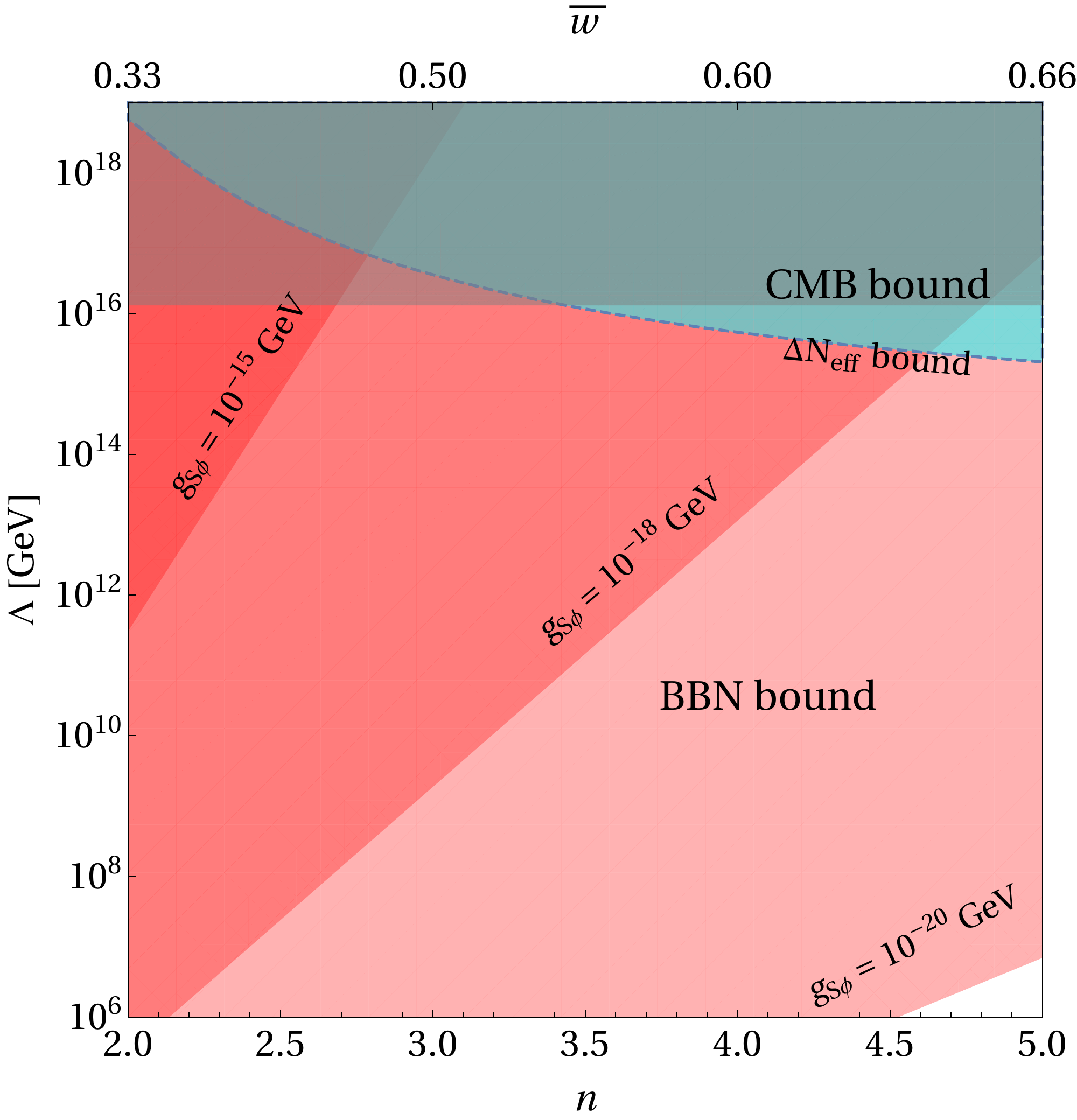}~~\includegraphics[scale=0.34]{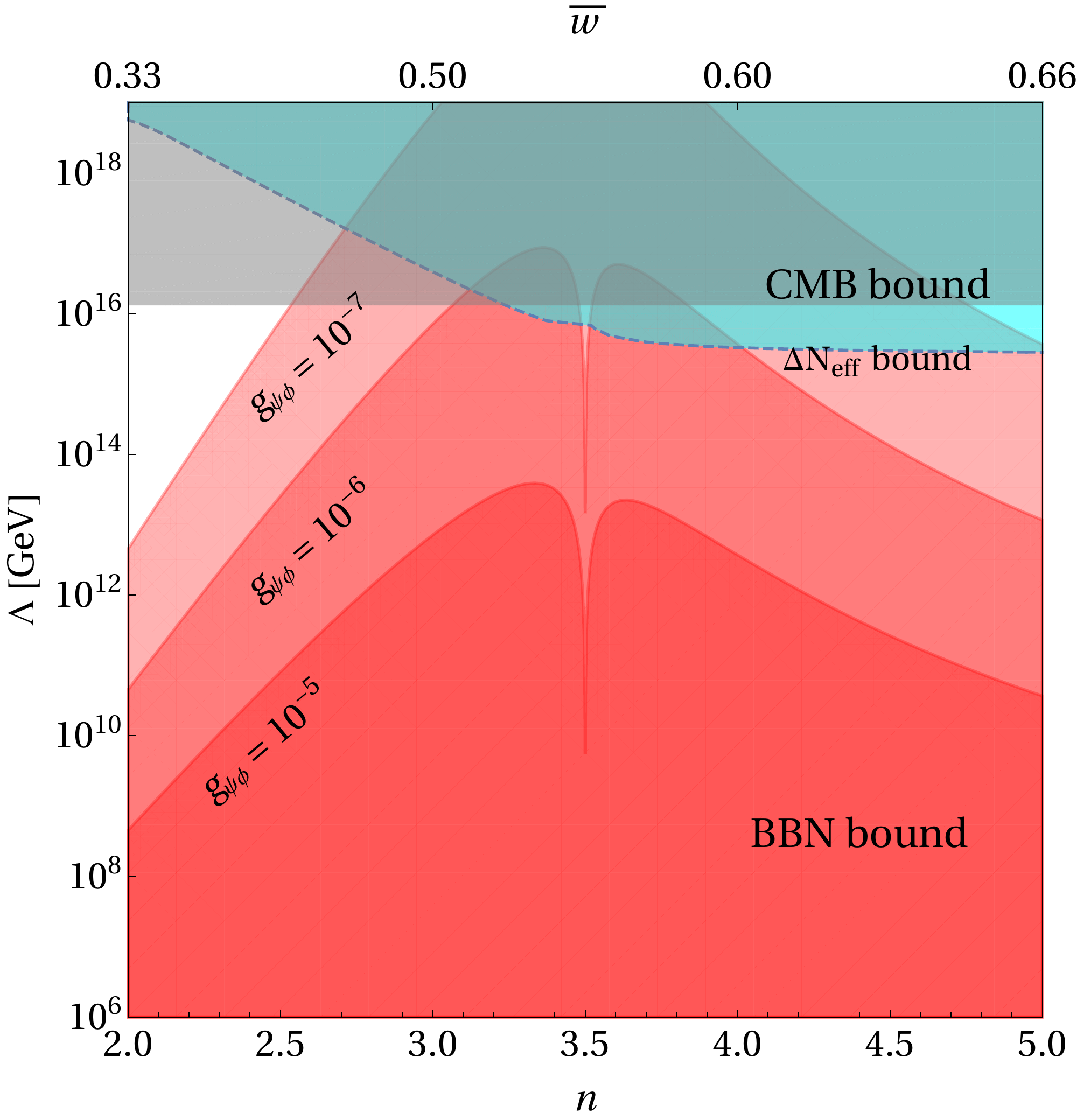}\\[10pt]
    \includegraphics[scale=0.34]{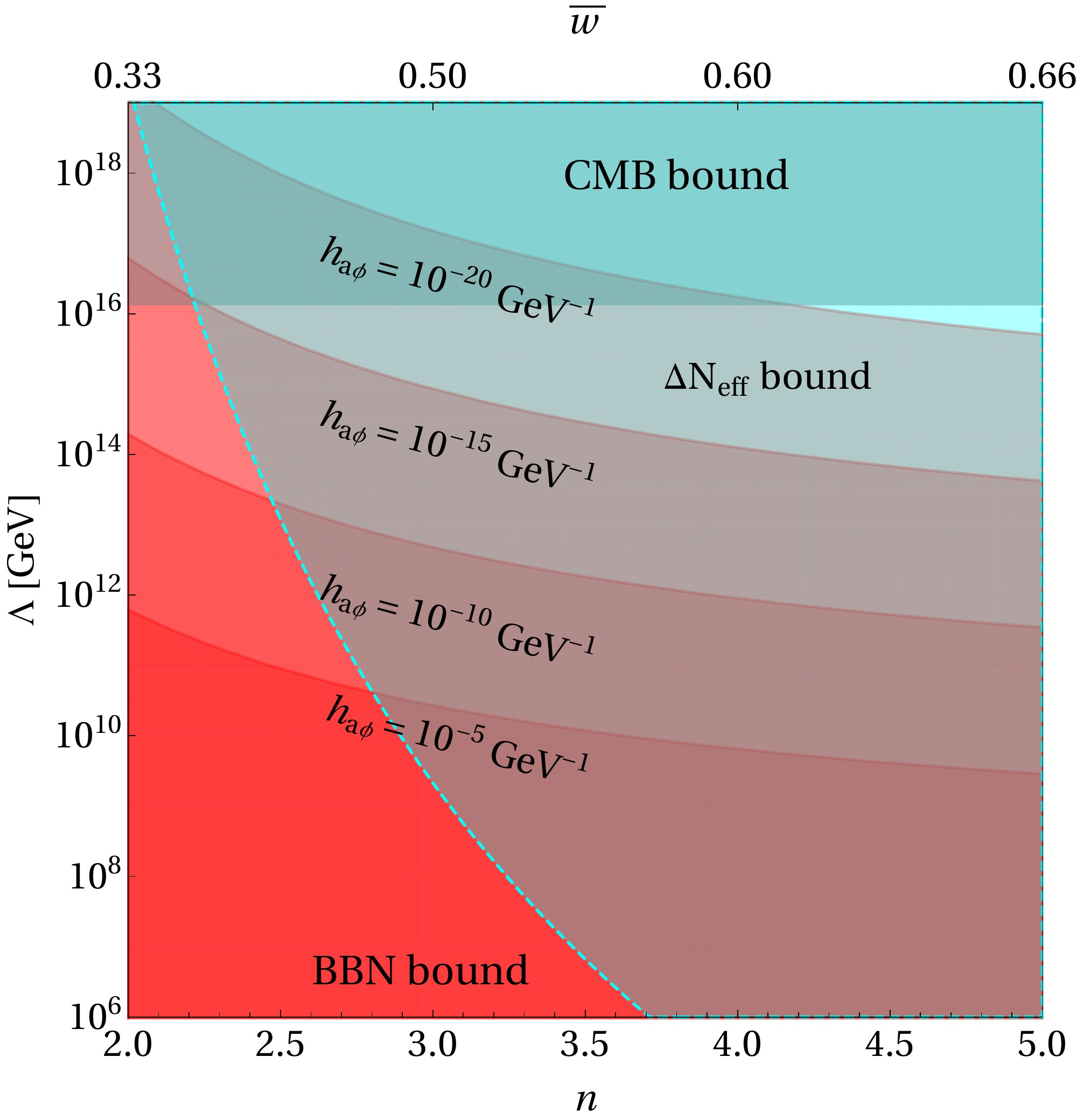}
    \caption{$\phi\to SS$ (top left), $\phi\to\psi\psi$ (top right) and $\phi\to aa$ (bottom) scenario. In all cases the red shaded region is disallowed from BBN bound on reheating temperature $\Trh<4$ MeV, the cyan dashed region is forbidden from PLANCK observed $\DNeff$ bound on GW overproduction at $f=\fmax$, and the ``CMB bound" discards scale of inflation $\Lambda>1.34\times 10^{16}$ GeV from constraint on tensor to scalar ratio.
    }
    \label{fig:summary}
\end{figure}

%%%%%%%%%%%%%%%
In Fig.~\ref{fig:summary}, we present the parameter space that remains allowed after imposing the limits discussed in the above subsections.  The red shaded regions are disallowed from BBN bound on reheating temperature, that requires $\Trh>4$ MeV [cf. subsection.~\ref{sec:trh-bbn}], for different choices of the couplings. The ``CMB bound", shown via the gray shaded region parallel to the horizontal axes, implies bound on the scale of inflation $\Lambda\lesssim 10^{16}$ GeV, from CMB observables [cf. subsection.~\ref{sec:inf-lambda}]. The cyan shaded regions are forbidden from $\DNeff$ bound from PLANCK on overproduction of GW energy density [cf. subsection.~\ref{sec:neff}]. Note that, in case of derivative interaction, the cyan region corresponds only to $\hap=10^{-20}\,\text{GeV}^{-1}$, as for other values of $\hap$ this bound is absent. Before moving on we would like to clarify that limits on tri-linear inflaton couplings from CMB observables have been derived in, for example, Ref.~\cite{Ueno:2016dim,Drewes:2013iaa}, considering $\alpha$-attractor potential for the inflaton  and taking non-perturbative effects into account. These analyses assume $n=1$ reheating scenario. We, on the other hand, are typically interested in $n> 2$ in order to include bounds from $\DNeff$, while considering only perturbative reheating.

%%%%%%%%%%%%%%%%%%%%%%%%%%%%%
%%%%%%%%%%%%%%%%%%%%%%%%%%%

%%%%%%%%%%%%%%%%%%%%%%%%%
\section{UV freeze-in with time-dependent inflaton decay}
\label{sec:uv-fi}
%%%%%%%%%%%%%%%%%%%%%%%%%
In this section, we discuss Dark Matter (DM) production via UV freeze-in in a model-agnostic way, i.e., adopting the time-dependent inflaton decay rate as shown in Eq.~\eqref{eq:decay-gen}. The evolution of the DM number density $\ndm$ is governed by the Boltzmann equation (BEQ), which can be written in a generalized form as
\begin{equation} \label{eq:BE0}
    \frac{d\ndm}{dt} + 3\, H\, \ndm = \gamma(T)\,,
\end{equation}
where $\gamma(T)$ corresponds to the DM production rate out of SM particles as a function of the bath temperature $T$.
Note that the equation \eqref{eq:BE0} is coupled through $H$ to the Friedmann equation, so, in this case, we are solving the full set of equations together with  Eqs.~\eqref{eq:rhophi_rhord}.
When the SM entropy is conserved, it is instructive to rewrite Eq.~\eqref{eq:BE0} in terms of the DM yield defined as a ratio of DM number to entropy density of the Universe, $Y\equiv \ndm /s$, where $s(T) \equiv \frac{2\pi^2}{45}\gss(T)\, T^3$, with $\gss(T)$ being the number of relativistic degrees of freedom contributing to the SM entropy. Then Eq.~\eqref{eq:BE0} reads\footnote{Here we have considered the inflaton decay products thermalize instantaneously, exception of which can lead to DM production from the scattering
of non-thermal high energy particles as well~\cite{Chowdhury:2023jft}.}
\begin{equation}\label{eq:BEQ-Y}
    \frac{dY}{dT} = -\frac{\gamma(T)}{H(T)\, T\, s(T)}\,.
\end{equation}
For the UV freeze-in, we assume that DM communicates with the visible sector through non-renormalizable operators suppressed by an appropriate power of the scale of new physics (NP) $\lNP$ and that the maximal temperature of the bath is well below $\lNP$, so $T_\text{max} \ll \lNP$. If, in addition,
we assume that the temperature of the thermal bath at the end of reheating is large enough to neglect radiation and DM masses ($m_\text{SM}, m_\text{DM} \ll \Trh$), then the production rate of DM from the SM bath can be parametrized as~\cite{Elahi:2014fsa,Bernal:2019mhf,Kaneta:2019zgw,Barman:2020plp}
\begin{equation}\label{eq:gamma-SM}
    \gamma(T) = \frac{T^\kappa}{\lNP^{\kappa-4}}\,,
\end{equation}
where $\kappa = 2\,(d-2)$ with $d$ being  mass dimension of the relevant effective operator, ($d \geq 5$). Note that this scale of new physics is uncorrelated with the scale of inflation $\Lambda$ and can be larger, smaller, or equal to $\Lambda$. Finally, to match the whole observed abundance, the present DM yield $Y_0$ has to be fixed so that $\mdm\, Y_0 = \Omega_\text{DM} h^2 \frac{1}{s_0}\, \frac{\rho_c}{h^2} \simeq 4.3 \times 10^{-10}$~GeV, where $\mdm$ is the DM mass in GeV, $\rho_c \simeq 1.1 \times 10^{-5} h^2$~GeV/cm$^3$ is the present critical energy density, $s_0 \simeq 2.9 \times 10^3$~cm$^{-3}$ is the entropy density at present, and $\Omega_\text{DM} h^2\simeq 0.12$~\cite{Planck:2018vyg}.

%%%%%%%%%%%
\subsection {Case-I: $\Trh \gg \mdm$}
%%%%%%%%%%%
Utilizing the solutions for $T(a)$ and $H(a)$, obtained in the previous section, one finds that the DM comoving number $N\equiv\ndm\,a^3$ during the non-standard reheating evolves as
\begin{align}\label{eq:dm-beq}
& \frac{dN}{da} = \frac{a^2\,\gamma}{H} =  \left[\frac{(30/\pi^2\,\gsr)^{\kappa/4}}{H_e\,\lNP^{\kappa-4}}\right]\,\frac{a^{\frac{5\,n+2}{n+1}}}{a_e^{\frac{3\,n}{n+1}}}\,\rSM(a)^{\kappa/4}\,,
\end{align}
which, after on integration between $a_e$ and $\arh$ with the standard freeze-in initial condition $N(a_e)=0$, leads to DM number at the end of reheating
\begin{align}
& N(\arh) \simeq a_e^3\, \frac{\left(\frac{90\,(1+\bar w)}{\pi^2\,\gsr}\,\Gamma_\phi^e\,H_e\,M_P^2\right)^\frac{\kappa}{4}}{H_e\,\lNP^{\kappa-4}}
\begin{cases}
\frac{1}{\zeta}\left(\frac{2}{5-3\bar w-2\beta}\right)^{\kappa/4}\,\left[1-\left( \frac{\arh}{a_e}\right)^{- \zeta}\right]\,,&\beta<\frac{n+4}{n+1},\\
\Gamma \left(\frac{\kappa+4}{4},0\right)-\mathcal{I}+\left(\log\left[\frac{\arh}{a_e}\right]\right)^{1+\frac{\kappa}{4}}\,,&\beta=\frac{n+4}{n+1},\\
\frac{1}{\xi} \left[1- \left( \frac{\arh}{a_e}\right)^{-\xi} \right], &\beta > \frac{n+4}{n+1}\,,
\end{cases} \label{eq:Narh-Light}
\end{align}
where
\begin{align}
    &\zeta \equiv \frac{\kappa\,\beta+n \left[(\beta +3)\,\kappa -24\right]-12}{4\,(n+1)}\,,&&\xi \equiv\frac{\kappa+(\kappa-12)\,n-6}{2 \,(n+1)}\,,
\end{align}
and
\begin{align}
& \mathcal{I} = E_{-\frac{n}{4}}\left[\frac{n^2-5 n-3}{n+1}\,\left(\log\left[\frac{\arh}{a_e}\right]\right)\right]\,,
\end{align}
is the exponential integral function defined as
\begin{align}
& E_i(z) = \int_{-\infty}^z\,dt\,\frac{e^t}{t}\,,
\end{align}
whereas $\Gamma[a,z]$ is the incomplete gamma function. Note that the above result \eqref{eq:Narh-Light} is valid for light DM, i.e., with mass lower than the reheating temperature. Moreover, the UV freeze-in dark species are mainly produced just after the cosmic inflation, when the thermal bath temperature is the highest. Hence, the contribution from the late times, i.e., $a > a_{\rm rh}$, is negligible. Although the SM entropy density is not conserved when the inflaton is decaying, one can calculate the DM yield at $\arh$ as $ Y(\arh) =\ndm(\arh)/s(\Trh) = N(\arh)/\left(\arh^3\,s(\Trh)\right)$. After the end of reheating, the SM entropy is conserved and therefore $Y(\arh)$ remains constant. Thus we obtain
\begin{align}\label{eq:Yrhl}
Y(\arh)\simeq \frac{45}{2\,\pi^2\,\gss}\, \left(\frac{a_e}{\arh} \right)^3 \frac{\left(\frac{180\,n}{\pi^2\,\gsr\,(n+1)}\,\Gamma_\phi^e\,H_e\,M_P^2\right)^\frac{\kappa}{4}}{H_e\,\Trh^3\,\lNP^{\kappa-4}}
\begin{cases}
\frac{1}{\zeta}\left(\frac{n+1}{n-\beta\, (n+1)+4}\right)^{\kappa/4}\,\left[1-\left( \frac{\arh}{a_e}\right)^{- \zeta}\right]\,,\beta<\frac{n+4}{n+1},\\
\Gamma \left(\frac{\kappa+4}{4},0\right)-\mathcal{I}+\left(\log\left[\frac{\arh}{a_e}\right]\right)^{1+\frac{\kappa}{4}}\,,\beta=\frac{n+4}{n+1}\,,\\
\frac{1}{\xi} \left[1- \left( \frac{\arh}{a_e}\right)^{-\xi} \right], \beta > \frac{n+4}{n+1}\,,
\end{cases}
    \end{align}
where $a_e/\arh$ can be read from Eq.~\eqref{eq:arh}, whereas $\Trh$ is given by Eq.~\eqref{eq:Treh}. To obtain the right abundance for DM, $\mdm\,Y(\arh)$ needs to match the present DM abundance, which leads to
\begin{align}\label{eq:rel-dm}
& \frac{\lNP^{\kappa-4}}{\mdm}\simeq\frac{\text{GeV}^{-1}}{4.3\times 10^{-10}} \frac{45}{2\,\pi^2\,\gss}\,\frac{\left(\frac{180\,n}{\pi^2\,\gsr\,(n+1)}\,\Gamma_\phi^e\,H_e\,M_P^2\right)^\frac{\kappa}{4}}{H_e\,\Trh^3}
\nonumber\\&\times
\begin{cases}
\frac{1}{\zeta}\left[\frac{n+1}{n-\beta\, (n+1)+4}\right]^{\frac{\kappa}{4}+\frac{3\,(n+1)}{n\,(\beta+1)-3\,n}}\,\left(1-\left( \frac{\arh}{a_e}\right)^{- \zeta}\right)\, \left(2^\frac{n-1}{2}\,\frac{\Lambda^2\,(n+1)}{2\,M_P\,\Gamma_\phi^e} \right)^\frac{3\,(n+1)}{n\,(\beta+1)-3\,n}\,,\beta<\frac{n+4}{n+1},\\
\Gamma \left(\frac{\kappa+4}{4},0\right)-\mathcal{I}+\left(\log\left[\frac{\arh}{a_e}\right]\right)^{1+\frac{\kappa}{4}}\,\left[\left(\frac{2^\frac{n-1}{2}\,(n-2)\,\Lambda^2}{n^{1-n}\,M_P\,\Gamma_\phi^e}\right)^{\frac{n+1}{2\,n-4} } \mathcal{W}^{\frac{n+1}{4-2n}}\left(\frac{2^\frac{n-1}{2}\,(n-2)\,\Lambda^2}{n^{1-n}\,M_P\,\Gamma_\phi^e}\right)\right]^{-3}\,,\beta=\frac{n+4}{n+1}\,,\\
\frac{1}{\xi}\,\left(1- \left( \frac{\arh}{a_e}\right)^{-\xi} \right)\,\left[\frac{2^\frac{n-1}{2}\,\Lambda^2}{n^{1-n}\,M_P\,\Gamma_\phi^e}\,\frac{\beta-4+n\,(\beta-1)}{2} \right]^{\frac{3+3\,n}{4-2\,n}}\,, \beta > \frac{n+4}{n+1}\,,
\end{cases}
\end{align}
which shows that for a given decay channel, and $n$, $\Lambda$ the ratio $\lNP^{\kappa-4}/\mdm$ is fixed.

%%%%%%%%%%%%%%%%%%%%%%%%%%%
\subsection {Case-II: $\Trh \ll \mdm \ll \Tmax$}
%%%%%%%%%%%%%%%%%%%%%%%%%%
Production of heavy DM particles with mass exceeding the reheating temperature, is Boltzmann suppressed in the region where $\mdm > \Trh$. The present DM number density in that case could be estimated by integrating Eq.~\eqref{eq:dm-beq} from $a_e$ to $\adm\equiv a(T=\mdm)$, where
\begin{align}\label{eq:adm}
    \adm &= \arh \begin{cases}
        \left(\frac{\Trh}{\mdm} \right)^{\frac{4\,(n+1)}{3\,n+\beta\,(n+1)}}, &\beta <\frac{n+4}{n+1}\,,\\
       \frac{a_e}{\arh}\,\exp\left(-\frac{1}{4}\,\mathcal{W}\left[-4\,\left(\frac{a_e}{\arh}\right)^4\,\left(\frac{\mdm}{\Trh}\right)^4\,\log \left(\frac{\arh}{a_e}\right)\right]\right), &\beta = \frac{n+4}{n+1}\,, \\
       \frac{\Trh}{\mdm}, &\beta>\frac{n+4}{n+1}\,.
    \end{cases}
\end{align}
The DM yield $Y(\adm)=N(\adm)/\left(\adm^3\,s(\mdm)\right)$ then reads
\begin{align}\label{eq:Yrhm}
 Y(\adm)&\simeq
\frac{45}{2\,\pi^2\,\gss}\, \left(\frac{a_e}{\adm} \right)^3 \frac{\left(\frac{180\,n}{\pi^2\,\gsr\,(n+1)}\,\Gamma_\phi^e\,H_e\,M_P^2\right)^\frac{\kappa}{4}}{H_e\,\mdm^3\,\lNP^{\kappa-4}} \\
&\times
\begin{cases}
\frac{1}{\zeta}\left(\frac{n+1}{n-\beta  (n+1)+4}\right)^{\kappa/4}\,\left[1-\left( \frac{\adm}{a_e}\right)^{- \zeta}\right]\,,&\beta<\frac{n+4}{n+1}, \non\\
\Gamma \left(\frac{\kappa+4}{4},0\right)-\mathcal{I}+\left(\log\left[\frac{\adm}{a_e}\right]\right)^{1+\frac{\kappa}{4}}\,, &\beta=\frac{n+4}{n+1},\\
\frac{1}{\xi} \left[1- \left( \frac{\adm}{a_e}\right)^{-\xi} \right], &\beta > \frac{n+4}{n+1}\,.
\end{cases}
\end{align}
While the DM yield is conserved {\it after} reheating and not {\it during} reheating due to the entropy injection, the final yield at the end of reheating reads
\begin{align}
& Y(\arh) = Y(\adm)\, \left(\frac{\adm}{\arh} \right)^3\frac{s(\adm)}{s(\arh)} \simeq Y(\adm)\,\left(\frac{\mdm}{\Trh}\right)^3\,\left(\frac{\adm}{\arh}\right)^3\,,
\end{align}
where $\adm/\arh$ is given by Eq.~\eqref{eq:adm}.  As before requiring the right abundance for the DM, one can show, for a given channel, $\mdm/\lNP^{\kappa-4}$ is a constant for a given inflaton decay channel for a fixed $n\,,\Lambda$ [cf. Eq.~\eqref{eq:rel-dm}]. Note that, in determining the UV freeze-in yield, we introduce three more free parameters here, namely, the DM mass $\mdm$, the scale of effective interaction $\lNP$, and the mass dimension of the DM-SM operator, $\kappa$.

%%%%%%%%%%%%%%%%%%
\subsection{Dark matter production}
\begin{figure}[htb!]
    \centering
    \includegraphics[scale=0.32]{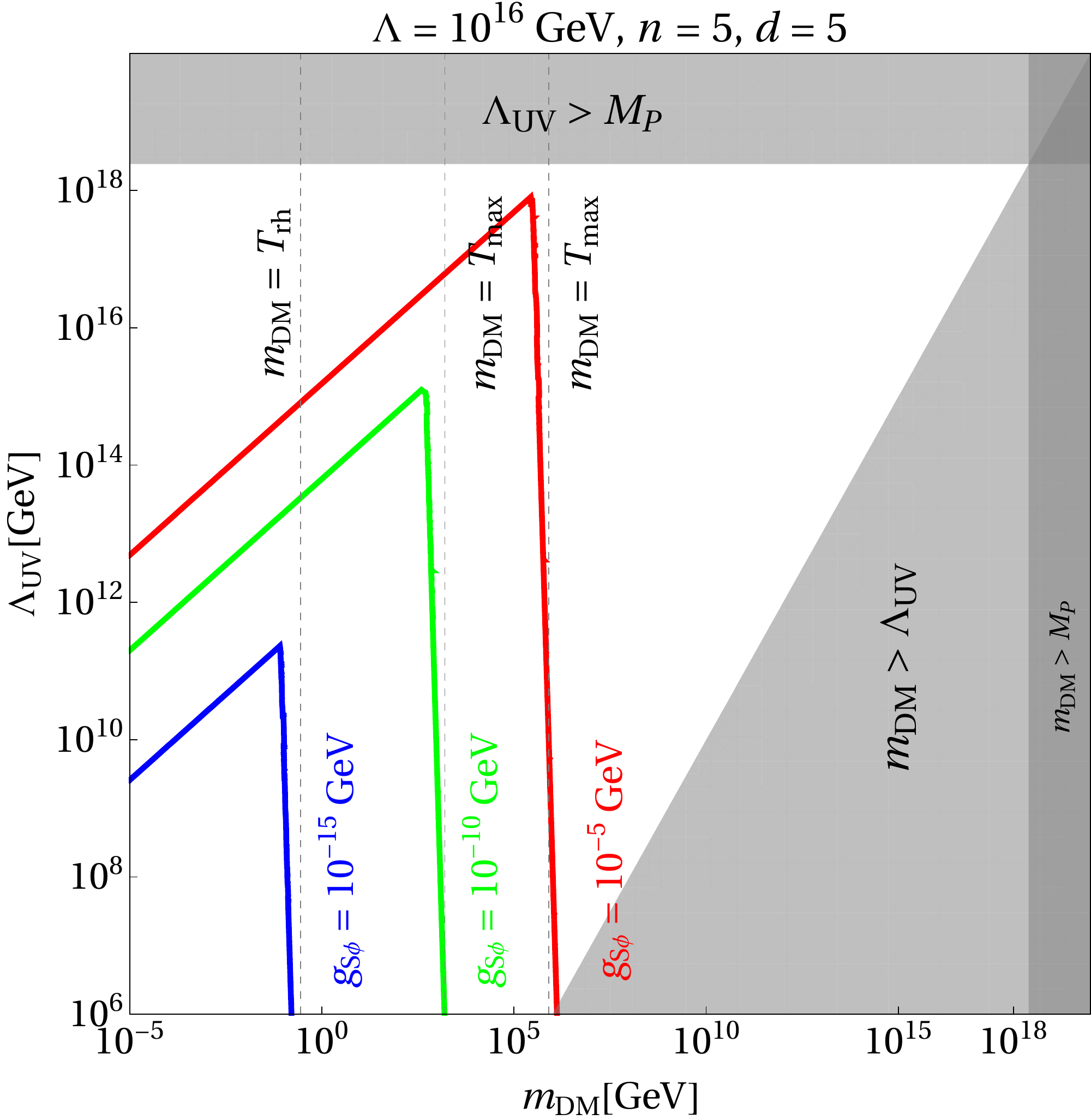}~~~~
    \includegraphics[scale=0.32]{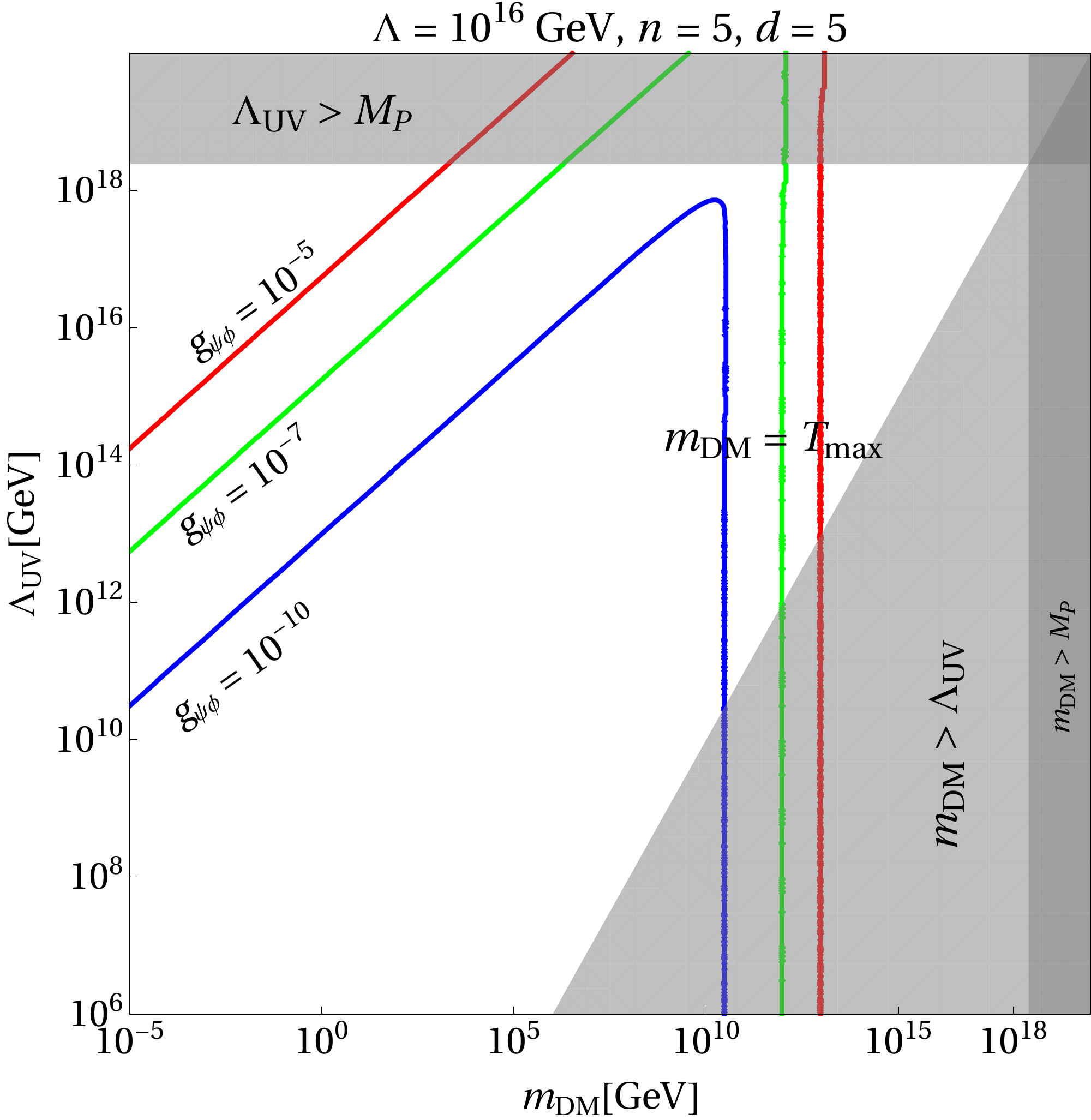}\\[10pt]
    \includegraphics[scale=0.32]{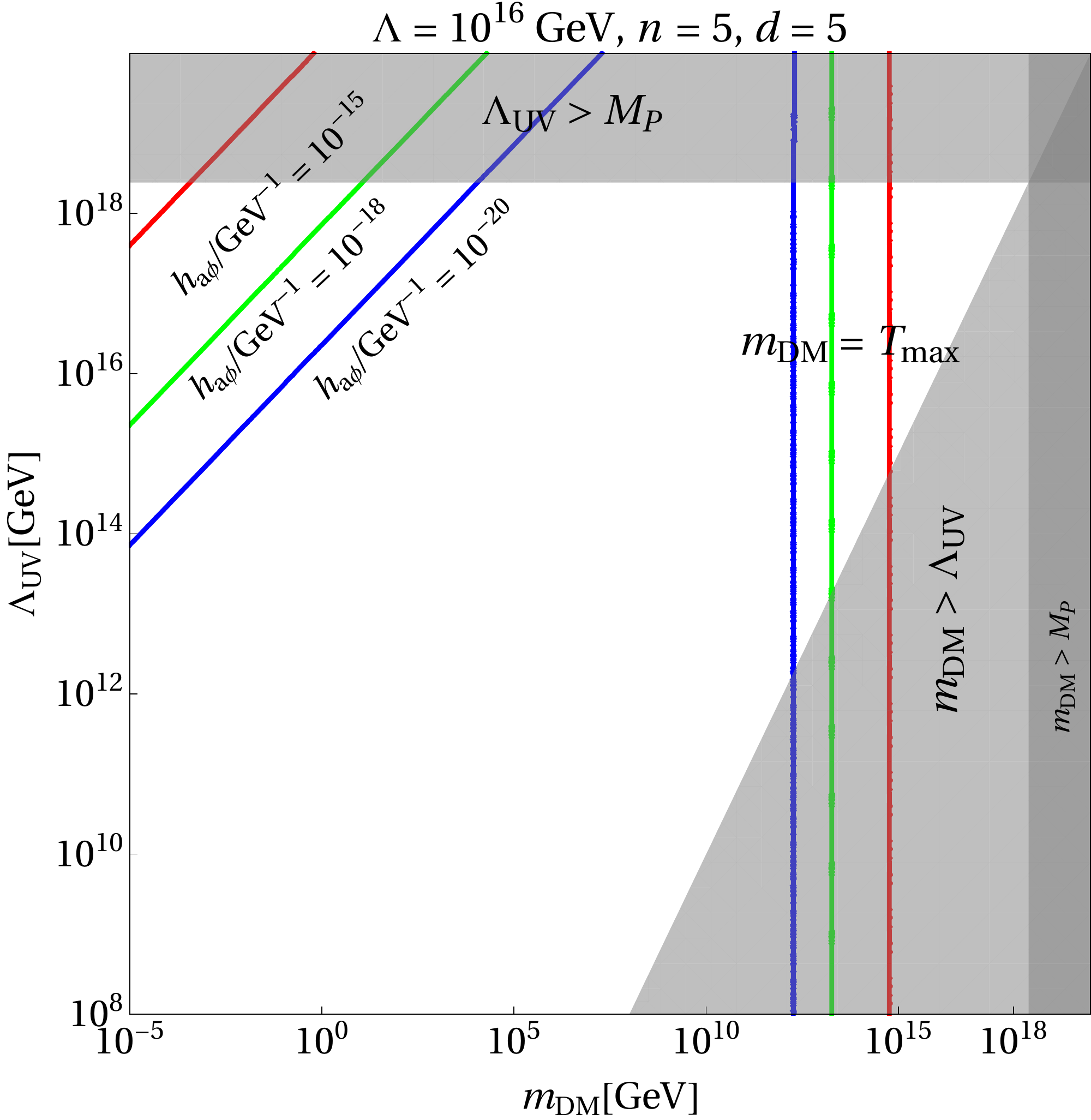}~~~~\includegraphics[scale=0.32]{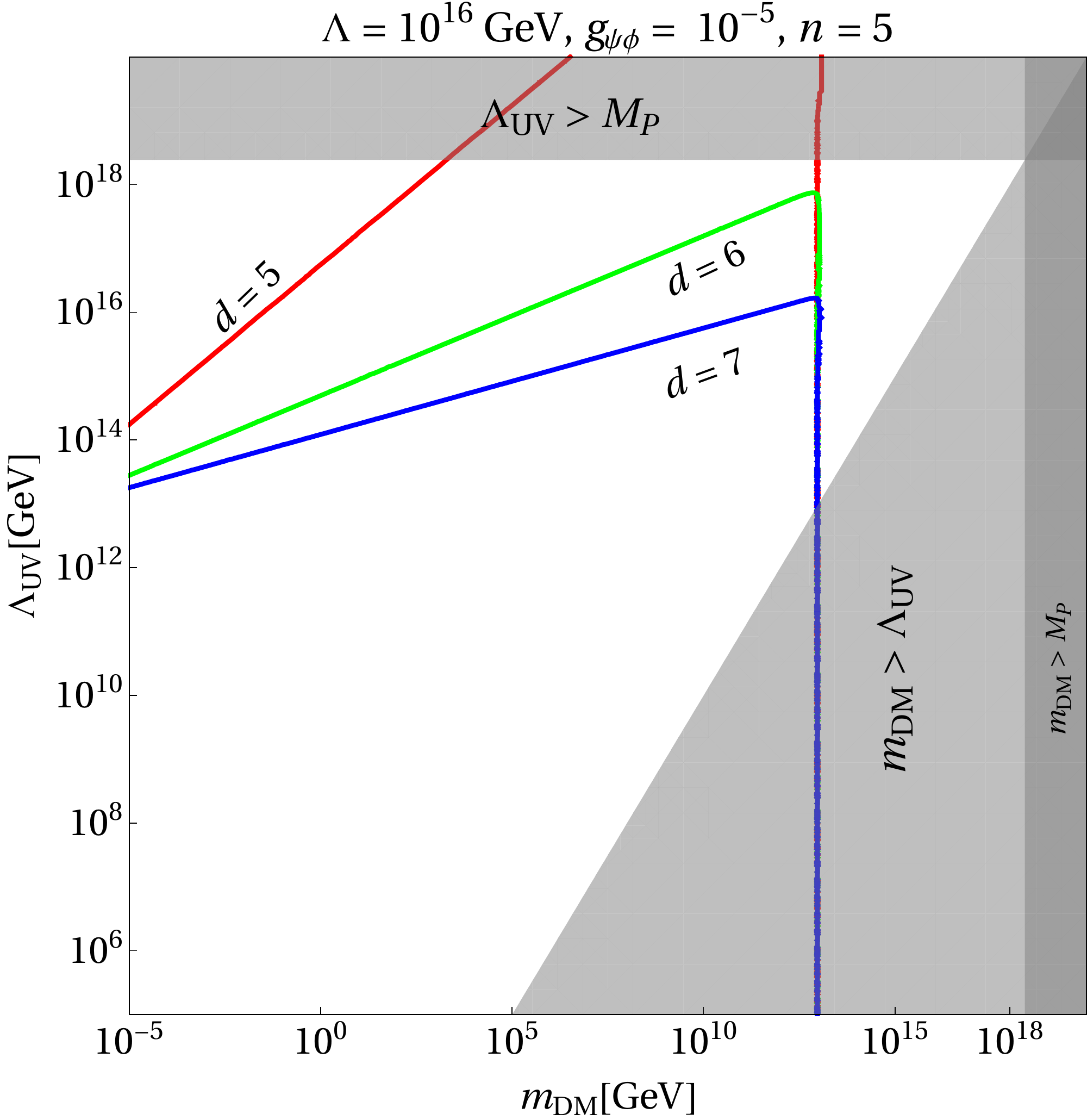}
    \caption{Dark Matter parameter space producing right relic density in $\lNP-\mdm$ plane, considering reheating via $\phi\to SS$ (top left), $\phi\to \psi\psi$ (top right) and $\phi\to aa$ (bottom left), for a fixed $\Lambda=10^{15}$ GeV, $n=10$ and DM-SM operator of dimension $d=5$. In the bottom right panel we compare parameter space for DM -SM operators of dimension $d=\{5\,,6\,,7\}$, considering fermionic reheating as a benchmark scenario.}
    \label{fig:relic}
\end{figure}

%%%%%%%%%%%%%%%%%%%
In Fig.~\ref{fig:relic}, we show the contours that reproduce the observed DM relic abundance. We consider different reheating scenarios, e.g., $\phi\to SS\,,\psi\psi\,,aa$ and project the parameter space in $\lNP-\mdm$ plane by fixing the scale of inflation $\Lambda=10^{16}$ GeV, $n=5$ and $\kappa=6$ (or equivalently, $d=5$). We choose the strength of the inflaton-matter coupling in such a way that they give rise to observable GW spectrum as discussed in Sec.~\ref{sec:result}. From the top left panel, we first note that in all cases, the parameter space shows the existence of a cut-off for certain $\mdm$, beyond which DM can no longer be produced. This cut-off depends on the two cases discussed before. In case where $\mdm\ll\Trh$, DM species heavier than $\Trh$ cannot be created from the thermal bath due to the Boltzmann suppression. On the other hand, for dark particles with mass $\Trh\ll\mdm\ll\Tmax$, the production ceases beyond $\mdm=\Tmax$. In both cases, UV freeze-in production becomes inefficient when the SM temperature drops below dark matter mass. The slope of each contour is exactly given by $1/(\kappa-4)$, which can easily be obtained from Eq.~\eqref{eq:rel-dm} once all parameters are fixed except for $\mdm$ and $\lNP$. Next, we note that in all cases, a larger inflaton coupling $g_{i\phi}$ requires larger $\lNP$ to produce the correct DM abundance for a given DM mass (keeping all other parameters fixed). This can be understood from Eq.~\eqref{eq:Yrhl}, where we find, for $n=5\,,\kappa=6\,,\Lambda=10^{16}$ GeV, the DM yield approximately reads $Y(\arh)\propto \gsp^{5/2}/\lNP^2$ in case of inflaton decaying into scalars, while for fermionic decay $Y(\arh)\propto \gpp^{3/2}/\lNP^2$ and for the derivative interaction $Y(\arh)\propto \hap^{3/2}/\lNP^2$ (similar dependence can also be found from Eq.~\eqref{eq:Yrhm}, for a fixed DM mass). Therefore, irrespective of the inflaton decay products, a larger coupling leads to a larger abundance that can be compensated by a higher $\lNP$ for a given DM mass. Moreover, for our choice of parameters, we find that the decay width into pairs of scalars is small compared to the fermionic or derivative scenario. Since the DM yield in all cases is proportional to $(\Gamma_\phi^e)^x/\lNP^{\kappa-4}$ ($x$ is some function of $n$, the exact form of which depends on the decay mode), hence we see the scale $\lNP$ is more suppressed for the scalar case compared to other two cases.
%%%%%%%%%%%%%%%%%%%%%%%%%%%%%
%%%%%%%%%%%%%%%%%%%%%%%%%%%
%%%%%%%%%%%%%%%%%%%%%%%%%%%%%
\begin{figure}[htb!]
    \centering
    \includegraphics[scale=0.34]{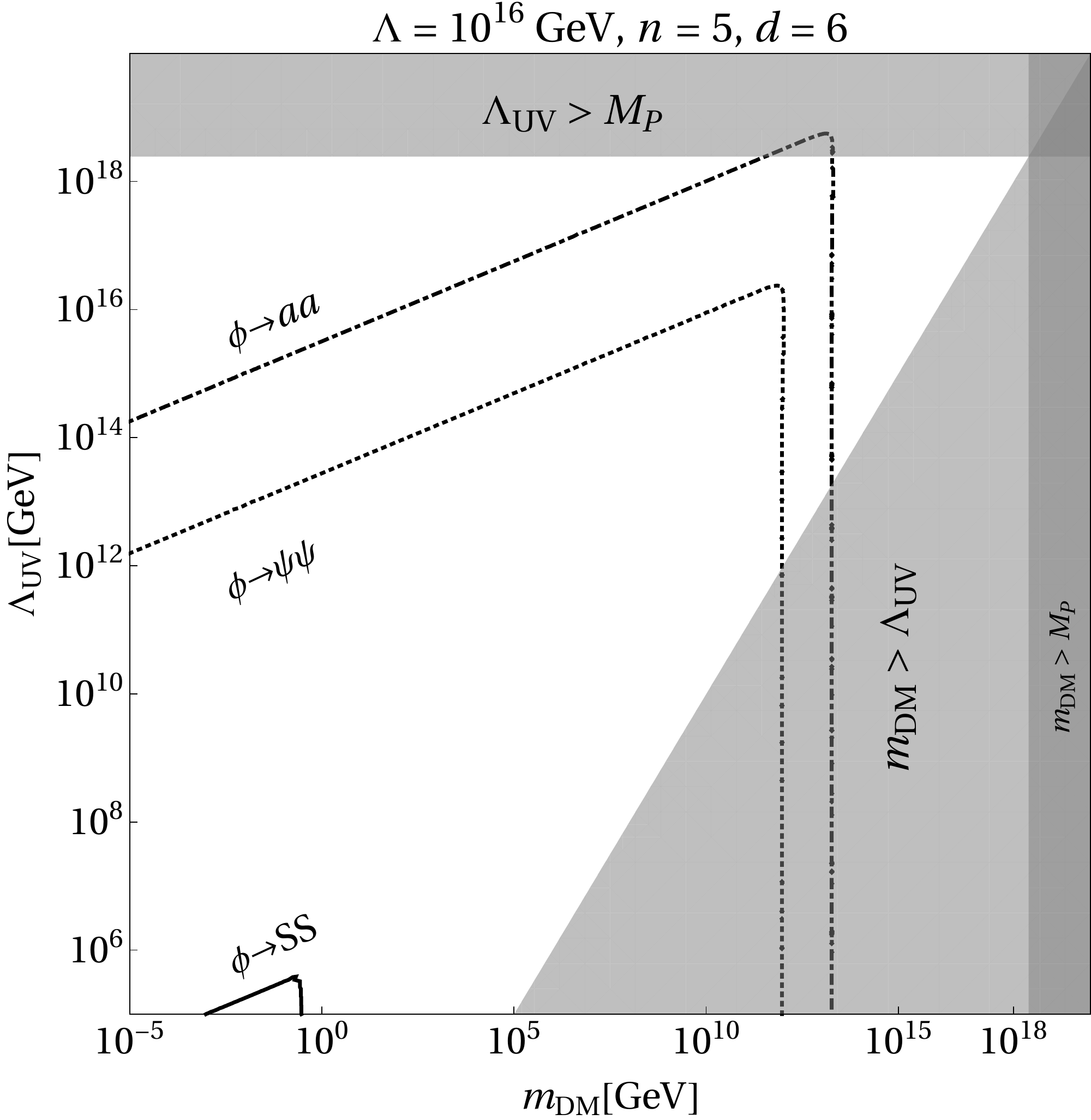}~~~~\includegraphics[scale=0.34]{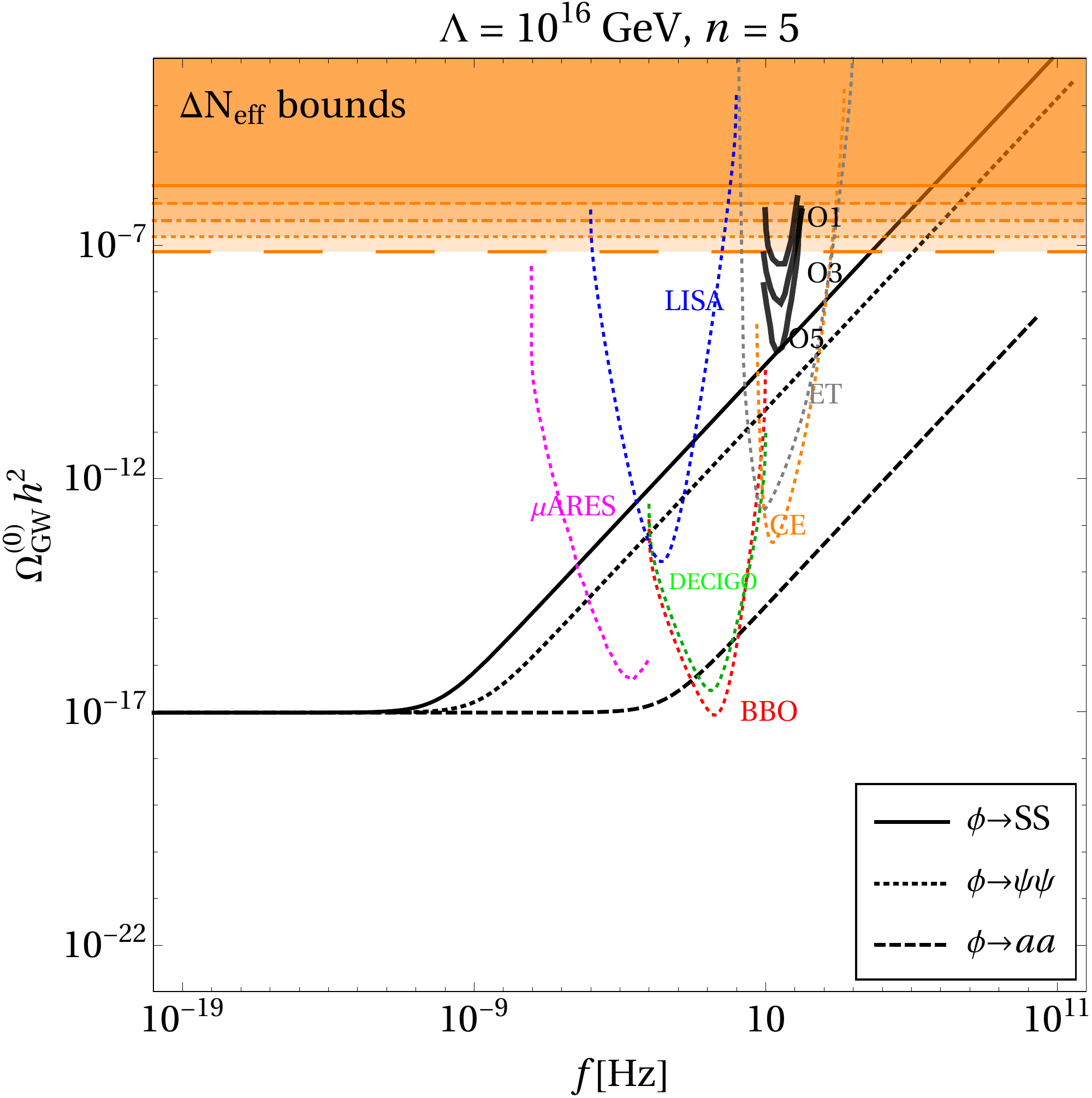}
    \caption{Left: Contours satisfying relic abundance for $\phi\to SS$ (black solid), $\phi\to\psi\psi$ (black dotted) and $\phi\to aa$ (black dot-dashed) scenarios. Right: GW spectrum for the three cases. We fix $\Lambda=10^{15}$ GeV and $n=10$, while choose different coupling strengths (see text).}
    \label{fig:gw-dm}
\end{figure}
%%%%%%%%%%%%%%%%%%%%%%%%%%%
Finally, in the bottom right panel, we show the effect of DM-SM operators of different dimensions on the DM parameter space. As an illustrative example, we choose the fermionic reheating scenario and fix all the relevant parameters. Now, we see, a larger $d\,(\equiv\kappa)$ relaxes the bound on $\lNP$ for a given DM mass since $Y(\arh)\propto \mdm\,\lNP^{8-2\,d}$. Hence, operators of larger dimensions lead to more suppression, allowing lower $\lNP$. 

Since the same inflaton-matter coupling controls DM abundance and GW spectrum, it is, therefore, possible to make a one-to-one correspondence between DM abundance and observable PGW spectrum through such interactions. This is what is portrayed in Fig.~\ref{fig:gw-dm}. In the left panel, we show the parameter space that produces right relic abundance considering all possible inflaton-matter interactions and choosing $\{\gsp\,,\gpp\,,\hap\}=\{10^{-15}\,\text{GeV}\,,10^{-7}\,,10^{-18}\,\text{GeV}^{-1}\}$, for a fixed $\lNP\,,n$, and considering DM-SM operator of mass dimension $d=6$. For the same choice of parameters, we show the GW spectrum in the right panel, where each curve corresponds to a range of $\mdm$ that gives rise to the right abundance. This shows, for the fermionic reheating scenario, that it is possible to probe the relic density allowed parameter space with GW detectors like BBO or DECIGO, for $\mdm\simeq\{10^{-5}-10^{11}\}$ GeV. This is also true for the bosonic (scalar) reheating scenario, where the DM mass lies in the range $\mdm\simeq\{10^{-5}-10^0\}$ GeV. Although derivative interactions can provide the right relic abundance, the prediction remains beyond the reach of future GW detectors, as shown by the black dashed curve. However, for this set of parameters, $\phi\to SS$ coupling could potentially be ruled out from future constraints on $\Delta N_\text{eff}$ from CMB. It is understandable that by choosing the inflaton-matter coupling and the scale of inflation appropriately, it is always possible to have a detectable GWs spectrum, satisfying BBN constraint, while accommodating the proper abundance for the DM requires tuning $\lNP$ and $\mdm$. Thus, a future GW detection can be regarded as a potential freeze-in DM signal, originating from an underlying new physics leading to a DM-SM operator of a given dimension.

%%%%%%%%%%%%%%%%%%%%%%
\section{Conclusions}
\label{sec:concl}
%%%%%%%%%%%%%%%%%%%%%%
This work investigates the possibility of probing inflaton couplings to the visible sector by primordial gravitational waves (GWs) of inflationary origin. We consider the inflaton field oscillating with a time-dependent amplitude assuming the $\alpha$-attractor T-model potential. Such a scenario inevitably implies a time-dependent decay width of the inflaton for potential different from quadratic (for small field strength). The inflaton is assumed to couple to a pair of bosonic or fermionic matter within a perturbative regime that can give rise to successful reheating. We then compute the primordial power spectrum and the spectral energy density of primordial GW. The energy density of the Universe turns out to be a function of the inflaton decay width at the end of inflation, which, in turn, is related to the inflaton couplings to matter. Satisfying bounds from the requirement of successful BBN and CMB, we find the inflaton coupling as small as $\gsp\sim 10^{-15}$ GeV becomes sensitive to future gravitational wave detectors in case the inflaton decays into a pair of scalar bosons via an operator of dimension three. For inflaton, Yukawa interactions with fermions, the corresponding coupling that falls within the detector sensitivity turns out to be $\gpp\sim 10^{-5}$. For derivative interaction involving a mass dimension five operator, coupling strength $\gap\sim 10^{-17}\,\text{GeV}^{-1}$ becomes sensitive to the experiments. It is important to note that our result strongly depends on other free parameters of the model, for example, the shape of the potential or scale of inflation which is currently bounded from PLANCK data. However we envisage that once we got more information on inflation from next generation CMB experiments, like measuring the scale of inflation, uncertainties of our prediction would be reduced.

We finally discuss the production of dark matter (DM) during reheating from the radiation bath, assuming the DM communicates with the SM via non-renormalizable operators suppressed by large-scale $\lNP$. In this scenario, the UV freeze-in from the thermal bath is the primary mechanism of DM production. The freeze-in scenario requires tiny couplings between the bath and DM particles.  Therefore it is very challenging if not impossible to probe such very feebly coupled dark sector. Here we promote GW as a novel observable to probe such dark sectors and identify the parameter space that satisfies the observed relic density. Because of the time-dependent inflaton decay width, the standard DM yield also gets modified in this framework. It turns out to be possible to tune the scale of new physics $\lNP$ and the DM mass to satisfy the PLANCK observed relic DM density. The interesting point here is that depending on the nature of the inflaton decay channel, one can produce DM over a large range of mass. As the same inflaton decay width influences DM production and controls the GW spectrum, hence a detectable GW may be interpreted as an indication of the DM detection and provide information about the scale of new physics.

Finally we comment on the fact that in our analysis we have only considered one GW detector running at a time. However once more than one GW detector starts operating then sensitivities to GW spectral shapes are expected to improve so that much larger regions of the parameter space could be probed. Such an analysis is beyond the scope of the current manuscript but
we envision precision measurements that GW astronomy promises due to the planned global network of GW detectors, which can fulfill the dream of testing high-scale and weakly-coupled fundamental BSM scenarios a reality in very near future.

%%%%%%%%%%%%%%%%%%%%%%%%%
\section*{Acknowledgements}
%%%%%%%%%%%%%%%%%%%%%%%
This project is supported in part by the National Science Centre (Poland) as a research project no 2020/37/B/ST2/02746. BB would like to thank Riajul Haque for fruitful discussions.\\

{\underline{{\it Note}}} : During the completion of this work Ref.~\cite{Chakraborty:2023ocr} appeared, where constraints on inflaton-matter couplings derived from primordial gravitational waves were also studied. Although the strategy of Ref.~\cite{Chakraborty:2023ocr} is similar to the one we have adopted, the numerical analysis is different.

%%%%%%%%%%%%%%%
\appendix
%%%%%%%%%%%%%%%%%
\section{Inflaton decay width calculations}
\label{sec:app-inf-decay}
%%%%%%%%%%%%%%%%%
In this work, the inflaton field is treated as a time-dependent external and classical background field that coherently oscillates in time. Its evolution can be parametrized by a product of a slowly varying envelope $\varphi(t)$ and a fast oscillating function $\mathcal{P}(t)$~\cite{Shtanov:1994ce,Garcia:2020wiy,Clery:2021bwz},
\begin{align}
    \phi(t) = \varphi(t) \cdot \mathcal{P}(t).
\end{align}
The envelope is defined by the condition $\rho_\phi = V(\varphi)$ \cite{Shtanov:1994ce}, and for the $\alpha-$attractor T model one finds
\begin{align}
   \varphi(t) = M_P \left( \frac{\rho_\phi(t)}{\Lambda^4} \right)^{\frac{1}{2n}}\,.
\end{align}
The final states are produced in a quantum process in the classical time-varying background. More specifically, we consider the transition from the vacuum, i.e., $\ket{\alpha}= \ket{0}$, to two-particle final states $\ket{\beta}= \ket{FF}$, where $F$ denotes the final state particle. The leading order S-matrix element describing this process is 
\begin{align}
& S_{\beta \alpha}^{(1)} = -i\int d^{4}x\, \Big\langle\beta \Big|T\Big[\mathcal{L}_{\rm int}(x)\Big]\Big|\alpha \Big\rangle \nonumber \\ &= -i g_{\phi F} \int d^{4}x\, \phi(t) \cdot  \Big\langle 0 \Big| \hat{a}_F(p_1) \hat{a}_F(p_2) \hat{F} \hat{F}|0 \Big\rangle
-i h_{\phi F} \int d^{4}x\, \phi(t) \cdot  \Big\langle 0 \Big| \hat{a}_F(p_1) \hat{a}_F(p_2) (\partial_\mu \hat{F}) (\partial^\mu \hat{F})|0 \Big\rangle\,,
\end{align}
where $\hat{a}_F^\dagger(p_1), \hat{a}_F^\dagger(p_2)$ are the creation operators for the final state particles, and $\hat{F}$ denotes the quantum field operator for the $F$ state. We will either consider contribution $\propto g_{\phi F}$ or $\propto h_{\phi F}$, but not both simultaneously. The form of $\hat{F}$ depends on the spin of $F$ particles. In what follows, we assume that the time scale of $\varphi$ variation is much longer than the time scale of interactions. Moreover, we decompose the $\mathcal{P}$ function into the Fourier modes
\begin{align}
    \mathcal{P}(t) = \sum_{l=-\infty}^\infty \mathcal{P}_l e^{- i l \omega t},
    \label{Fourier}
\end{align}
with $\omega$ being the frequency\footnote{In fact, since $\mathcal{P}(t)$ is exactly periodic only for $n=1$ the representation (\ref{Fourier}) of the oscillatory function $\mathcal{P}(t)$ is an approximation unless $n=1$.} of the inflaton oscillations.  Hence, the S-matrix element can be written as
\begin{align}
S_{\beta \alpha}^{(1)} &= -i g_{F\phi} M_P \left( \frac{\rho_\phi}{\Lambda^4}\right)^{\frac{1}{2n}} \sum_{l}  \mathcal{P}_l \int dt \, e^{- i l \omega t} \int d^{3}x\, \Big\langle 0 \Big| \hat{a}_F(p_1) \hat{a}_F(p_2) \hat{F} \hat{F}|0 \Big\rangle  \nonumber \\
&-h_{F\phi} M_P \left( \frac{\rho_\phi}{\Lambda^4}\right)^{\frac{1}{2n}}  \mathcal{P}_l \int dt \, e^{- i l \omega t} \int d^{3}x\, \Big\langle 0 \Big| \hat{a}_F(p_1) \hat{a}_F(p_2) (\partial_\mu \hat{F})(\partial^\mu \hat{F})|0 \Big\rangle \nonumber \\
&\equiv - i M_P \left( \frac{\rho_\phi}{\Lambda^4}\right)^{\frac{1}{2n}} \sum_{l}  \mathcal{P}_l  \times  (1 + \delta_{F}) \mathcal{F}(F) (2 \pi)^4 \delta(l \omega - p_1^0 - p_2^0) \delta^{(3)} (\vec{p}_1 + \vec{p}_2),
\end{align}
where $\delta_F=1$ for identical particles in the final state, and $\delta_F=0$ otherwise. The function $\mathcal{F}(F)$ depends on form of the inflaton interactions,
\begin{align}
    \mathcal{F}(F) &= \begin{cases}
    g_{S \phi}, &F=S,\\
    g_{\psi \phi} \cdot \bar{u}^{s_1}(p_1)v^{s_2}(p_2), &F=\psi,\\
    g_{V \phi} \cdot g^{\mu \nu} \epsilon_\mu^{\sigma, *}(p_1)  \epsilon_\nu^{\sigma^\prime, *}(p_2),  &F=V,\\
    h_{a \phi} \cdot (p_1 \cdot p_2), &F=a,
    \end{cases}
\end{align}
with $\bar{u}^{s_1}(p_1)$, $v^{s_2}(p_2)$ denoting the Dirac spinors, and $ \epsilon_\mu^{\sigma, *}(p_1)$ being the polarization vectors of the gauge spin-1 field. The probability $P$ for the production of two $F$ particles with momenta $p_1$ and $p_2$ in the oscillating background of the inflaton field is given by
\begin{align}
P(p_1, p_2) &= \frac{\big\lvert S_{\phi \rightarrow FF}^{(1)} \big\rvert^2}{\braket{0|0} \braket{p_1|p_1} \braket{p_2|p_2}}, \label{eq:trans_prob}
\end{align}
where $\braket{0|0}=1$ and $\braket{p_{i}|p_{i}}=(2 \pi)^3 2 p_{i}^0 \delta^{(3)}(0)$ for $i=1,2$. The modulus squared of the S-matrix element takes the form
\begin{align}
\big\lvert S_{\phi \rightarrow FF}^{(1)} \big\rvert^2 &=  M_P^2 \left(\frac{\rho_\phi}{\Lambda^4} \right)^{\frac{1}{n}} \sum_{\rm{final}} \Big[\sum_l \mathcal{P}_l   \times  (1+\delta_F) \mathcal{F}(F) (2 \pi)^4 \delta(l \omega - p_1^0 - p_2^0) \delta^{(3)}(\vec{p}_1+\vec{p}_2) \Big]^2, \label{eq:s1}
\end{align} 
where we have summed over spin/polarization states of $F$ particles. We can further simplify Eq.\eqref{eq:s1}, obtaining
\begin{align}
\big\lvert S_{\phi \rightarrow FF}^{(1)} \big\rvert^2 &= M_P^4 \left(\frac{\rho_\phi}{\Lambda^4} \right)^{\frac{1}{n}} \sum_l |\mathcal{P}_l|^2 (1+ \delta_F)^2 |\overline{\mathcal{M}}^{(1)}_{\phi \rightarrow FF}|^2 \times  (2 \pi)^4 \delta(l \omega - p_1^0 - p_2^0) \delta^{3}(\vec{p}_1+\vec{p}_2) V T.
\end{align}
The square of the dimensionless \textit{amplitude} summed over spin/polarization states of particles $F$ is defined as
\begin{align}
&|\overline{\mathcal{M}}^{(1)}_{\phi \rightarrow FF}|^2 \equiv \frac{1}{M_P^2}\sum_{\rm{final}} |\mathcal{F}(F)|^2 = \begin{cases}
(g_{S \phi}/M_P)^2, & F=S,\\
4 g_{\psi \phi}^2 [p_1 \cdot p_2 - m_\psi^2]/M_P^2, &F=\psi,\\
\left(\frac{g_{V \phi}}{M_P}\right)^2 \left[2+ \frac{(p_1 \cdot p_2)^2}{m_V^4} \right], & F=V \text{ and } m_V \neq 0,\\
4 \left(\frac{g_{V \phi}}{M_P}\right)^2, & F=V \text{ and } m_V = 0,\\
(h_{a \phi}/M_P)^2 \cdot (p_1 \cdot p_2)^2& F=a,
\end{cases}
\end{align}
where 
\begin{align}
    p_1 \cdot p_2 = \frac{s}{2} - m_F^2,
\end{align}
with $s\equiv (p_1+ p_2)^2$ being the Mandelstam variable and $m_\psi$, $m_V$  denoting the mass of fermions and vectors in the final state, respectively. Now, the probability (Eq.~\eqref{eq:trans_prob}) takes the form
\begin{align}
P(p_1, p_2) &= M_P^4 \left(\frac{\rho_\phi}{\Lambda^4} \right)^{\frac{1}{n}}  \sum_l \lvert \mathcal{P}_l \rvert^2 \cdot (1+\delta_F)^2  \frac{\big\lvert \overline{\mathcal{M}}^{(1)}_{\phi \rightarrow FF} \big\rvert^2  }{2p_{1}^0 V \,2 p_{2}^0 V}  (2 \pi)^4 \delta (l \omega - p_{1}^0 - p_{2}^0) \delta^{(3)}(\vec{p}_1+\vec{p}_2) V T\,.
\end{align}
The total probability can then be found by summing over each outgoing momenta. In the continuum limit, this reduces to multiplying $P (p_1, p_1)/T$ by a factor $ V  d^3 \vec{p}_1/(2 \pi)^3\,  V d^3 \vec{p}_2/(2 \pi)^3$. \\
The energy gain from created particles in volume $V$ and time ${\rm d} t$ is 
\begin{align}
{\rm d}E (p_1,p_2)= (p_1^0 + p_2^0)\frac{V {\rm d}^3 \vec{p}_1}{(2 \pi)^3} \frac{V {\rm d}^3 \vec{p}_2}{(2 \pi)^3} \frac{P(p_1, p_1)}{T} {\rm d}t.
\end{align}
Thus, the total energy gain per volume and time for the $F$ particles is given by
\begin{align}
\frac{{\rm d}E}{V {\rm d} t}&= M_P^4 \left(\frac{\rho_\phi}{\Lambda^4} \right)^{\frac{1}{n}}  \sum_l \lvert \mathcal{P}_l \rvert^2 \cdot (1+\delta_F)^2 \big\lvert \overline{\mathcal{M}}^{(1)}_{\phi \rightarrow FF}(k) \big\rvert^2  \nonumber \\
&\times \int \frac{d^3 \vec{p}_1}{(2 \pi)^3 2 p_1^0} \int  \frac{d^3 \vec{p}_2}{(2 \pi)^3 2 p_2^0} (p_1^0 + p_2^0)  (2 \pi)^4 \delta (l \omega - p_{1}^0 - p_{2}^0) \delta^{3}(\vec{p}_1+\vec{p}_2) \nonumber \\
&=\frac{M_P^4}{8 \pi} \left(\frac{\rho_\phi}{\Lambda^4} \right)^{\frac{1}{n}}  \sum_{l=1}^\infty l\omega \lvert \mathcal{P}_l \rvert^2 \cdot ( 1+\delta_F)^2 \big\lvert \overline{\mathcal{M}}^{(1)}_{\phi \rightarrow FF} \big\rvert^2  \sqrt{1\!-\! \frac{4 m_F^2}{(l \omega)^2}}
\end{align}
The total energy density gained from production of $F$ particles must be compensated by the energy loss of the inflaton field, which we {\it parametrize} as $(1+ \bar{w})\cdot \rho_\phi  \Gamma_{\phi \rightarrow FF}$.
Consequently, the \textit{decay width} accounting for the $\phi \rightarrow FF$ interaction can be calculated from the following
\begin{align}
    \Gamma_{\phi \rightarrow FF} & = \frac{1}{1+ \bar{w}} \frac{1}{\rho_\phi} \frac{{\rm d}E}{V {\rm d} t}\nonumber \\
    &= \frac{1}{1+ \bar{w}} \frac{1}{8 \pi} \frac{\omega M_P^4}{\rho_\phi} \left(\frac{\rho_\phi}{\Lambda^4} \right)^{\frac{1}{n}}  \sum_{l=1}^\infty  l \lvert \mathcal{P}_l \rvert^2 \cdot ( 1+\delta_F)^2 \big\lvert \overline{\mathcal{M}}^{(1)}_{\phi \rightarrow FF} \big\rvert^2  \sqrt{1\!-\! \frac{4 m_F^2}{(l \omega)^2}}.
\end{align}
The corresponding decay widths read
\begin{align}\label{eq:decay-msv}
     \Gamma_{\phi \rightarrow SS} &= \frac{1+n}{2n} \frac{4 }{8 \pi}  \left(\frac{g_{S \phi}}{M_P} \right)^2\frac{\omega M_P^4}{\rho_\phi} \left(\frac{\rho_\phi}{\Lambda^4} \right)^{\frac{1}{n}}  \sum_{l=1}^\infty  l \lvert \mathcal{P}_l \rvert^2 \sqrt{1\!-\! \frac{4 m_S^2}{(l \omega)^2}}\,,
     \nonumber\\
      \Gamma_{\phi \rightarrow \psi \bar{\psi}} &= \frac{1+n}{2n} \frac{1}{4\pi}  \left(\frac{g_{\psi \phi}}{M_P} \right)^2\frac{\omega M_P^4}{\rho_\phi} \left(\frac{\rho_\phi}{\Lambda^4} \right)^{\frac{1}{n}}  \sum_{l=1}^\infty  l \lvert \mathcal{P}_l \rvert^2 (l \omega)^2 \left(1\!-\! \frac{4 m_\psi^2}{(l \omega)^2}\right)^{3/2}\,,
       \nonumber\\
       \Gamma_{\phi \rightarrow VV} &= \frac{1+n}{2n} \frac{1}{32\pi}  \left(\frac{g_{V \phi}}{M_P} \right)^2\frac{\omega M_P^4}{\rho_\phi} \left(\frac{\rho_\phi}{\Lambda^4} \right)^{\frac{1}{n}} \\ 
       &\times \sum_{l=1}^\infty  l \lvert \mathcal{P}_l \rvert^2 \left(\frac{l \omega}{m_V} \right)^4 \left[1- 4 \left(\frac{m_V}{l \omega} \right)^2 + 12 \left(\frac{m_V}{l \omega} \right)^4 \right] \sqrt{1\!-\! \frac{4 m_V^2}{(l \omega)^2}}\,,
       \nonumber\\
        \Gamma_{\phi \rightarrow aa} &= \frac{1+n}{2n} \frac{1 }{8 \pi}  \left(\frac{h_{a \phi}}{M_P} \right)^2\frac{\omega M_P^4}{\rho_\phi} \left(\frac{\rho_\phi}{\Lambda^4} \right)^{\frac{1}{n}}  \sum_{l=1}^\infty  l \lvert \mathcal{P}_l \rvert^2 (l \omega)^4 \left( 1- \frac{2 m_a^2}{(l \omega)^2} \right)^2 \sqrt{1\!-\! \frac{4 m_a^2}{(l \omega)^2}}, \nonumber
\end{align}
In the limit $m_F \rightarrow 0$, these equations simplify 
\begin{align}\label{eq:decay-massls}
     \Gamma_{\phi \rightarrow SS} &= \frac{1+n}{2n} \frac{\omega }{2\pi}  \left(\frac{g_{S \phi} M_P}{\Lambda^2} \right)^2 \left(\frac{\rho_\phi}{\Lambda^4} \right)^{\frac{1-n}{n}}  \sum_{l=1}^\infty  l \lvert \mathcal{P}_l \rvert^2\,, 
     \nonumber\\
      \Gamma_{\phi \rightarrow \psi \bar{\psi}} &= \frac{1+n}{2n}  \omega \cdot  \frac{g_{\psi \phi}^2}{4\pi}  \left(\frac{\omega M_P}{\Lambda^2} \right)^2 \left(\frac{\rho_\phi}{\Lambda^4} \right)^{\frac{1-n}{n}}  \sum_{l=1}^\infty  l^3 \lvert \mathcal{P}_l \rvert^2\,, 
      \nonumber\\
      \Gamma_{\phi \rightarrow\mathcal{V}\mathcal{V}} &= 2 \cdot  \Gamma_{\phi \rightarrow SS}  (g_{S \phi} \leftrightarrow g_{V \phi})\,,
      \nonumber\\
      \Gamma_{\phi \rightarrow aa} &= \frac{1+n}{2n} \frac{\omega }{8 \pi}  \left(\frac{ \omega^2 M_P h_{a \phi}}{\Lambda^2} \right)^2 \left(\frac{\rho_\phi}{\Lambda^4} \right)^{\frac{1-n}{n}}  \sum_{l=1}^\infty  l^5 \lvert \mathcal{P}_l \rvert^,,
\end{align}
where $V$ and $\mathcal{V}$ denote the massive and massless gauge fields, respectively. For the T-model, the frequency $\omega$ is related to the effective mass (\ref{eq:sr-param}) of the inflaton field through
\begin{align}
    \omega = m_\phi\,\sqrt{\frac{ \pi n}{2n-1}}\,\frac{\Gamma \left( \frac{n+1}{2n}\right)}{\Gamma \left( \frac{1}{2n}\right)}\,.
\end{align}
Its time-evolution is described by the power-law solution with the scale factor
\begin{align}
    &\omega = \omega_e \cdot  \left(\frac{\rho_\phi}{\Lambda^4}\right)^{\frac{n-1}{2n}} \simeq \omega_e \left( \frac{\rho_e}{\Lambda^4}\right)^{\frac{n-1}{2n}} \left(\frac{a_e}{a}\right)^{\frac{3(n-1)}{n+1}}, 
    &\omega_e \equiv \sqrt{2\,n^2\,\pi }\, \frac{\Gamma \left( \frac{n+1}{2n}\right)}{\Gamma \left( \frac{1}{2n}\right)}\, \frac{\Lambda^2}{M_P}.
\end{align}
Hence, the time dependency of the inflaton decay rate can be parametrized as 
\begin{align}
    \Gamma_{\phi \rightarrow FF} \equiv \Gamma_{\phi \rightarrow FF}^e \left(\frac{a_e}{a} \right)^{\beta_F}, 
\end{align}
where 
\begin{align}
  &\Gamma_{\phi \rightarrow SS}^e= \frac{1+n}{2n} \frac{\omega_e }{2\pi}  \left(\frac{g_{S \phi} M_P}{\Lambda^2} \right)^2 \left(\frac{\rho_e}{\Lambda^4} \right)^{\frac{1-n}{2n}}  \sum_{l=1}^\infty  l \lvert \mathcal{P}_l \rvert^2\,, \nonumber\\ &\Gamma_{\phi \rightarrow \psi \bar{\psi}}^e= \omega_e \cdot \frac{1+n}{2n}   \frac{g_{\psi \phi}^2}{4\pi}  \left(\frac{\omega_e M_P}{\Lambda^2} \right)^2 \left(\frac{\rho_e}{\Lambda^4} \right)^{\frac{n-1}{2n}}  \sum_{l=1}^\infty  l^3 \lvert \mathcal{P}_l \rvert^2\, \nonumber\\
  &\Gamma_{\phi \rightarrow\mathcal{V}\mathcal{V}}^e= 2 \cdot \Gamma_{\phi \rightarrow SS}^e (g_{S \phi} \leftrightarrow g_{V \phi}) \nonumber\\  &\Gamma_{\phi \rightarrow aa}^e= \frac{1+n}{2n} \frac{\omega_e }{8 \pi}  \left(\frac{ \omega_e^2 M_P h_{a \phi}}{\Lambda^2} \right)^2 \left(\frac{\rho_e}{\Lambda^4} \right)^{\frac{3n-3}{2n}}  \sum_{l=1}^\infty  l^5 \lvert \mathcal{P}_l \rvert^2, \label{eq:Gammae}
\end{align}
and 
\begin{align}
  &\beta_S= \frac{3(1-n)}{1+n}, &&\beta_\psi=\frac{3(n-1)}{1+n}, &&&\beta_v=\beta_S,  &&&&\beta_a=\frac{9(n-1)}{1+n}\,.   
\end{align}
%%%%%%%%%%%%%%%%%%%%%%%%%%%%%%
\begin{table}[t!]
\centering
$
\begin{array}{|c||c|c|c|c|c|c|c|c|c|c|c|}
\hline
n & 1 & 2 & 3 & 4 & 5  \\
\hline
\sum_l l |\mathcal{P}_l|^2 & 1/4 & 0.2294 & 0.2101 & 0.2177 & 0.2047  \\
\hline
\sum_l l^3 |\mathcal{P}_l|^2 & 1/4 & 0.2406 & 0.2442 & 0.2504 & 0.2571  \\
\hline
\sum_l l^5 |\mathcal{P}_l|^2 & 1/4 & 0.3433 & 0.5254 & 0.762 & 1.0492  \\
\hline
\end{array}
$\vspace{7pt}
\caption{Numerical values of the summation factors appearing in Eq.~\eqref{eq:Gammae},  for different values of $n$.}
\vspace{7pt}
\label{tab:PkSums}
\end{table}
%%%%%%%%%%%%%%%%%%%%%%%%%%%%%%%%%
%%%%%%%%%%%%%%%%%
\section{Constraints on reheating from inflation}
\label{sec:app-inf}
%%%%%%%%%%%%%%%%%%%%
Here we derive constraints of the $\alpha$-attractor T-model from the combined WMAP, Planck, and BICEP/Keck data \cite{Planck:2018jri, BICEP:2021xfz, Tristram:2021tvh}. 
Let us start with revisiting some of the necessary inflationary parameters. 
The so-called potential slow-roll parameters are defined as
\begin{align}\label{eq:cmb-SR}
&  \epsilon_V(\phi) \equiv \frac{M_P^2}{2}\,\left(\frac{V_{,\phi}(\phi)}{V(\phi)}\right)^2, &\eta_V(\phi) \equiv M_P^2 \frac{V_{,\phi \phi}(\phi)}{V(\phi)}\,.   
\end{align}
Thus, for the $\alpha$-attractor T-model, we find
\begin{align}
    &\epsilon_V^T (\phi) = \frac{4 n^2}{3 \alpha } \operatorname{csch}^2\left( \sqrt{\frac{2}{3 \alpha}} \frac{\phi}{M_P}\right), \label{eq:epsilon_T} \\
    &\eta_V^T (\phi) = \frac{4 n}{3 \alpha } \left[2n-\operatorname{cosh} \left( \sqrt{\frac{2}{3 \alpha}} \frac{\phi}{M_P} \right)\right] \operatorname{csch}^2\left(\sqrt{\frac{2}{3 \alpha}} \frac{\phi}{M_P} \right)\,. \label{eq:eta_T}
\end{align}
Note that at the end of inflation $\ddot{a}=0$, which roughly corresponds to $\epsilon_V(\phi_e)\simeq 1$. This condition allows us to find the inflaton field value at the end of inflation
\begin{align}
    \phi_e \simeq \sqrt{\frac{3 \alpha}{2}} M_P \sinh ^{-1}\left(\frac{2 n}{\sqrt{3\alpha}}\right). \label{eq:phi_e}
\end{align}
The inflationary number of e-folds 
between the horizon crossing of the perturbation with a comoving wave number $k_\star$ and the end of inflation is
\begin{align}
& N_\star \simeq \frac{1}{M_P}\int_{\phi_e}^{\phi_\star} \frac{d\phi}{\sqrt{2 \epsilon_V^T(\phi)}} = \frac{3\,\alpha}{4\,n}\,\left[\cosh\left(\sqrt{\frac{2}{3\,\alpha}}\,\frac{\phi_\star}{M_P}\right)-\cosh\left(\sqrt{\frac{2}{3\,\alpha}}\,\frac{\phi_e}{M_P}\right)\right]\,,
\end{align}
with $\phi_\star \equiv \phi(a_\star)$ being the field value at the moment when the Planck pivot scale $k_\star = 0.05 \, {\rm{Mpc}}^{-1}$ crosses the comoving Hubble radius, i.e., $k^{-1}_\star = (a_\star H_\star)^{-1}$. 

The two principal CMB observables, namely, the spectral index, $n_S$, and the tensor-to-scalar ratio, $r$, in the slow-roll approximation are defined as
\begin{align}
    &r(\phi_\star)= 16 \epsilon_V(\phi_\star), &n_S(\phi_\star) -1 = 2 \eta_V(\phi_\star) - 6 \epsilon_V(\phi_\star). 
\end{align}
Using \eqref{eq:epsilon_T} and \eqref{eq:eta_T}, we get
\begin{align}
    r^T(\phi_\star) &= \frac{64 n^2}{3 \alpha } \operatorname{csch}^2\left( \sqrt{\frac{2}{3 \alpha}} \frac{\phi_\star}{M_P}\right), \\ n_S^T(\phi_\star) -1 &= -\frac{8 n}{3 \alpha } \operatorname{csch}^2\left(\sqrt{\frac{2}{3 \alpha}} \frac{\phi_\star}{M_P}\right) \left[\operatorname{cosh} \left(\sqrt{\frac{2}{3 \alpha}} \frac{\phi_\star}{M_P}\right)+n\right]\, \label{eq:nS_T}.
\end{align}
Next, one can use the above expression for the spectral tilt $n_S^T$ \eqref{eq:nS_T} to find the value of $\phi_\star$ as a function of $\alpha, n$, and $n_S^T$
\begin{align}
    \phi_\star &=\sqrt{\frac{3 \alpha}{2}} M_P \left\{- \log\left[3 \alpha (1- n_S^T)\right] + \log \left[4n + \sqrt{9 \alpha^2 (1-  n_S^T)^2 + 8n^2(2 + 3 \alpha (1- n_S^T) )} \right. \right. \nonumber \\
   &+\left.\left. 
    \sqrt{8n\left[ n (4+3\alpha  (1- n_S^T)) + \left.\sqrt{8 n^2 (2+3 \alpha(1 -  n_S^T))+9 \alpha ^2 (1- n_S^T)^2}\right)\right]} \,
   \right] \right\}\,. \label{eq:phi_star}
\end{align}
The measured value of the spectral tilt $n_S$ by the Planck satellite for the pivot scale $k_\star$ is $n_S^{\rm{obs}} = 0.9649 \pm 0.0042$ \cite{Planck:2018jri}. The numerical values of $\phi_\star$ \eqref{eq:phi_star} and $\phi_e$ \eqref{eq:phi_e} for different values of $n$ and fixed value of $\alpha=1/6$ are collected in Table \eqref{tab:phiVal}.
%%%%%%%%%%%%%%%%%%%%%%%%%%%%%%
\begin{table}[t!]
\centering
$
\begin{array}{|c||c|c|c|c|c|c|c|c|c|c|c|}
\hline
n & 1 & 2 & 3 & 4 & 5  \\
\hline 
\phi_e/M_P & 0.88 & 1.21 & 1.41 & 1.56 & 1.67  \\
\hline
\phi_\star/M_P & 3.40 & 3.75 & 3.95  & 4.10  & 4.21  \\
\hline
N_\star & 56.73 & 56.74 & 56.74  & 56.74  & 56.75  \\
\hline
\end{array}
$\vspace{1pt}
\caption{Numerical values of the inflaton field amplitude and the corresponding number of e-folds $N_\star$ for different values of $n$ and fixed $\alpha=1/6$.}
\vspace{1pt}
\label{tab:phiVal}
\end{table}
%%%%%%%%%%%%%%%%%%%%%%%%%%%%%%%%%

The CMB bound on $r$ allows us also to constrain the scale of inflation $\Lambda$ from above. The tensor-to-scalar ratio is defined as
\begin{align}
    r \equiv \frac{\Delta_t^2(k_\star)}{\Delta_s^2(k_\star)},
\end{align}
with 
\begin{align}
    &\Delta_t^2 (k_\star) = \frac{2}{\pi^2}\,\frac{H_\star^2}{M_P^2}, & \Delta_s^2 (k_\star) = \frac{1}{8 \pi^2}\,\frac{H_\star^2}{M_P^2} \frac{1}{\epsilon_\star} \label{eq:PS},
\end{align}
denoting the dimensionless tensor and scalar power-spectra, respectively. Above, $\epsilon_\star \equiv - \dot{H}_\star/H^2_\star$ is the (Hubble) slow-roll parameter. 
The amplitude of the scalar power spectrum measured by Planck at $k=k_\star$ is $\Delta_s^2(k_\star) = 2.1 \times 10^{-9}$ \cite{Planck:2018jri}, which, in turn, implies \mbox{$\Delta_t^2(k_\star) \leq 6.7 \times 10^{-11}$.} Utilizing \eqref{eq:PS}
one gets the upper bound on the Hubble rate
\begin{align}
    H_\star \simeq H_e \leq 4.4 \times 10^{13} \, \rm{GeV},\
\end{align}
which, in turn, allow us to constraint the inflaton energy density at the end of inflation
\begin{align}
    \rho_\phi^e \equiv \rho_\phi(a_e) = 3 M_P^2 H_e^2 \leq 3.4 \times 10^{64} \, \rm{GeV^4}.
\end{align}
Moreover, the inflaton potential at $a=a_\star$ can be expressed in terms of the scalar power-spectrum and tensor-to-scalar ratio as
\begin{align}
    V(\phi_\star) = \frac{3 \pi^2}{2} M_P^4\, \Delta_s^2 r,
\end{align}
which implies
\begin{align}
    \Lambda = \frac{M_P}{\tanh^{n/2}\left[\phi_\star/(\sqrt{6 \alpha}M_P)\right]} \cdot \left( \frac{3 \pi^2}{2} \Delta_s^2 r \right)^{1/4}
\end{align}
Since, at $a=a_\star$ the field value $\phi_\star > M_P$, the inflaton potential can be approximated by a constant value $V(\phi_\star) \approx \Lambda^4$, which in turn sets an upper bound on the inflationary scale, $\Lambda\leq 1.34\times 10^{16}$ GeV, which has a feeble dependence on $n, \alpha$. 

% %%%%%%%%%%%%%%%%%
% \section{Signal-to-noise Ratio for GW Detectors}
% \label{sec:SNR}
% %%%%%%%%%%%%%%%%%
% Interferometers as Gravitational Wave experiments measure displacements in terms of a so called dimensionless strain-noise $h_\text{GW}(f)$ which is related to the GW amplitude and can be converted into the corresponding  energy density \cite{Garcia-Bellido:2021zgu}
% \begin{align}
%     \Omega_\text{exp}(f) h^2 = \frac{2\pi^2 f^2}{3 H_0^2} h_\text{GW}(f)^2 h^2,
% \end{align}
% with $H_0 = h\times 100 \;\text{(km/s)}/\text{Mpc}$ being the Hubble rate today. 
% We compute the signal-to-noise ratio (SNR) for a given or projected experimental sensitivity $\Omega_\text{exp}(f)h^2$ in order to assess the detection probability of the primordial GW background via the following prescription~\cite{Thrane:2013oya,Caprini:2015zlo}
% \begin{align}
%      \text{SNR}\equiv \sqrt{\tau \int_{f_\text{min}}^{f_\text{max}} \text{d}f \left(\frac{ \Omega_\text{GW}(f) h^2}{\Omega_\text{exp}(f) h^2}\right)^2 } \label{eq:SNR},
% \end{align}
% where $h=0.7$ and  $\tau = 4\; \text{years}$ is the observation time we have taken. 

% %%%%%%%%%%%%%%%%%%%%%%%%%%
% \begin{figure}[htb!]
%     \centering
%     \includegraphics[scale=0.35]{figure/SNR-v1.pdf}
%     \caption{Plot of SNR versus $\Gamma_\phi^e$ for BBO and DECIGO detectors. If we consider $\text{SNR}\geq 1$ as the  detection threshold then we see the correcpnding inflaton decay widths which will be within the reach of each detector. }
%     \label{fig:stiff}
% \end{figure}
% %%%%%%%%%%%%%%%%%%%%%%%%%%%%%%%%

%%%%%%%%%%%%%%%%%%%%
\bibliographystyle{JHEP}
\bibliography{Bibliography}
%%%%%%%%%%%%%%%%%%

\end{document}